\documentclass[a4paper,twoside,prd,showpacs,preprintnumbers,superscriptaddress,nofootinbib]{revtex4-2}
\usepackage{amssymb,amsmath,bm,natbib}
\usepackage{color}
\usepackage{slashed}
\usepackage{graphics}
\usepackage{graphicx}
\usepackage[utf8]{inputenc}
\usepackage[caption=false]{subfig}
\usepackage{hyperref}
\usepackage{url}
\usepackage{cleveref}
\usepackage{makecell}
\usepackage{ulem}

\newcommand{\Lag}{{\mathcal L}}

\definecolor{coolblack}{rgb}{0.0, 0.18, 0.39}
\newcommand\cbc[1]{{\color{coolblack}{#1}}}

\begin{document}

\title{Deconstructing resonant Higgs pair production at the LHC:  \\ effects of coloured and neutral scalars in the NMSSM test case}

\author{Stefano Moretti}
\email{s.moretti@soton.ac.uk; stefano.moretti@physics.uu.se}
\affiliation{School of Physics and Astronomy, University of Southampton, Highfield, Southampton SO17 1BJ, UK}
\affiliation{Department of Physics and Astronomy, Uppsala University, Box 516, 751 20 Uppsala, Sweden}

\author{Luca Panizzi}
\email{luca.panizzi@unical.it}
\affiliation{Dipartimento di Fisica, Universit\`a della Calabria, 8703 Arcavacata di Rende, Cosenza, Italy}

\author{J\"orgen Sj\"olin}
\email{sjolin@fysik.su.se}
\affiliation{Department of Physics, Stockholm University, 10691, Stockholm, Sweden}

\author{Harri Waltari}
\email{harri.waltari@physics.uu.se}
\affiliation{Department of Physics and Astronomy, Uppsala University, Box 516, SE-751 20 Uppsala, Sweden}
\date{\today}

\begin{abstract}
We study resonant production of pairs of Standard Model (SM)-like Higgs bosons, in the presence of new neutral Higgs states together with new coloured scalars (stops or sbottoms) in loops within the Next-to-Minimal Supersymmetric SM (NMSSM). This is used as a test case to prove that the Large Hadron Collider has sensitivity to a variety of effects stemming from interferences between resonant (heavy) Higgs diagrams and/or among these and non-resonant topologies involving loops of both tops and stops. These effects can alter significantly the naive description of individual $s$-channel Breit-Wigner resonances, leading to distortions of the latter which, on the one hand, may mask their presence but, on the other hand, could enable one to extract features of the underlying new physics scenario. This last aspect is made possible through a decomposition of the $gg\to hh$  signal process into all its amplitude components, each of which has a well-defined coupling structure. Ultimately, such effects can be traced back to the relevant Feynman diagrams and can enable a detailed interpretation of this process. To illustrate this, we introduce various Benchmark Points that exhibit potentially observable features during the current and/or upcoming runs of the LHC in one or more of the three customary di-Higgs decay channels: $b\bar bb\bar b$, $b\bar b \tau^+\tau^-$ and $b\bar b\gamma\gamma$.
\end{abstract}

\maketitle

\section{Introduction}

Within the Standard Model (SM), Higgs pair, or di-Higgs, ($hh$) production is at Leading Order (LO) a one-loop process with destructive interference between the box and the triangle diagrams \cite{Glover:1987nx,Dicus:1987ic}, which thus makes it particularly sensitive to Beyond the SM (BSM) physics effects, which enter at the same perturbative order. The largest of such effects are modifications to the trilinear Higgs 
self-coupling \cite{Kanemura:2002vm,Noble:2007kk,Rodejohann:2012px,Wu:2015nba,DiLuzio:2017tfn,Bahl:2022jnx}, the possibility of new colored particles in the loops  \cite{Batell:2015koa,Huang:2017nnw,Moretti:2023dlx,DeCurtis:2023pus} or the existence of new Higgs bosons, which can give an $s$-channel resonant contribution to di-Higgs production \cite{Plehn:1996wb,Dolan:2012ac,No:2013wsa,Dawson:2015haa}. In principle, the cross section can also be modified by a deviation of the top quark Yukawa coupling from its SM value but, in practice, such effects are already somewhat constrained as the $t\overline{t}h$ production cross section does not show a large deviation from the SM prediction \cite{CMS:2020cga,ATLAS:2020ior,ATLAS:2024gth}.

The phenomenological importance of SM Higgs pair  production at the Large Hadron Collider (LHC) is  primarily related to the ability of accessing the aforementioned trilinear Higgs 
self-coupling, $hhh$, and modifications thereof (or indeed additional effects emerging in BSM scenarios). This is of paramount importance, as it gives insights into the shape of the  Higgs potential, thereby enabling one to potentially pinpoint the underlying Higgs mechanism of spontaneous symmetry breaking.  The latter ought to be embedded in some viable theory of the Electro-Weak (EW) sector, by which one intends a theoretical construct able to remedy the so-called `hierarchy problem' of the SM. A prevalent theory in this connection is supersymmetry (SUSY).

In a previous work \cite{Moretti:2023dlx}, we looked at non-resonant SM-like di-Higgs production in the context of SUSY. In this work we concentrate instead on the aforementioned $s$-channel resonant contributions and their interferences with other diagrams that contribute to di-Higgs production in the same theoretical framework. To gain better understanding of all effects, we use the deconstruction approach we previously used to describe squark effects in non-resonant SM-like di-Higgs production \cite{Moretti:2023dlx}, now extended to include the corresponding resonant case. Interference effects are known to be potentially important in various processes, with recent studies also including di-Higgs production \cite{Dawson:2016ugw,Carena:2018vpt,Feuerstake:2024uxs}. Our approach allows us to show the interference patterns clearly and, in addition, we shall discuss what kind of interpretations can be done within specific models if an interference pattern is observed.

We study specifically the NMSSM \cite{Ellwanger:2009dp} (see Ref.~\cite{Moretti:2019ulc} for a review of alternative realisations of it) for a number of reasons. Concerning  the aforementioned hierarchy (or fine-tuning) problem of the SM, SUSY is a well-motivated framework for BSM physics in relation to the Higgs sector as its symmetry between fermions and bosons produces a cancellation of loop contributions to the Higgs mass making a light Higgs boson natural, thereby elegantly solving this flaw of the SM. In fact, another appealing future of SUSY is that the Higgs potential is therein generated dynamically (instead, in the SM, it is put in by hand).  Concerning model realisations of SUSY, it should be noted that 
the parameter space of the MSSM, the simplest realisation of it, is constrained when it comes to resonant di-Higgs production as a $125$~GeV Higgs state requires a large value of $\tan\beta$. At large $\tan\beta$ the constraints from single Higgs boson searches force the heavier Higgs states to be beyond a TeV \cite{ATLAS:2020zms,CMS:2022goy}, including the heavy CP-even one, $H$, which will then suppress the corresponding signal $gg\to H\to hh$. The NMSSM has both a singlet Higgs field, for which the constraints are weaker, and the possibility of having a $125$~GeV Higgs at lower values of $\tan\beta$ \cite{Drees:1988fc}, which is also a less constrained region. Finally, a feature of SUSY models is that their Higgs self-couplings have less freedom than in non-SUSY scenario, which will help in the interpretation of di-Higgs results, as discussed below.

Higgs pair production in the NMSSM has been studied before. Refs.~\cite{Ellwanger:2013ova,Heng:2018kyd} tackled the case where the Higgs pairs $H_{1}H_{1}$, $H_{1}H_{2}$ and $H_{2}H_{2}$ were produced, when $H_{2}$ is the SM-like Higgs boson (the aforementioned $h$ state) and $H_{1}$ is dominantly a singlet state. In \cite{Cao:2013si,Heng:2013wia,Cao:2014kya}, the authors studied Higgs pair production with both mass orderings but only at the total cross section level without specifying resonant and non-resonant contributions. The non-resonant, resonant and interference contributions were addressed in \cite{Huang:2019bcs}, but the interference patterns were not studied therein. Here, we will concentrate on the case where the $125$~GeV Higgs boson pairs will be produced and our key addition to  existing studies is to deconstruct the BSM effects to the differential cross section and to show what are the interference patterns between the different contributions.

We stress, however, that the NMSSM scenarios addressed  here are only some of the possible examples which can be studied with the deconstruction method. With the same dataset we have produced, it is possible to study scenarios involving {\it any} number of coloured scalars (of any representation of $SU(3)$) and neutral scalars. A subset of the MC event samples we used for this analysis corresponds to those used in \cite{Moretti:2023dlx} and, since the samples involving the neutral scalars have been modularly added to the dataset, it is possible to select only these ones and study scenarios involving only neutral scalars, again, in any number, thereby allowing for full flexibility within this particle content. (New coloured fermions will also be  included in due course, {\it e.g.}, to allow for the effects of Vector-Like Quarks (VLQs) in Compositeness \cite{DeCurtis:2023pus}.)

The plan of the paper is as follows. In the next section we explain our deconstruction of the $gg\rightarrow hh$ signal. We then proceed to review triple-Higgs couplings in the NMSSM, following which we will discuss our results. Conclusions will then follow.

\section{Deconstruction of the signal cross section}
\label{sec:deconstruction}

The di-Higgs signal can be deconstructed in a finite number of kinematically independent contributions, which can be simulated individually and recombined a posteriori to obtain the full signal. This strategy, adopted also in \cite{Moretti:2023dlx}, allows one to assess the role of each contribution in the determination of the kinematical properties of the final state. At the same time, it shows how specific choices of coupling values weight differently the various amplitude contributions for a given set of kinematically relevant parameters (masses, widths and spins). In this analysis we consider new coloured and new neutral scalars propagating in the loops at the same time as SM objects.

The starting point is to write the simplified Lagrangians, \Cref{eq:LagNP}, which minimally extend the SM with modifications of the trilinear $hhh$ coupling and of the Yukawa couplings of the Higgs with SM top and bottom quarks, alongside the contribution of any number of coloured scalars $\tilde s_i$ and of new neutral scalars $S^0_I$,
\begin{subequations}
\begin{align}
\Lag_{M} &= -(\lambda^{\rm SM} + \kappa_{hhh}) v h^3 - {1\over\sqrt2}(y_f^{\rm SM} + \kappa_{hff}) h\bar f f\;, \\
\Lag_{\tilde s} &= \sum_i \kappa_{h\tilde s \tilde s}^{ii} v~h \tilde s_i^* \tilde s_i +
\kappa_{hh\tilde s \tilde s}^{ii}~h h \tilde s_i^* \tilde s_i + \left(\sum_{i>j} \kappa_{h\tilde s \tilde s}^{ij} v~h \tilde s_i^* \tilde s_j + h.c.\right) \;,\\
\Lag_{S} &= \sum_I \kappa_{Shh}^I v S^0_I h h + \kappa_{Sff}^I S^0_I \bar f f \;,\\
\Lag_{S\tilde s} &= \sum_{I,i} \kappa_{S\tilde s \tilde s}^{Ii} v S^0_I \tilde s_i^* \tilde s_i\;,
\end{align}
\label{eq:LagNP}
\end{subequations}
\noindent where all couplings $\kappa$ are independent dimensionless parameters (a Higgs 
Vacuum Expectation Value (VEV) has been factorised for all dimension-3 operators) and we have not included couplings between a neutral scalar and two different coloured scalars because this vertex does not appear in the di-Higgs process at leading order. It is important to notice that in this Lagrangian we are working with mass eigenstates and that the coloured scalars do not necessarily need to be in the fundamental representation of $SU(3)$. Our analysis will indeed  consider squarks but the method we use can be straightforwardly generalised to other cases, such as scalar sextets, octets and so on.

The signal and interferences contributions to the cross section $\sigma$ can be parameterised in terms of a sum of different terms, proportional to unique functions of the couplings and to {\it reduced} cross sections $\hat\sigma$ depending exclusively on the masses of all scalars and on the width of the (potentially resonant) neutral scalar. The topologies into which the signal can be deconstructed are represented in \Cref{tab:deconstructedtopologies}, where the proportionality of the amplitude with respect to the couplings is indicated. The explicit expressions of all unique contributions to the cross-section are provided in \Cref{app:xsdec}.

\begin{table}[]
\centering\small
{\setlength\arraycolsep{5pt}
\hspace*{-25pt}\begin{tabular}{cccc}
\hline\hline
\noalign{\vskip 3pt}
& \textbf{Topology type} & \textbf{Feynman diagrams} & \textbf{Amplitude} \\
\noalign{\vskip 3pt}

\hline
\hline
\noalign{\vskip 5pt}
\textbf{1} & \textbf{Modified $hhh$ coupling} &
\raisebox{-.5\height}{\includegraphics[width=.16\textwidth]{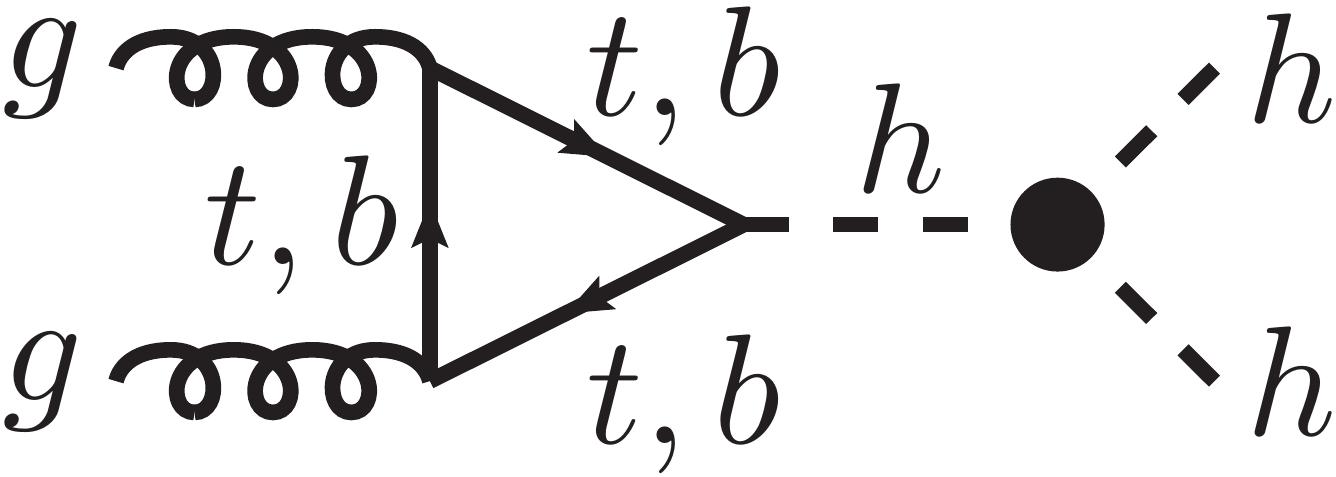}}
&
$\mathcal A_i \propto \kappa_{hhh}$\\
\noalign{\vskip 3pt}
\hline
\noalign{\vskip 3pt}
\textbf{2} & \textbf{One modified $hff$ coupling} &
\raisebox{-.35\height}{\includegraphics[width=.16\textwidth]{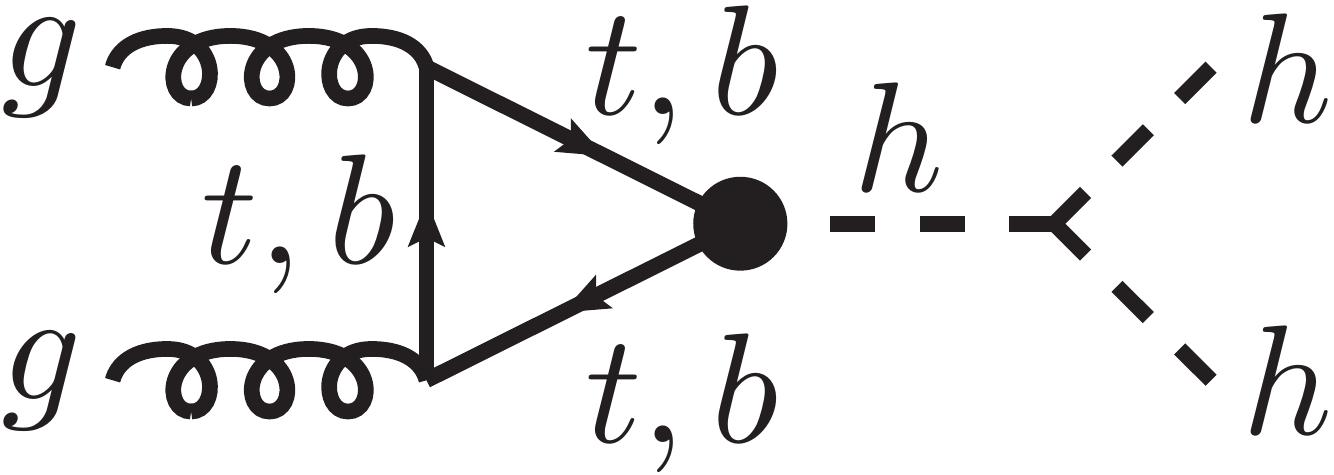}}
\raisebox{-.45\height}{\includegraphics[width=.16\textwidth]{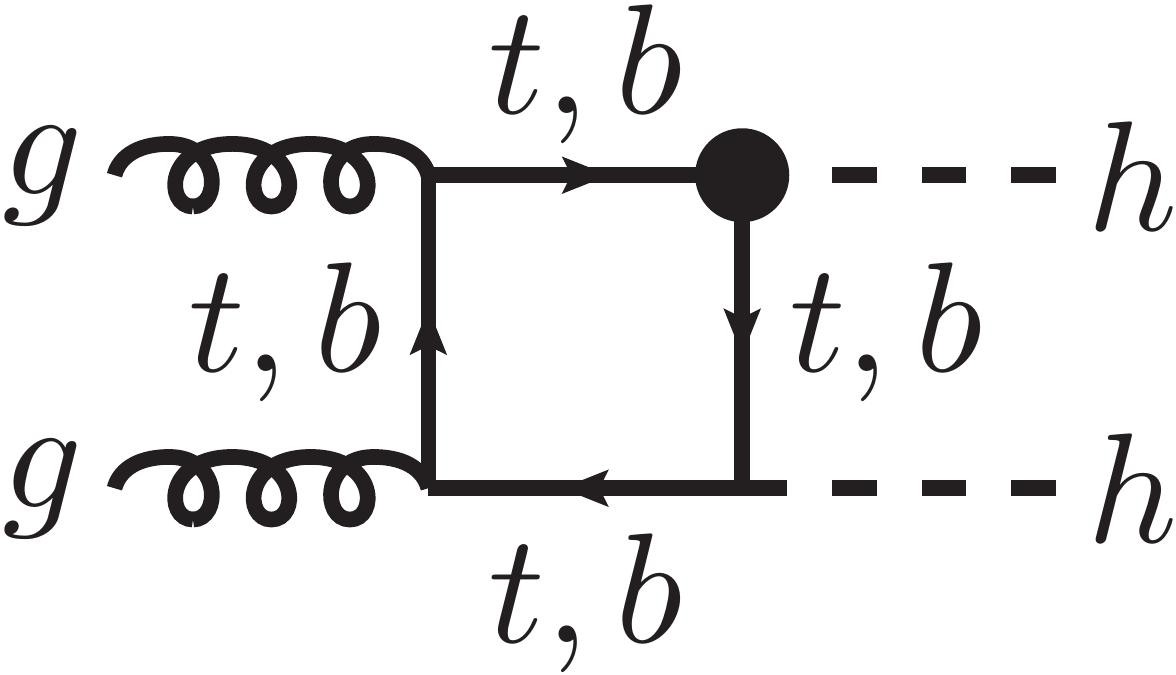}}
&
$\mathcal A_i \propto \kappa_{hff}$ \\
\noalign{\vskip 3pt}
\hline
\noalign{\vskip 5pt}
\textbf{3} & 
$\begin{array}{c}
\text{\bf Modified $hhh$ coupling}\\
\text{\bf and modified $hff$ coupling}
\end{array}$ 
&
\raisebox{-.4\height}{\includegraphics[width=.16\textwidth]{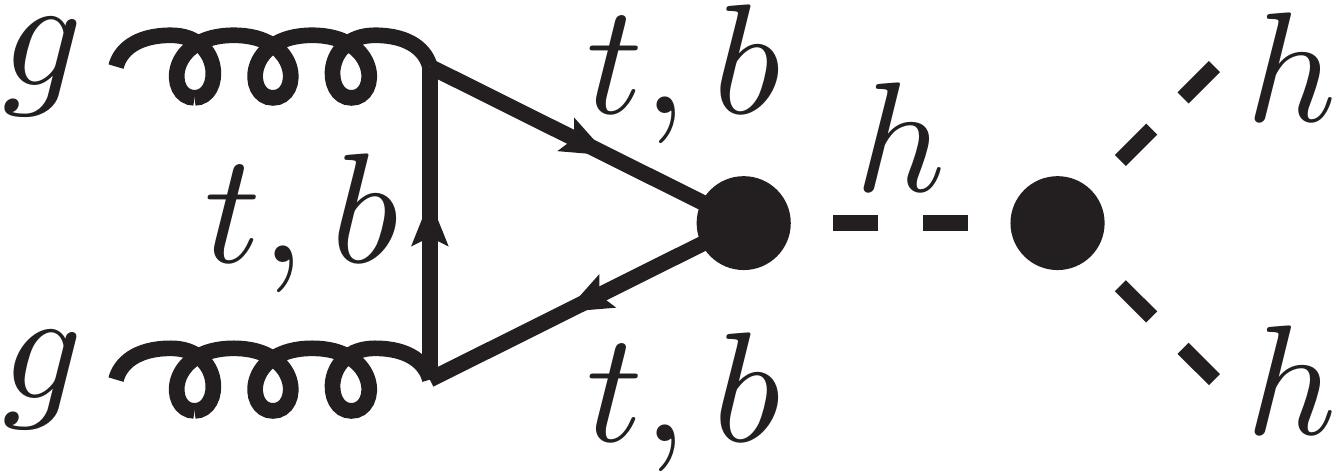}} &
$\mathcal A_i \propto \kappa_{hhh} \kappa_{hff}$\\
\noalign{\vskip 5pt}
\hline
\noalign{\vskip 3pt}
\textbf{4} & \textbf{Two modified $hff$ couplings}&
\begin{minipage}{.16\textwidth}
\includegraphics[width=\textwidth]{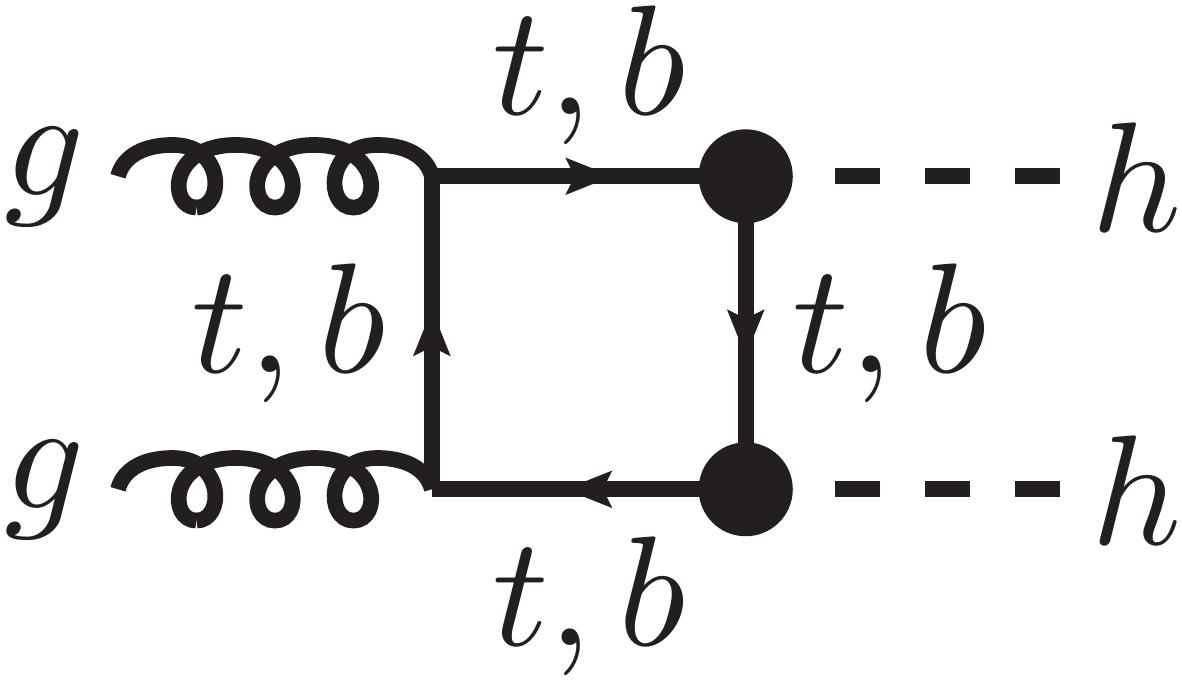} 
\end{minipage} &
$\mathcal A_i \propto \kappa_{hff}^2$ \\
\noalign{\vskip 2pt}

\hline
\hline

\noalign{\vskip 3pt}
\textbf{5} & 
$\begin{array}{c}
\text{\bf Scalar bubble and triangle}\\
\text{\bf with $h\tilde s\tilde s$ couplings}
\end{array}$ 
&
\raisebox{-.5\height}{\includegraphics[width=.16\textwidth]{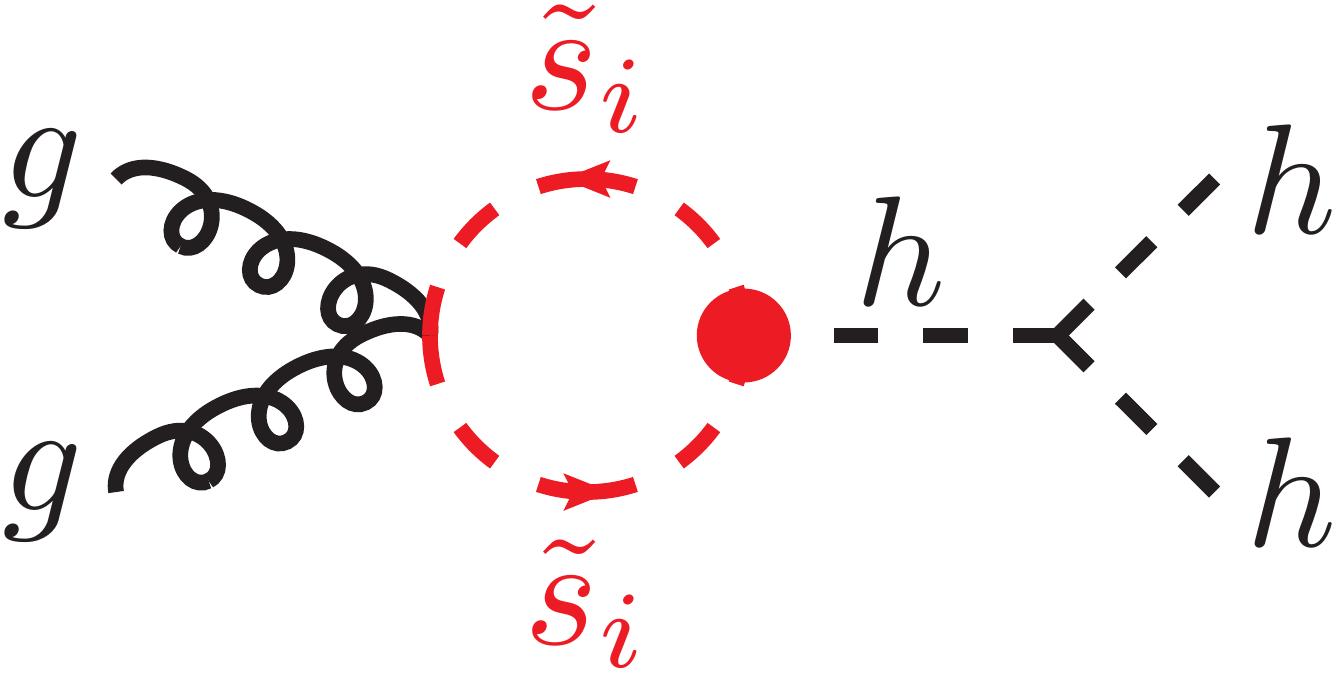}}
\raisebox{-.52\height}{\includegraphics[width=.16\textwidth]{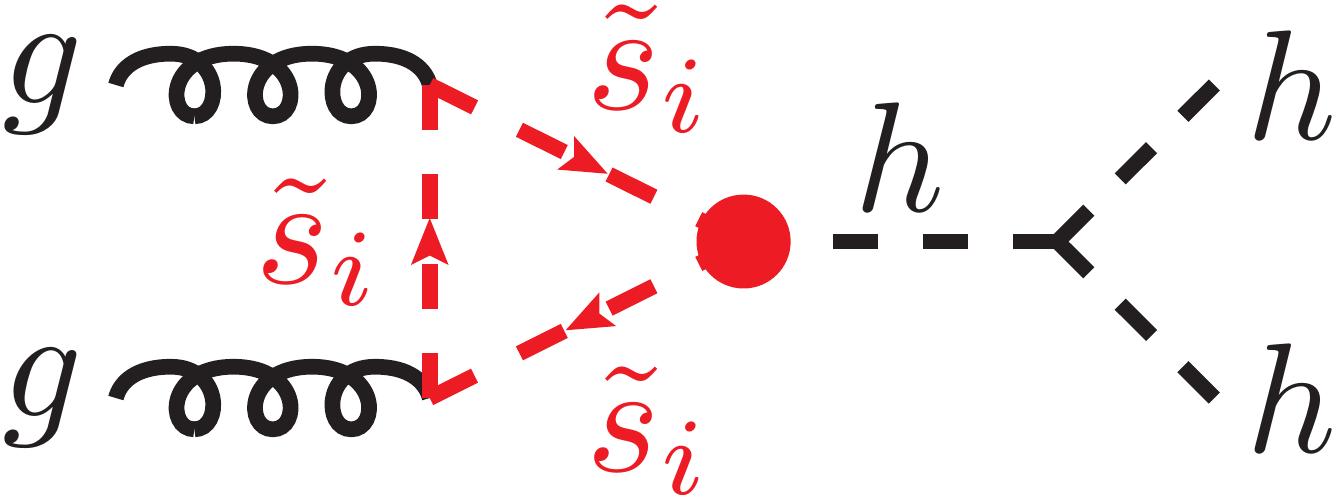}} &
$\mathcal A_i \propto \kappa_{h\tilde s\tilde s}^{ii}$ \\
\noalign{\vskip 3pt}
\hline
\noalign{\vskip 3pt}
\textbf{6} & 
$\begin{array}{c}
\text{\bf Modified $hhh$ coupling +}\\[-3pt]
\text{\bf Scalar bubble and triangle}\\[-3pt]
\text{\bf with $h\tilde s\tilde s$ coupling}
\end{array}$ & 
\raisebox{-.45\height}{\includegraphics[width=.16\textwidth]{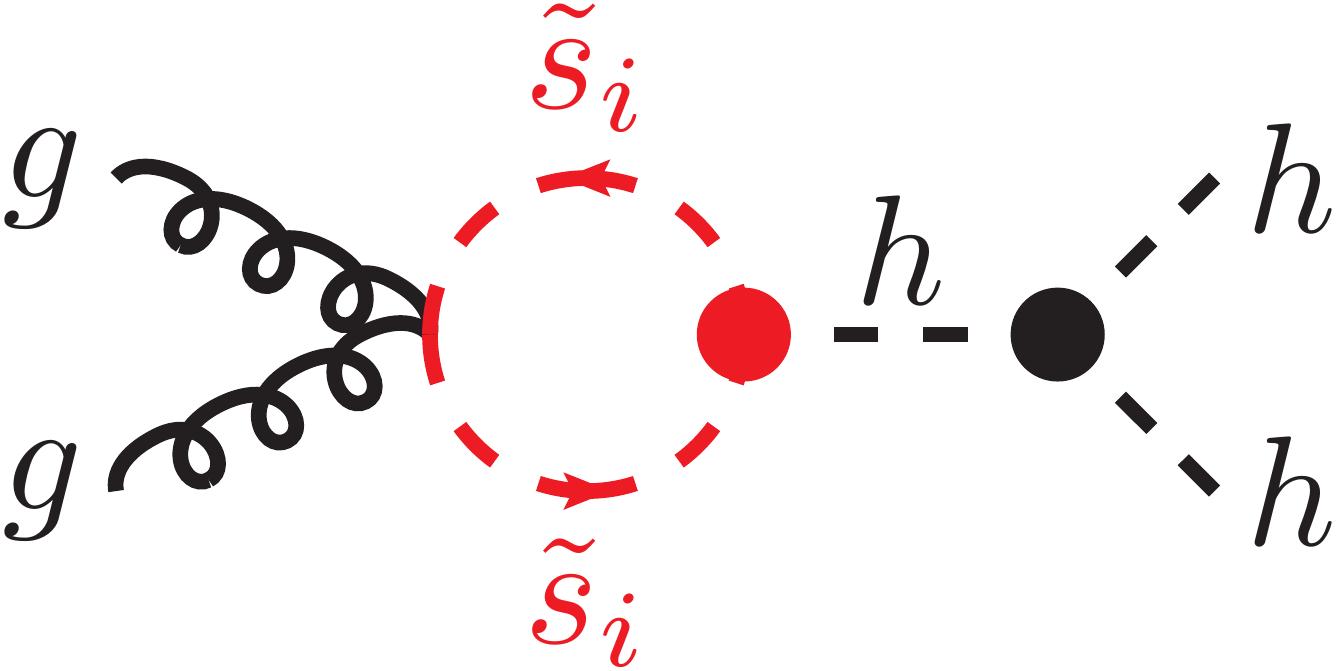}}
\raisebox{-.47\height}{\includegraphics[width=.16\textwidth]{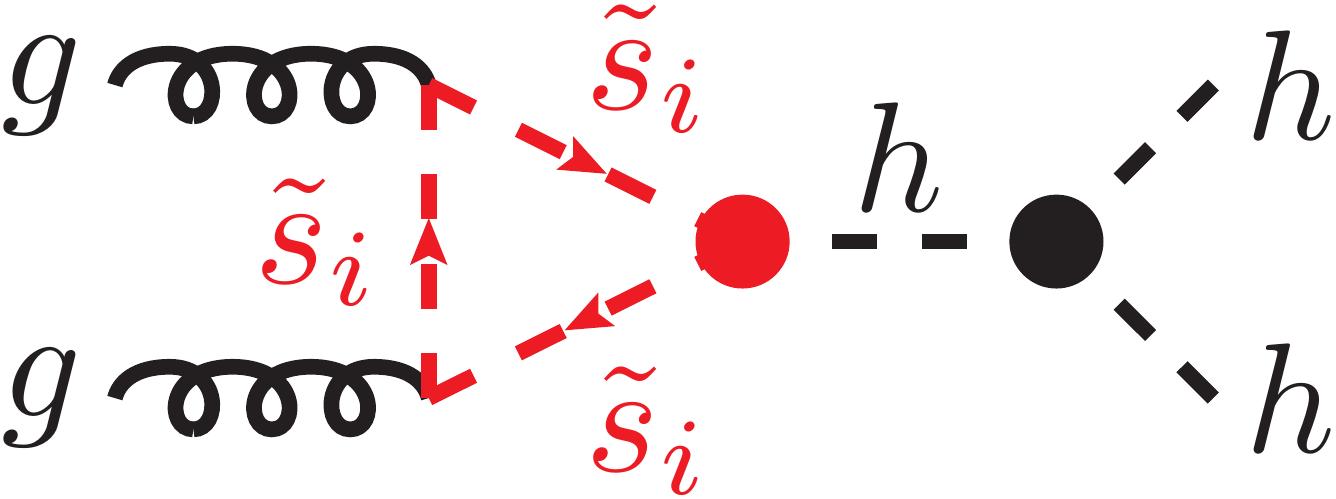}} &
$\mathcal A_i \propto \kappa_{hhh}\kappa_{h\tilde s\tilde s}^{ii}$ \\
\noalign{\vskip 3pt}
\hline
\noalign{\vskip 3pt}
\textbf{7} & 
$\begin{array}{c}
\text{\bf Scalar triangle and box}\\
\text{\bf with two $h\tilde s\tilde s$ couplings}
\end{array}$ &
\begin{tabular}{c}
\raisebox{-.45\height}{\includegraphics[width=.16\textwidth]{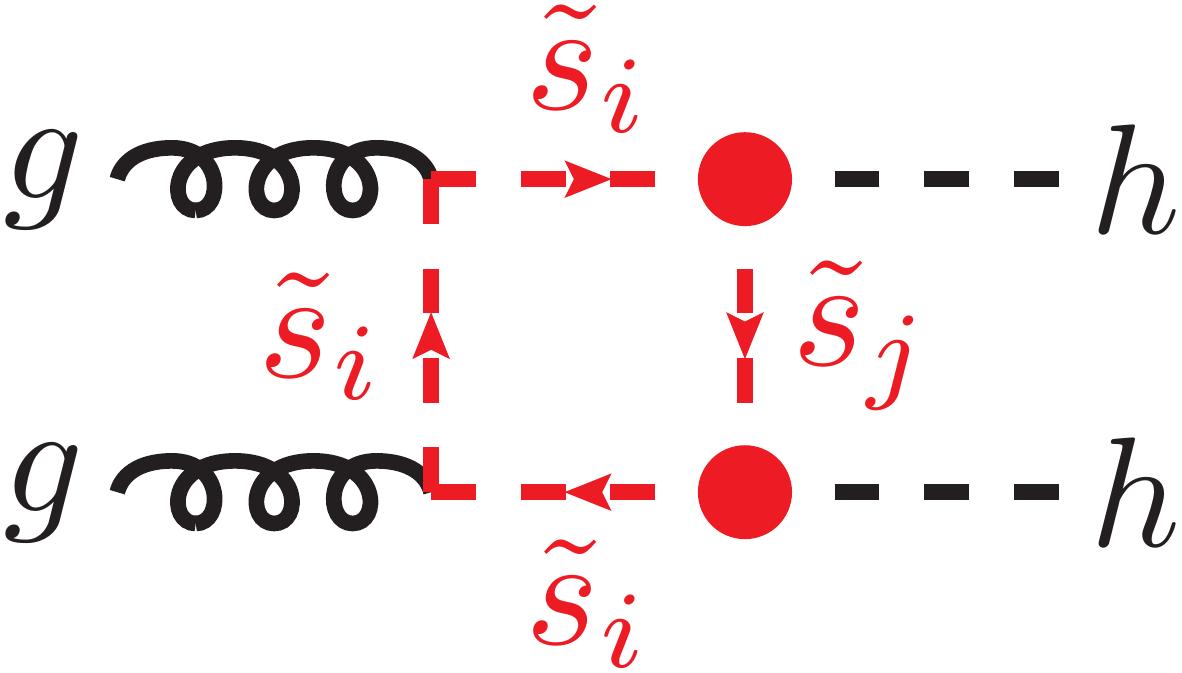}}
\raisebox{-.48\height}{\includegraphics[width=.16\textwidth]{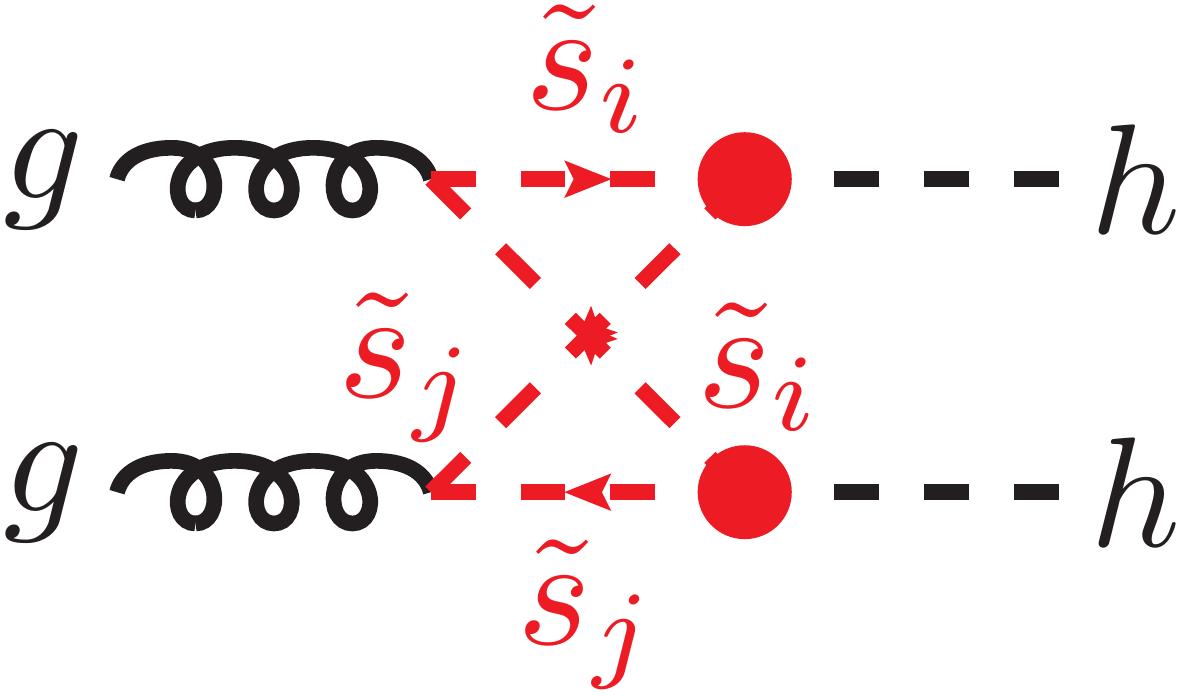}}
\raisebox{-.45\height}{\includegraphics[width=.16\textwidth]{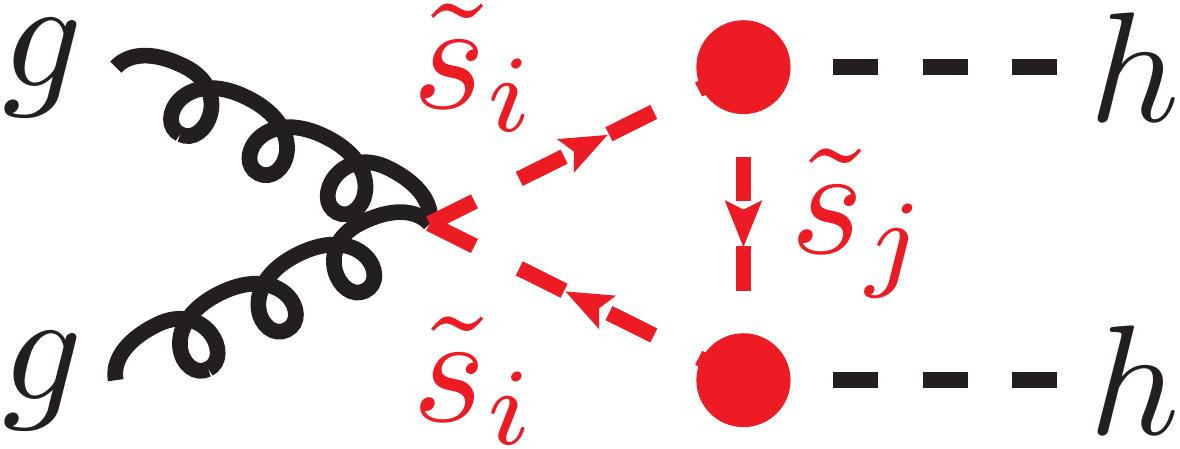}} 
\end{tabular} &
$\mathcal A_i \propto |\kappa_{h\tilde s\tilde s}^{ij}|^2$ \\
\noalign{\vskip 3pt}
\hline
\noalign{\vskip 3pt}
\textbf{8} & 
$\begin{array}{c}
\text{\bf Scalar bubble and triangle}\\
\text{\bf with $hh\tilde s\tilde s$ coupling}
\end{array}$ &
\raisebox{-.46\height}{\includegraphics[width=.16\textwidth]{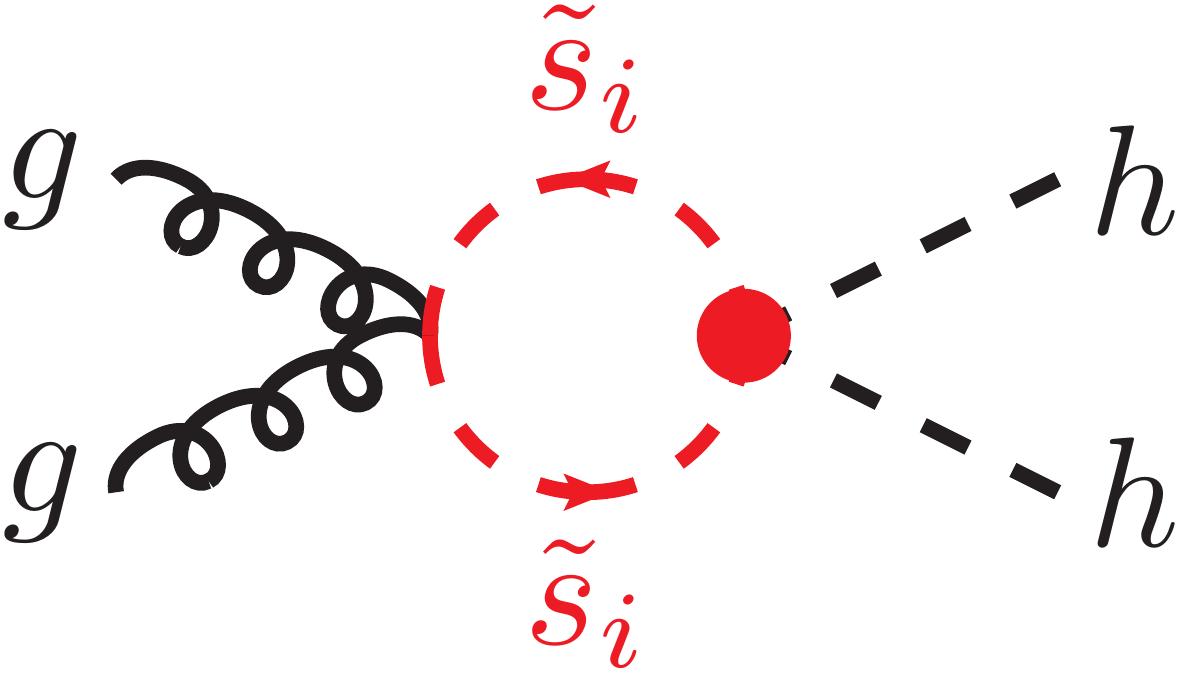}}
\raisebox{-.5\height}{\includegraphics[width=.16\textwidth]{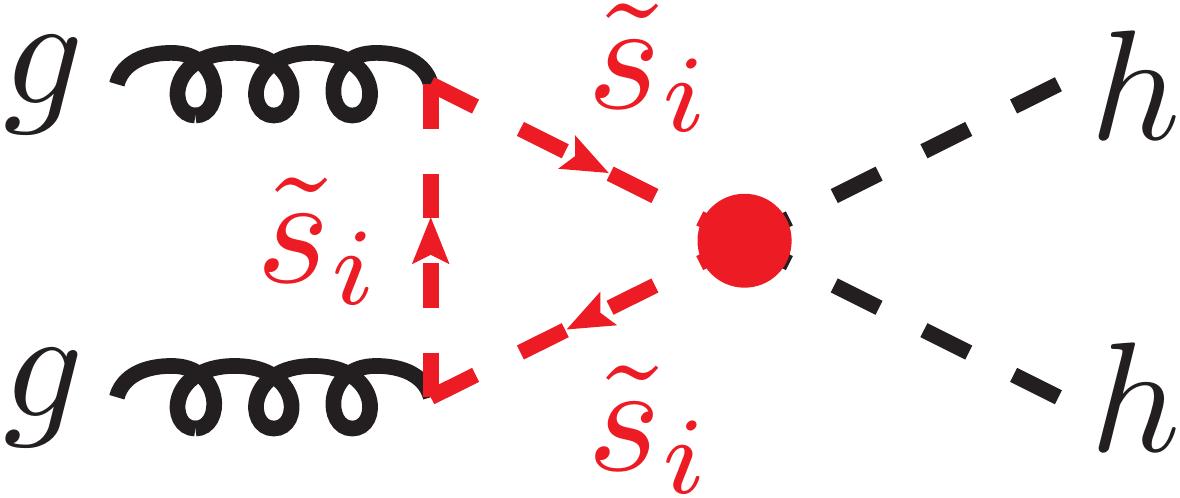}} &
$\mathcal A_i \propto \kappa_{hh\tilde s\tilde s}^{ii}$
\\
\noalign{\vskip 3pt}

\hline
\hline

\noalign{\vskip 5pt}
\textbf{9} & \textbf{Neutral scalar} &
\raisebox{-.5\height}{\includegraphics[width=.16\textwidth]{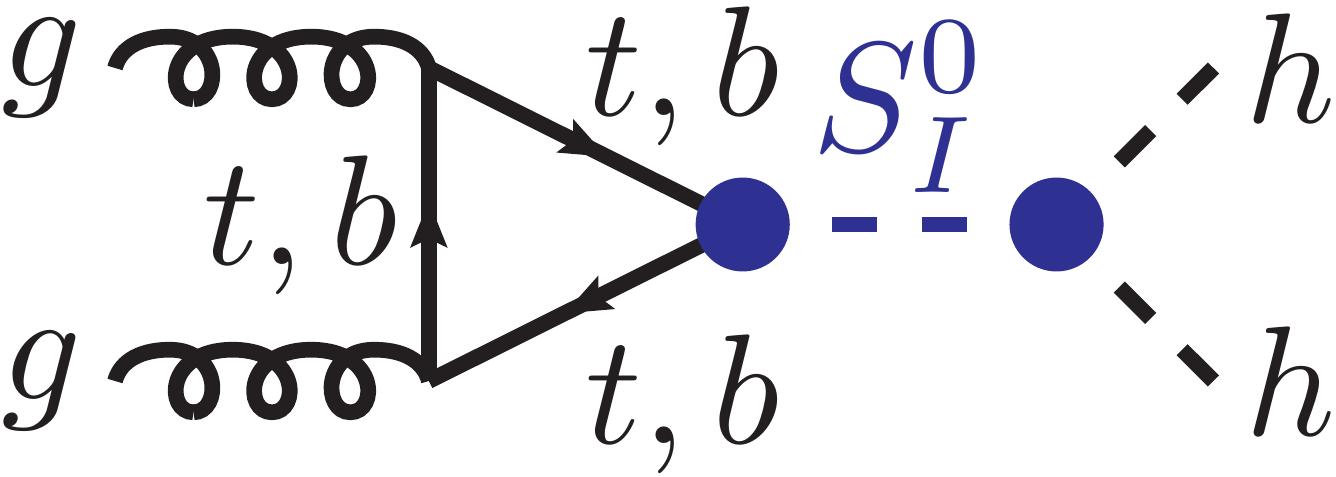}} &
$\mathcal A_i \propto \kappa_{Shh}^I\kappa_{Sff}^I$ \\
\noalign{\vskip 3pt}

\hline
\hline

\noalign{\vskip 5pt}
\textbf{10} & 
$\begin{array}{c}
\text{\bf Neutral scalar +}\\[-3pt]
\text{\bf coloured scalar}
\end{array}$ & 
\raisebox{-.45\height}{\includegraphics[width=.16\textwidth]{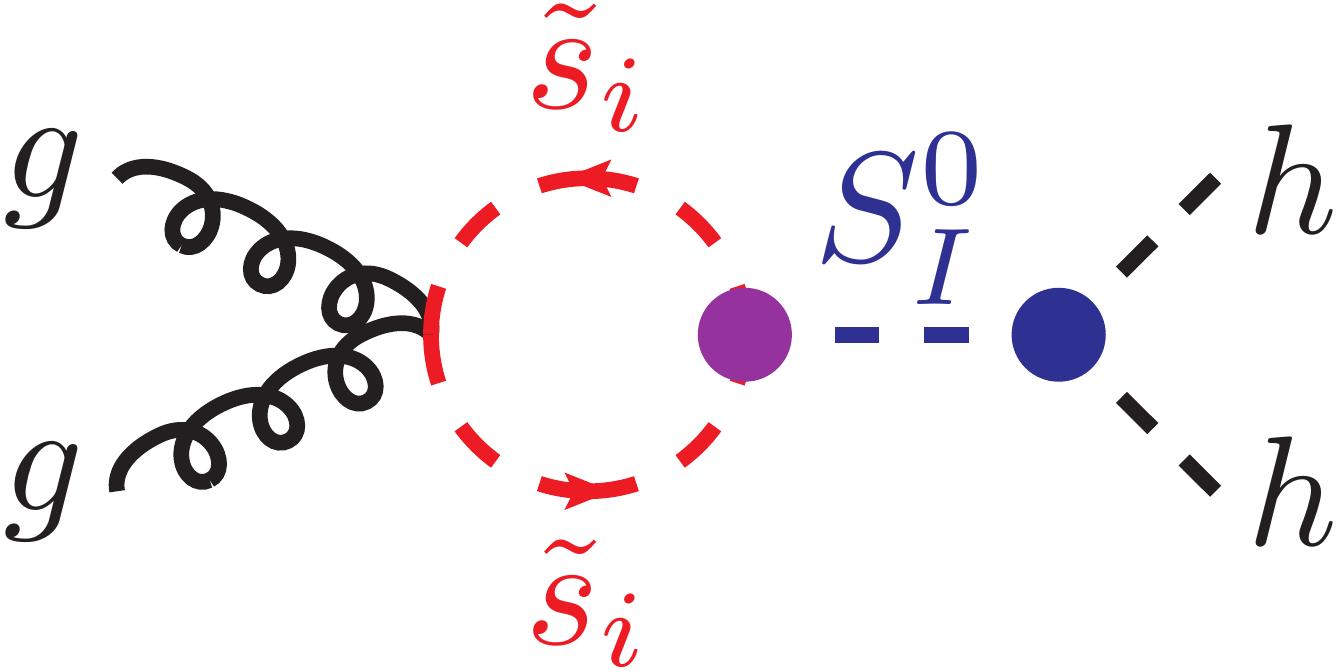}}
\raisebox{-.47\height}{\includegraphics[width=.16\textwidth]{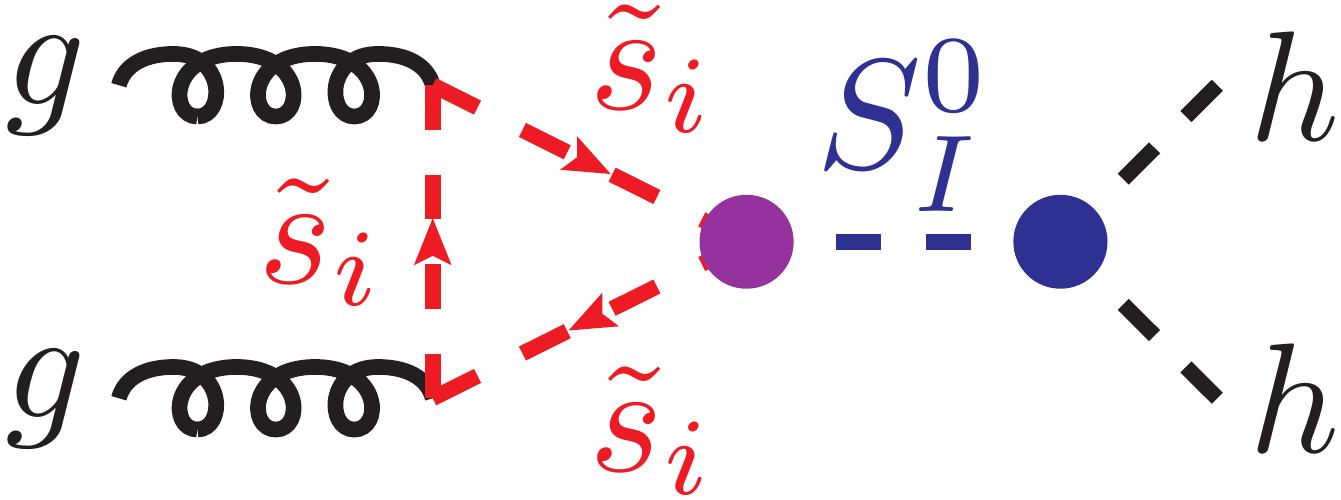}} &
$\mathcal A_i \propto \kappa_{Shh}^I\kappa_{S\tilde s\tilde s}^{Ii}$ \\
\noalign{\vskip 3pt}

\hline
\hline
\end{tabular}
}
\caption{\label{tab:deconstructedtopologies} Complete list of topologies describing gluon-initiated di-Higgs production with modified SM couplings $hhh$ and $htt$ and propagation of any number of coloured scalars ($\tilde s_i$) and/or neutral scalars ($S^0_I$). The topologies are classified according to the different products of new couplings, defined in the Lagrangians of \Cref{eq:LagNP}, to which the amplitudes are proportional.}
\end{table}

From a practical point of view, we have performed Monte Carlo (MC) simulations corresponding to each term of \Cref{eq:sigmahats} in a grid of masses and, only for the neutral scalar, total widths, such that the total cross sections and differential distributions can be built by summing each individual result with weights given as external numerical inputs. The advantages of this procedure are multiple.
\begin{enumerate}
\item The event samples can be recycled to account for multiple coloured or neutral scalars appearing in different models of new physics. In fact, the numerical model used for the simulations, available as auxiliary data, contains only the {\it minimal} number of particles needed to simulate all combinations. In particular, the largest amount of neutral scalars which can appear in any amplitude squared for the di-Higgs final states is two (corresponding to interference terms between topologies involving two different scalars): if more neutral scalars are present in a specific theory, the same grid of samples can be used, one just needs to pick the closest masses and total widths and rescale by the appropriate (only coupling-related) weights. For what concerns the squarks, one can immediately realise that the electric charge does not play any role in the determination of the cross sections, and that the minimal amount of squarks needed to determine the cross section is 4, corresponding to interferences between amplitudes of topology type 7 in \Cref{tab:deconstructedtopologies} where all squark indices are different.
\item Reinterpreting the results for different scenarios does not require new simulations: since event samples are combined according to \Cref{eq:sigmahats}, different benchmark points (BPs) characterised by different masses, total widths and/or  couplings can be analysed using the same dataset. However, while couplings can be continuously modified as they enter as multiplicative parameters of the linear combination, masses and widths are linked to the grid of values used for the simulations. Interpolation for intermediate values is not trivial, as the kinematical features associated to resonances and threshold effects result in non-linear behaviours of the differential distributions, but results are reliable for points sufficiently close to the grid. (Interpolation methods are currently being studied to overcome this limitation.) 
\item Non-linear features of differential distributions related to the interplay of interferences or threshold effects can be inspected with unprecedented detail by isolating their sources and straightforwardly relating these to the underlying physics. This is especially relevant when the mass scales of such non-linear effects are similar for different new particles, so that determining the shape of distributions and their dependence upon variations of the input parameters would be a difficult task using dedicated simulations, especially if such new particles are many. We stress here that the individual deconstructed samples are not physical {\it per se}, they become physical only once they have been all combined through  \Cref{eq:sigmahats} for a given set of input parameters. However, they provide essential information about their role in the determination of the final result.
\end{enumerate}

The key element to achieve the deconstruction of the signal is to build a numerical model where each new interaction is labelled by a specific string, so that, when doing the simulations, interactions can be counted at the level of amplitudes squared. A dedicated model has thus been build in {\sc UFO} format~\cite{Degrande:2011ua} and individual coupling orders have been assigned to each BSM interaction of the Lagrangians in \Cref{eq:LagNP} (see \Cref{app:UFOmodel} for more details). {\sc MG5\_aMC}~\cite{Alwall:2014hca,Frederix:2018nkq} has then been used, as it allows to generate process by specifying a given number of couplings with a given coupling order in the amplitude squared. The MC simulations have been performed at 13 TeV, using the LO set of the {\sc NNPDF 3.0} PDFs~\cite{NNPDF:2014otw}. {\sc Pythia 8}~\cite{Sjostrand:2014zea} has been used to take care of Higgs decays, hadronisation/fragmentation and parton shower. The reconstruction of objects in the final state has been done using the anti-$k_T$ algorithm~\cite{Cacciari:2008gp} with jet cone radius $\Delta R=0.4$, implemented in {\sc FastJet}~\cite{Cacciari:2011ma}. {\sc MadAnalysis 5}~\cite{Conte:2012fm} has been used to generate the numerical distributions for the deconstructed samples, combined a posteriori with a dedicated script.

\subsection{Width corrections}

As previously described, the deconstruction method relies on combining event datasets on a grid of masses and widths of the new particles. As far as masses as concerned, if the studied benchmark points are close enough to the simulated grid, it is possible identify key features related to the interplay of contributions for different physical scenarios in a reasonable approximation. For the total widths, however, if the narrow-width approximation holds, it is possible to apply corrections for points which are not on the grid, allowing to accurately determine the size of peaks and the total cross-section for more flexible range of input parameters. 

We start by the consideration that the {\it partial} width describing the resonant decay of a neutral scalar to the final di-Higgs state is proportional to the trilinear $Shh$ coupling, used to compute the weights of the corresponding  contribution in \Cref{eq:sigmahats}: $\Gamma_{S \to hh} \propto \kappa_{Shh}^2$.

The resonant contribution to the  di-Higgs production process can be written as:
\begin{equation}
\sigma_{pp\to S \to hh}=\sigma_{pp\to S} \times BR_{S\to hh}=\sigma_{pp\to S} \times {\Gamma_{S \to hh} \over \Gamma_S^{\rm tot}}\;, 
\end{equation}

The cross-section to produce the neutral scalar, $\sigma_{pp\to S}$, does not depend on the scalar width, therefore it must be the same whether we consider the closest point on the widths grid, or the benchmark point (BP) under study. Under the assumption that the BP masses are close enough to the simulation grid, it follows that:
\begin{equation}
\sigma_{pp\to S \to hh}^{\rm BP} = \sigma_{pp\to S \to hh}^{\rm sim} \times {BR_{S\to hh}^{\rm BP} \over BR_{S\to hh}^{\rm sim}} \simeq \sigma_{pp\to S \to hh}^{\rm sim} \times \left(\kappa_{Shh}^{\rm BP}\over\kappa_{Shh}^{\rm sim}\right)^2 \times {\Gamma_S^{\rm tot,sim} \over \Gamma_S^{\rm tot, BP}}
\end{equation}
Therefore, to take into account widths which are not on the simulation grid, it is sufficient to do the approximate substitution:
\begin{equation}
\kappa_{Shh}^{\rm BP} \rightarrow \kappa_{Shh}^{\rm BP} \sqrt{\Gamma_S^{\rm tot,sim} \over \Gamma_S^{\rm tot, BP}}    
\end{equation}
for each neutral scalar with mass larger than $2m_h$ in the BP, using the closest width on the simulation grid. This substitution becomes less and less accurate as the total widths of the scalar increases. Notice also that this rescaling only affects the {\it height} of the resonant peak. The {\it shape} of the peak in our results is still determined by the simulated width of the scalar, so this rescaling is not ought to approximate the shape of the resonance, but only its relative contribution to the cross-section. In the narrow-width approximation, however, this should be sufficient in most practical situations, considering the limited experimental resolution after reconstructions and selections are applied.

\section{The NMSSM case}

The particle content we have at hand, coloured and neutral BSM scalars, exists naturally in SUSY models. The MSSM allows only heavy new Higgs bosons, so resonant di-Higgs production will be suppressed. Therefore we look at the NMSSM, where light singlets can evade the experimental constraints and the parameter space for the heavy doublet states also allows lighter Higgs states.

\subsection{Triple-Higgs couplings in the NMSSM}

Resonant di-Higgs production depends on the various triple Higgs couplings, so we shall have a look at them first. The MSSM is defined through the superpotential
\begin{equation}
    W_{\mathrm{MSSM}}=y_{u}\hat{Q}\hat{H}_{u}\hat{U}^{c}+y_{d}\hat{Q}\hat{H}_{d}\hat{D}^{c}+y_{\ell}\hat{L}\hat{H}_{d}\hat{E}^{c}+\mu \hat{H}_{u}\hat{H}_{d}.
\end{equation}
The NMSSM extends the field content with a singlet superfield $\hat{S}$ and generates the $\mu$-term dynamically through the VEV of the scalar component of the singlet field.
The superpotential of the NMSSM reads
\begin{equation}
W_{\mathrm{NMSSM}}=W_{\mathrm{MSSM}}(\mu=0)+\lambda \hat{S}\hat{H}_{u}\hat{H}_{d}+\frac{\kappa}{3}\hat{S}^{3}.
\end{equation}

The SUSY part of the tree-level scalar potential is CP-conserving and we assume here that also the soft SUSY  breaking terms respect CP invariance. In such a case only the CP-even states contributes to resonant di-Higgs production.

We write the scalar parts of the doublet Higgs superfields as
\begin{equation}
    H_{u}=\begin{pmatrix}
    H_{u}^{+}\\
    \frac{1}{\sqrt{2}}(H_{u}^{0}+iA_{u}^{0})
    \end{pmatrix},\quad
    H_{d}=\begin{pmatrix}
    \frac{1}{\sqrt{2}}(H_{d}^{0}+iA_{d}^{0})\\
    H_{d}^{-}
    \end{pmatrix}
\end{equation}
and we define the VEVs such that $\langle H_{u,d}^{0}\rangle = v_{u,d}$.

The scalar potential gets three types of contributions. The superpotential gives the F-term contribution
\begin{equation}
    V_{F}=\sum_{i}\left|\frac{\partial W}{\partial \varphi_i}\right|^{2},
\end{equation}
where the sum is over all superfields and you retain only the scalar part. The coupling of chiral superfields to vector superfields give the D-term contributions, which have the same form as in the MSSM
\begin{equation}
    V_{D}=\frac{1}{2}g^{2}\left( H_{d}^{\dagger}\frac{\sigma^{\mu}}{2}H_{d}+H_{u}^{\dagger}\frac{\sigma^{\mu}}{2}H_{u} \right)^{2}+\frac{1}{2}\left( \frac{g^{\prime}}{2}\right)^{2}(H_{u}^{\dagger}H_{u}-H_{d}^{\dagger}H_{d})^{2}.
\end{equation}
The third contribution are the soft SUSY breaking terms, which include mass terms for the scalar fields and trilinear scalar interactions, which are proportional to superpotential terms. For a superpotential coupling $\alpha$ the corresponding trilinear SUSY breaking coupling will be denoted as $A_{\alpha}$.

The scalar potential of the CP-even Higgs sector at tree-level will be ($H_{u}^{0}$, $H_{d}^{0}$ and $S$ being real scalar fields)
\begin{multline}\label{eq:scalarpotential}
V(H_{u}^{0},H_{d}^{0},S)=\frac{1}{2}m_{H_{u}}^{2}(H_{u}^{0})^{2}+\frac{1}{2}m_{H_{d}}^{2}(H_{d}^{0})^{2}+\frac{1}{2}m_{S}^{2}S^{2}+\frac{A_{\lambda}}{2\sqrt{2}}SH_{u}^{0}H_{d}^{0}+\frac{A_{\kappa}}{6\sqrt{2}}S^{3}
+\frac{1}{32}(g^{2}+g^{\prime 2})((H_{d}^{0})^{2}-(H_{u}^{0})^{2})^{2}\\+\frac{1}{8}g^{2}(H_{d}^{0}H_{u}^{0})^{2}
+\frac{\lambda^{2}}{4}(|H_{u}^{0}H_{d}^{0}|^{2}+|H_{u}^{0}S|^{2}+|H_{d}^{0}S|^{2})+\frac{\kappa\lambda}{4}H_{u}^{0}H_{d}^{0}S^{2}+\frac{\kappa^{2}}{4}S^{4},
\end{multline}
$S$ being here the CP-even component of the singlet scalar field having a VEV $\langle S \rangle = v_S$. In addition there will be loop corrections from the top-stop sector, which give a significant correction to the $H_{u}^{4}$ term, which we denote as 
\begin{equation}
V_{\mathrm{loop}}=\delta |H_{u}^{0}|^{4}.
\end{equation}

The Higgs coupling data is compatible with the SM predictions \cite{CMS:2022dwd,ATLAS:2024lyh}. This suggests that we are close to the so-called alignment limit \cite{Carena:2015moc}, in which one of the mass eigenstates is aligned with the VEV of the doublet states. Therefore we shall derive the triple Higgs couplings under the assumption that we are in the alignment limit, which means that the CP-even mass matrix is diagonalised by the angle $\beta$ for which $\tan\beta=v_{u}/v_{d}$. Hence, we can represent the doublet states as the linear combinations
\begin{eqnarray}
    h & = & \sin\beta H_{u}^{0} + \cos\beta H_{d}^{0},\\
    H & = & \cos\beta H_{u}^{0} - \sin\beta H_{d}^{0}.
\end{eqnarray}
We shall first neglect the mixing with the singlet state. If the latter is negligible, the trilinear Higgs couplings will be determined solely by the quartic couplings, \textit{i.e.}, gauge couplings and $\lambda$. The equations for $h$ and $H$ can be inverted to give
\begin{eqnarray}
H_{u}^{0} & = & \sin\beta h+\cos\beta H,\\
H_{d}^{0} & = & \cos\beta h -\sin\beta H.
\end{eqnarray}
When we insert this into the scalar potential, we get the following quartic interactions between the mass eigenstates in the alignment limit:
\begin{multline}
V(h,H)=\frac{\cos^{2}2\beta}{32}(g^{2}+g^{\prime 2})(hhhh-2hhHH+HHHH)
+\frac{g^{2}+2\lambda^{2}}{32}[\sin^{2}2\beta \,hhhh+4\sin 2\beta\cos 2\beta\, hhhH+\\(4\cos^{2}2\beta-2\sin^{2}2\beta) hhHH
- 4\sin 2\beta\cos 2\beta\, hHHH+\sin^{2}2\beta\, HHHH ]+\delta (\sin^{4}\beta\,hhhh+4\sin^{3}\beta\cos\beta\, hhhH\ldots).
\end{multline}
In the alignment limit, $\langle h \rangle = v$ and $\langle H \rangle = 0$, so the triple Higgs couplings relevant for di-Higgs production become
\begin{eqnarray}
\lambda_{hhh} & = &\left( \frac{g^{2}+g^{\prime 2}\cos^{2}2\beta}{8}+\frac{\lambda^{2}\sin^{2}2\beta}{4}+4\delta\sin^{4}\beta \right) v,\\
\lambda_{Hhh} & = &\left(\frac{3(g^{2}+2\lambda^{2})}{16}\sin 4\beta + 12\delta \sin^{3}\beta\cos\beta\right) v.
\end{eqnarray}
Both of these couplings are subject to the constraint that $m_{h}=125$~GeV. The first of these couplings governs  non-resonant di-Higgs production whereas the second one gives the strength of the resonant di-Higgs production via the heavy doublet Higgs.

If $\tan\beta$ is large, the term proportional to $\lambda^{2}$ in $\lambda_{hhh}$ gets suppressed and the remaining terms lead to $\lambda_{hhh}$ being close to its SM value as the part of the Higgs potential relevant for Higgs mass generation is of the form $\lambda_{\mathrm{eff}}|H_{u}^{0}|^{4}$, like in the SM. At low $\tan\beta$ the $\lambda^{2}$ term becomes relevant and, if it is greater then the SM gauge couplings, it can enhance $\lambda_{hhh}$ above its SM value, a feature well-known in the literature \cite{Wu:2015nba}.

One may immediately observe that $\lambda_{hhh}$ is positive whereas the sign of $\lambda_{Hhh}$ is not definite and depends on $\tan\beta$ as well as the relative size of $\lambda$ and $\delta$. Also it is easy to observe that at large $\tan\beta$ both terms in $\lambda_{Hhh}$ go to zero, so that the resonant contribution becomes small. While this result was derived in the context of the NMSSM, we notice that we have no other Higgs coupling structures in the MSSM. Hence, the general conclusion of small $\lambda_{Hhh}$ at large $\tan\beta$ applies also to the MSSM. However, since a large value of $\tan\beta$ is required to obtain a $125$~GeV Higgs mass in the MSSM (unless the SUSY breaking scale is extremely large), resonant effects are expected to be small, which is a further reason to use the NMSSM as our example instead of the MSSM.

\begin{figure}
    \centering
    \includegraphics[width=0.7\linewidth]{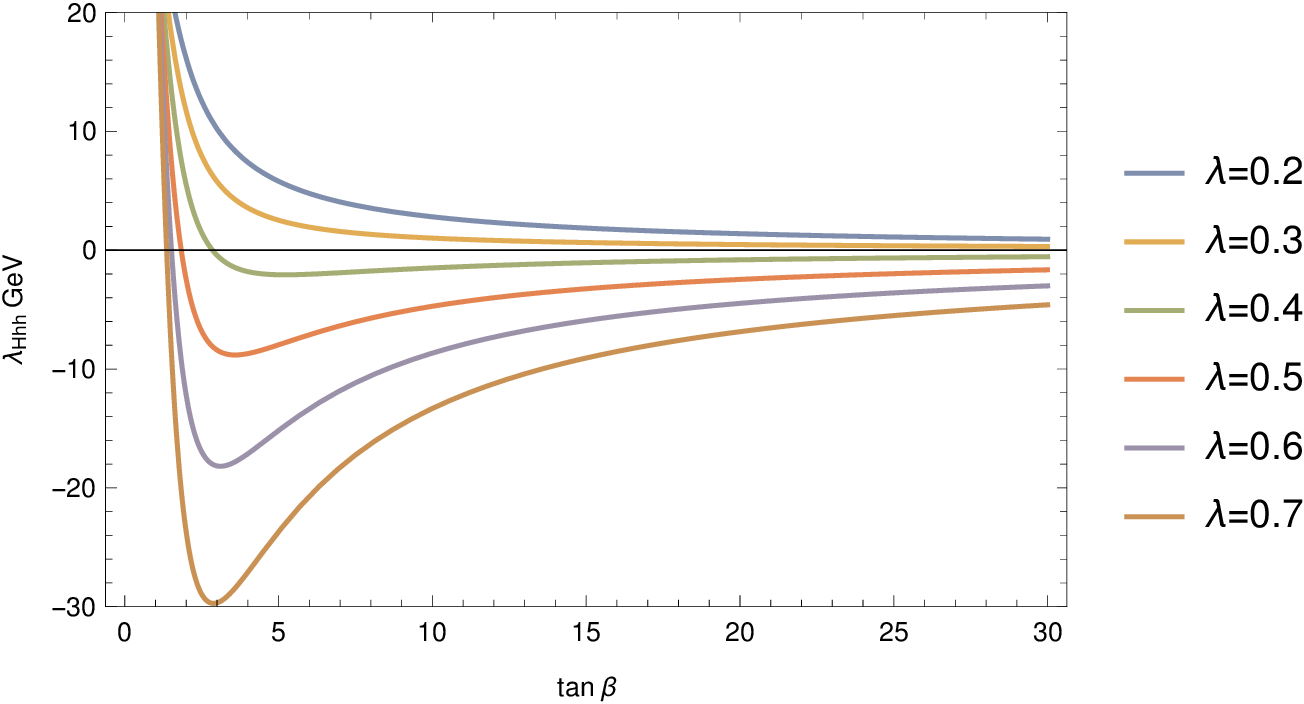}
    \caption{The schematic dependence of the $Hhh$ triple Higgs coupling on $\tan\beta$ for values of $\lambda$ between $0.2$ (top) to $0.7$ (bottom) with a fixed value of $\delta=0.04$. Large and small values of $\lambda$ lead to different interference patterns in di-Higgs production.}
    \label{fig:Hhhcoupling}
\end{figure}

We show schematically the dependence of $\lambda_{Hhh}$ on $\tan\beta$ in \Cref{fig:Hhhcoupling} using $\lambda=0.2\ldots 0.7$ and $\delta=0.04$. If we limit ourselves to $\tan\beta > 1$, for large $\lambda$ the coupling is negative except for a small range of values close to $\tan\beta=1$. Close to $\tan\beta=3$ the coupling gets its most negative value and at higher values of $\tan\beta$ it tends to zero (as mentioned above). If $\lambda$ is small, the second term will be larger and the coupling is positive at all $\tan\beta$, which can be seen from \Cref{fig:Hhhcoupling}.  Therefore the sign of this coupling and consequently the interference pattern of resonant di-Higgs production with the di-Higgs continuum will depend on whether the $125$~GeV Higgs mass is achieved through tree-level superpotential contributions or stop loops.

We then look at the CP-even singlet state in resonant di-Higgs production. In practice there will be a small mixing between the singlet and doublet states through the $\lambda$ term. Even in the absence of this mixing, there is a chance for resonant di-Higgs production via the singlet as the scalar potential contains a term $\frac{1}{2}y_{t}\lambda \tilde{t}_{L}\tilde{t}_{R}SH_{d}^{0}$, which will induce a non-negligible coupling between stops and the singlet. Unfortunately even if there is a relatively light stop and stop mixing is large, the resonant cross section is very small as we show below through some examples. Therefore, the singlet states are in practice produced through top loops resulting from the mixing between the singlet and doublet states.

The coupling between the singlet and the SM-like Higgs states in the perfect alignment limit is
\begin{equation}
\lambda_{Shh}=\frac{\lambda^{2}}{2}v_{S}+\left(\frac{A_{\lambda}}{4\sqrt{2}}+\frac{\kappa\lambda}{4}v_{S}\right)\sin 2\beta,\label{eq:Shhcoupling}
\end{equation}
where $v_{S}$ is the VEV of the singlet field and $A_{\lambda}$ is the soft SUSY-breaking trilinear coupling between $S$, $H_{u}$ and $H_{d}$. If $\lambda$ is large, the first term is large and positive, while the second term may have either sign. The second term has its largest absolute value at $\tan\beta=1$ and becomes small at large $\tan\beta$. If there is to be a large resonant signal from the singlet state, $\lambda_{Shh}$ must be positive and $A_{\lambda}$ should be of the same sign as the other terms. The non-zero mixing with the doublet states lead to some corrections but if the expression in equation (\ref{eq:Shhcoupling}) is not small, the correction is subleading.

Both singlet and doublet states can give a large contribution to the resonant di-Higgs production cross section in the so called $\lambda$SUSY limit, where $\lambda$ is large and $\tan\beta$ not much larger than one. In that case the signs of the $\lambda_{Hhh}$ and $\lambda_{Shh}$ couplings will be opposite, which has implications for the interference patterns as we will show below.

\subsection{Resonant di-Higgs signals}

We shall now show representative examples of the impact of new Higgs bosons onto the di-Higgs differential cross section. There are overall three classes of deformation of the cross section from the SM: the modification of the Higgs trilinear coupling or top Yukawa (denoted by {\sl M} in the figures), squark effects (denoted by {\sl s} in the figures) and effects of additional scalars (denoted by {\sl S} in the figures). Interferences between classes A and B are denoted by A$|$B and the class {\sl mix} arises from contributions where more than one type of classes ends up with the same coupling structure. The class {\sl Ss} denotes diagrams where BSM Higgs states are produced through squark loops. The grouping is done by the combination of couplings, so sometimes contributions are not where they would seem to belong by the naming of the groups\footnote{For instance the diagram $\mathbf{2}$ squared (which one would think belongs to the category M) is proportional to $\kappa_{hff}^{2}$ as is the interference of diagram $\mathbf{4}$ with the SM (which one would think belongs to the category M$|$B). Now they are both in the category 'mix' instead.}. The detailed containts of each category are given in \Cref{app:xsdec}. 

In all figures the black line denotes the non-resonant di-Higgs production in the SM, the red curve ({\sl Signal}) the sum of all BSM effects and the blue curve the total differential cross section. We stress that \textit{only the black and blue curves represent physical events}, and must therefore be positive-definite in each bin. The red curve, representing the contribution of the sum of all signal and interference terms, can have bins with negative event rates: this contribution is {\it unphysical} if taken alone, but it is correct as it only makes sense when summed to the irreducible background (black curve) to give the (physical) total number of events of the blue curve. Any curve with negative events indicates destructive interference between the diagrams involved, again unphysical if taken alone, but leads to the correct differential event rate when all contributions are summed. 
Prominent features of all plots will be the rise of the cross section from $\approx2m_h$ (where the phase space begins) to slightly above the $\approx2m_t$ threshold (produced by the top loops), followed by a nearly exponential decrease of the SM cross section, upon which the various BSM effects will install (as described in the forthcoming subsections), the latter having the common feature of a second  threshold at $\approx2m_{{\tilde t}_1}$ (produced by the lightest stop), which may be visible or not depending on the actual value of $m_{{\tilde t}_1}$ and other parameters specific to a given BP.

We show the values of some of the essential model parameters for our BPs in \Cref{tab:params}. The spectrum of masses, widths and couplings were computed with \textsc{SPheno v4.0.4} \cite{Porod:2003um,Porod:2011nf}, where we modified the \textsc{SPheno} output to include the various couplings needed by the UFO model, normalised these and fixed their signs according to \Cref{eq:LagNP}. The \textsc{SPheno} model was generated through \textsc{Sarah v4.14.5} \cite{Staub:2013tta}.

\begin{table}
~\\[10pt]
\begin{tabular}{c|ccccc|ccc|p{5.8cm}}
&  $\tan\beta$ & $\lambda$ & $\kappa$ & $A_{\lambda}/\mathrm{GeV}$  & $v_{S}/\mathrm{GeV}$  & $m_{H}/\mathrm{GeV}$ & $m_{S}/\mathrm{GeV}$ & $m_{\tilde{t}_1}/\mathrm{GeV}$ & \hskip 2cm{\bf Peculiar features}\\[2pt]
\hline
\hline
&&&&&&&&&\\[-7pt]
{\bf BP1} & 30 & 0.043 & 0.04 & 150 & 13150 & 2000 & 350 & 1400 & Light singlet on the SM top threshold\\[3pt]
{\bf BP2} & 1.38 & 0.69 & 0.43 & -340 & 1250 & 800 & 500 & 600 & Light squark and scalars (all sub-TeV), doublet heavier than singlet\\[3pt]
{\bf BP3} & 2.5 & 0.7 & 0.54 & -345 & 1210 & 1200 & 800 & 600 & Doublet scalar on the $\tilde t_1$ threshold\\[3pt]
{\bf BP4} & 2.31 & 0.65 & 0.68 & 220 & 1280 & 800 & 1200 & 600 & Light squark and doublet, singlet heavier than doublet \\[3pt]
{\bf BP5} & 7.0 & 0.21 & 0.16 & -550 & 943 & 800 & 100 & 1400 & 100 GeV singlet, sub-TeV doublet\\[1pt]
\end{tabular}

\caption{The summary of the relevant parameters for resonant di-Higgs production for our BPs. Here $m_{H}$ denotes the mass of the heavy (mostly) doublet Higgs and $m_{S}$ the mass of the (mostly) singlet Higgs.}\label{tab:params}
\end{table}

\subsubsection{BP1: light singlet scalar}

\begin{figure}
    \centering
    \includegraphics[width=0.7\linewidth]{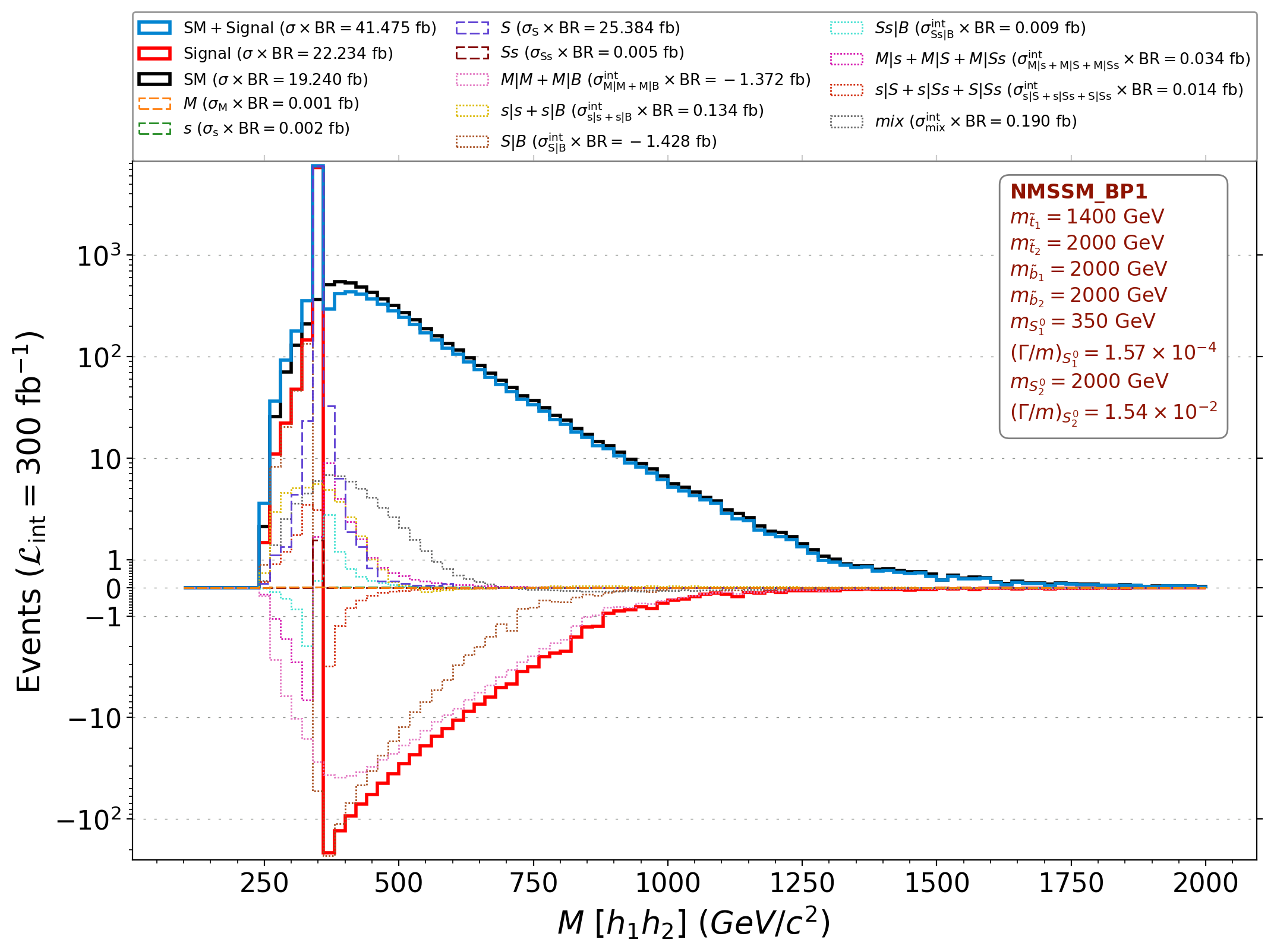}
    \caption{Differential distribution of the di-Higgs invariant mass for BP1, which features a singlet Higgs with $m_S= $350 GeV. In this and following figures of this section, the $y$ axis represents the number of events per bin, associated to an integrated luminosity of 300/fb. 
    }
    \label{fig:NMSSM350}
\end{figure}

Our first BP introduces a singlet scalar at $m_{S}=350$~GeV. Since the mixing between the singlet and the SM-like Higgs pushes the SM-like Higgs mass down, one needs a relatively small value for $\lambda$, a large value for $\tan\beta$ and heavy squarks to achieve a $125$~GeV Higgs mass. A large value for $\tan\beta$ forces the Higgs self-coupling to be SM-like. This case is shown in \Cref{fig:NMSSM350}.

The production of the singlet scalar relies here almost completely on the mixing with the SM-like Higgs, \textit{i.e.}, the inherited coupling with the top quark. Other contributions to resonant production are at a per mille level. The sign of the triple Higgs coupling is positive (in the convention of \Cref{eq:Shhcoupling}), which leads to an interference pattern producing an excess for $m_{hh}<m_{S}$ and a deficit for $m_{hh}>m_{S}$ with respect to the SM. The interference effects near the resonant peak are $\sim 30$--$40\%$ of the SM cross section at parton level. The singlet states always have a very small width, with $\Gamma/m \lesssim 0.001$, which limits the size of the interference effects.

\subsubsection{BP2: intermediate scale scalars with light squarks}

This BP is actually our second NMSSM one from \cite{Moretti:2023dlx}, where we had a $600$~GeV squark combined with a large modification of the Higgs trilinear self-coupling. This is achieved through a low value for $\tan\beta$ and a large value for $\lambda$. This BP had also a singlet scalar with a mass close to $500$~GeV and a doublet scalar with mass close to $800$~GeV. Here we show what happens when you include the effects of the new Higgs states.

\begin{figure}
    \centering
    \includegraphics[width=0.7\linewidth]{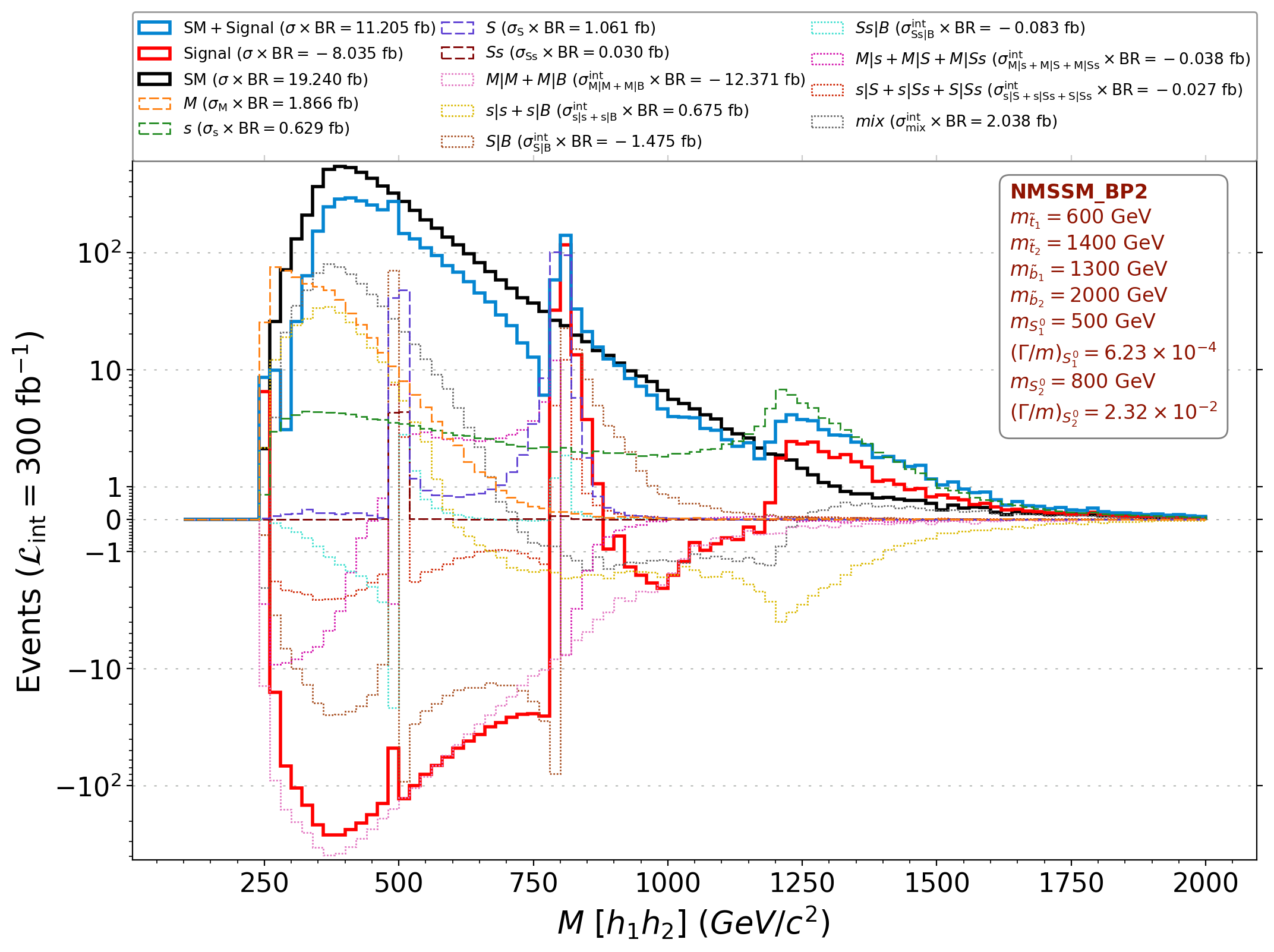}
    \caption{Differential distribution of the di-Higgs invariant mass for BP2, which features a singlet Higgs with $m_{S}=500$~GeV and a doublet Higgs with $m_{H}=800$~GeV. 
    }
    \label{fig:nonresBP3}
\end{figure}

The singlet Higgs is nearly invisible due to its very small couplings to the SM quarks, but even for this BP  the quark contribution is larger than that from squark loops. The main effect of the singlet state is to assist the doublet Higgs in creating a negative interference with the SM diagrams over the range $500$--$800$~GeV. This negative interference together with the resonant peak at $800$~GeV are the main modifications given by the two scalars. For doublet states we typically have $\Gamma/m\sim 0.01$, which means that the interference effects are larger than for singlet states. The destructive interference for BP2 just below the doublet peak is $\sim 75\%$ of the SM cross section. Interference effects between the squarks and new scalars are small but negative, slightly reducing the squark threshold effect.

There is a local minimum in the parton level cross section at low $m_{hh}$, which occurs if the triple Higgs coupling $\lambda_{hhh}$ is well above its SM value. The low event rate in that bin combined with the finite energy/momentum resolution (which leads to migration of events between adjacent bins with larger event rates) make this feature impossible to be observed experimentally.

\subsubsection{BP3: doublet scalar on top of the squark threshold}

Here we take a BP with two quite heavy scalars, $m_{S}=800$~GeV and $m_{H}=1200$~GeV, together with a $600$~GeV stop. In principle, the heavy Higgs is right on top of the squark threshold ($2m_{\tilde{t}_1}$). With these kinds of BPs there is a complementarity between resonant effects and squark effects. Resonant effects are large if we  have a large tree-level contribution to the SM-like Higgs mass (large $\lambda$ and low $\tan\beta$), while squark effects are large if we have a large loop-level contribution to the Higgs mass (small $\lambda$ and large $\tan\beta$, including large stop mixing). Our benchmark configuration is an example of the former.

\begin{figure}
    \centering
    \includegraphics[width=0.7\linewidth]{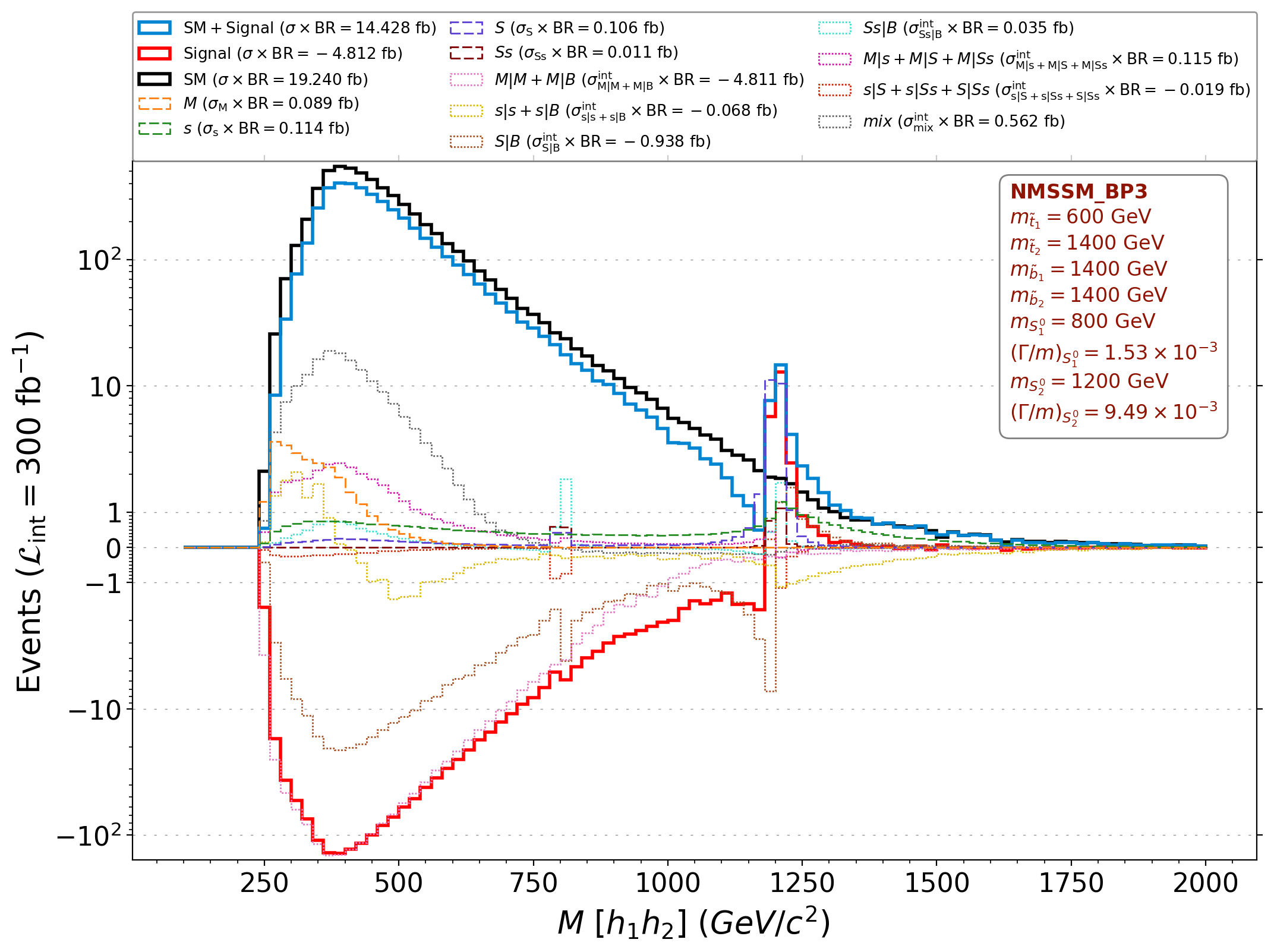}
    \caption{Differential distribution of the di-Higgs invariant mass for BP3, which features a singlet Higgs with $m_{S}=800$~GeV and a doublet Higgs with $m_{H}=1200$~GeV.}
    \label{fig:S800D1200}
\end{figure}

The differential cross section at parton level is given in \Cref{fig:S800D1200}. In this BP the tree-level effect to the SM-like Higgs mass dominates and therefore the behaviour at $1200$~GeV is dictated by a resonance structure. We see clearly the typical doublet interference pattern for large $\lambda$, \textit{i.e.}, destructive interference when $m_{hh}<m_{H}$ and constructive interference when $m_{hh}>m_{H}$. Interferences with the squarks come with an opposite sign with respect to the quarks, but are subdominant. 

There is a singlet Higgs at $800$~GeV but in this case it is completely invisible. In this BP the production through the lightest stop dominates, but the interference terms with the SM box diagram are larger than the resonant diagram squared. The interferences with the squark and quark diagrams come with opposite signs, which further suppresses the singlet contribution to become completely invisible.

\subsubsection{BP4: doublet lighter than  singlet}

Now we reverse the roles of the singlet and doublet compared to the previous BP, \textit{i.e.}, we set the doublet state to have $m_{H}=800$~GeV and the singlet state to have $m_{S}=1200$~GeV. We also have a large value of $\lambda$ and a small one of  $\tan\beta$ which lead to an enhanced value of $\lambda_{hhh}$ compared to the SM. As in the previous BP, the squark contribution becomes rather small.

\begin{figure}
    \centering
    \includegraphics[width=0.7\linewidth]{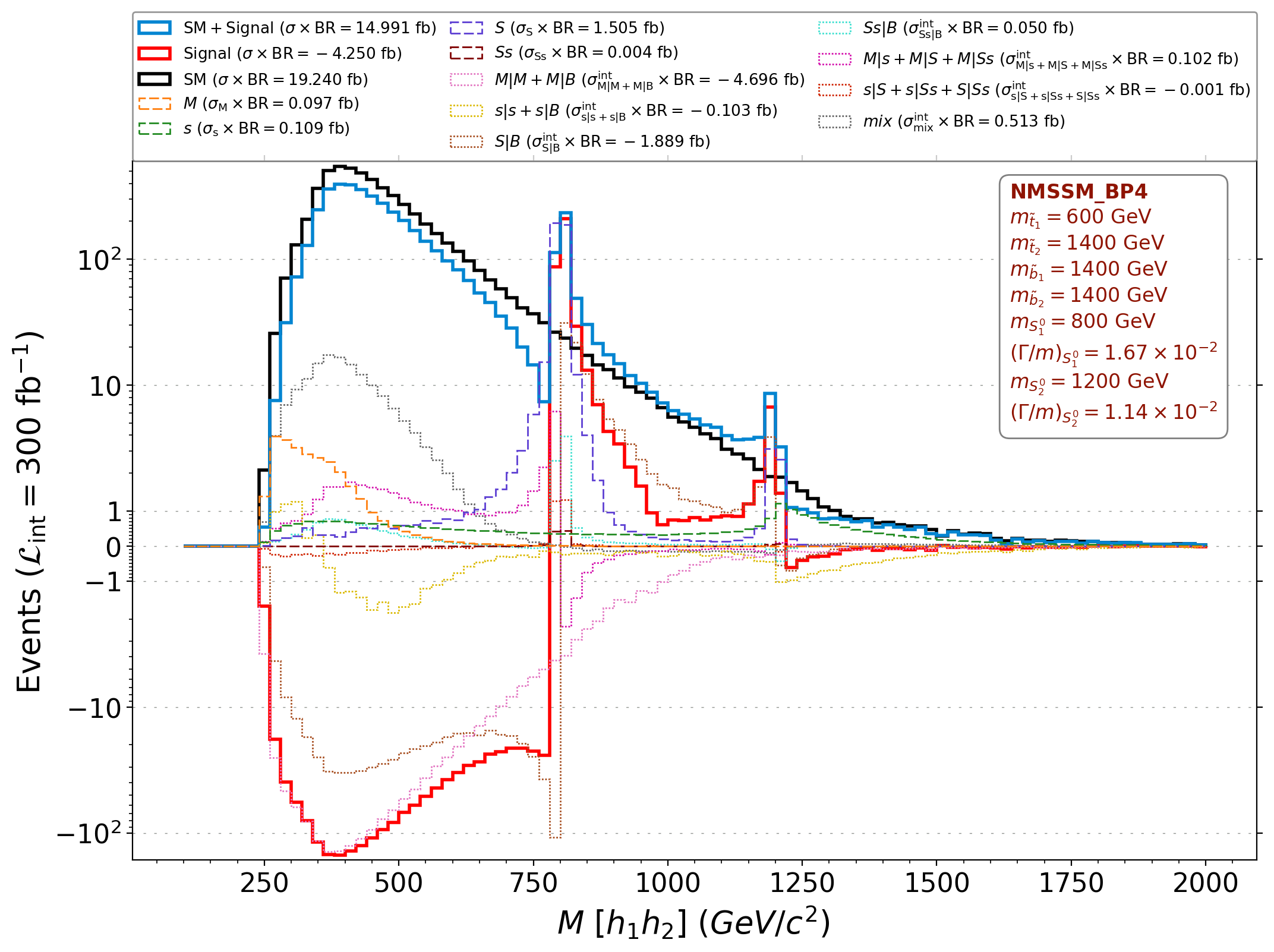}
    \caption{Differential distribution of the di-Higgs invariant mass for BP4, which features a doublet Higgs with $m_H=800$~GeV  and a singlet Higgs with $m_S=1200$~GeV. 
    }
    \label{fig:NMSSM_D800_S1200}
\end{figure}

This time both scalars give a resonant peak with their typical interference structures at large values of $\lambda$. There is constructive interference in the interval $m_{H}<m_{hh}<m_{S}$ and destructive interference over the range  $m_{hh}<m_{H}$ and $m_{hh}>m_{S}$. The negative interference above $m_{hh}=1200$~GeV is larger than the squark contribution, so the squark threshold becomes invisible. The destructive interference near the doublet state is the largest one among our BPs, reaching in excess of $75\%$ of the SM cross section at parton level.

Even though the singlet has a mass of $2m_{\tilde{t}}$, the squark diagrams do not contribute significantly to its production, in fact, only about $1\%$ of the resonant singlet contribution is due to squarks.

\subsubsection{BP5: 100 GeV singlet}

In singlet extensions of the Higgs sector it is possible that the $125$~GeV state is not the lightest Higgs boson. In fact, there are small excesses in various channels at LEP \cite{LEPWorkingGroupforHiggsbosonsearches:2003ing}, ATLAS \cite{ATLAS:2024bjr} and CMS \cite{CMS:2022goy,CMS:2018cyk}, which could be explained through a new scalar state with mass in the range $95$--$100$~GeV \cite{Biekotter:2022jyr,Azevedo:2023zkg,Belyaev:2023xnv,Ellwanger:2023zjc,Cao:2023gkc,Ellwanger:2024vvs}. We show what could be the typical impact of such a scalar onto SM-like di-Higgs production.

\begin{figure}
    \centering
    \includegraphics[width=0.7\linewidth]{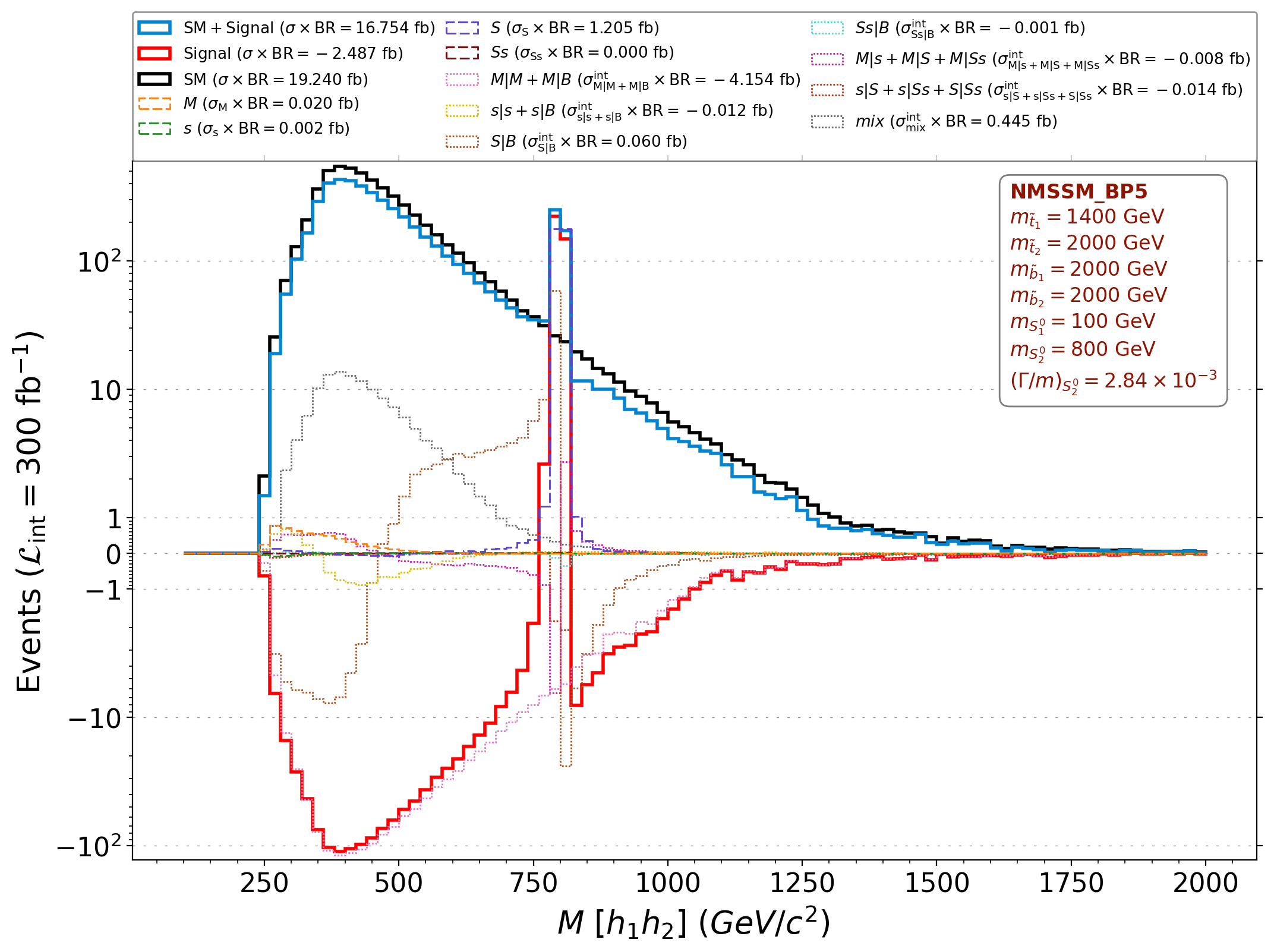}
    \caption{Differential distribution of the di-Higgs invariant mass for BP5, which features a singlet Higgs state with $m_S=100$~GeV and a doublet Higgs state with $m_H=800$~GeV. }
    \label{fig:NMSSM95}
\end{figure}

We present the differential cross section for such a BP in \Cref{fig:NMSSM95}. We have added a $100$~GeV mostly singlet scalar together with a $800$~GeV doublet. In order to get a $125$~GeV Higgs and a light singlet we must choose small values for $\lambda$ and $\kappa$, a moderate value for $\tan\beta$ and heavy stops. The small value of $\lambda$ together with $\tan\beta=7$ leads to a SM-like $hhh$ coupling.

The light singlet state leads a suppression of the di-Higgs signal at low $m_{hh}$.  This is due to some suppression in the SM-like Higgs coupling to top quarks, which could also be measured in $t\overline{t}h$ production. The singlet state itself has a negative interference with the SM diagrams, which has an impact at low $m_{hh}$, but this is subdominant compared to the modification of the top Yukawa. These effects together give a reduction in the low $m_{hh}$ region, similar to what we could have if $\lambda_{hhh}$ was above its SM value. This is again an example of where a simple coupling modifier approach could lead to an incorrect interpretation.

The doublet state at $800$~GeV has now an interference pattern with an excess for $m_{hh}<m_{H}$ and a deficit for $m_{hh}>m_{H}$. This is an example of the $Hhh$ coupling behaviour for small $\lambda$ shown in \Cref{fig:Hhhcoupling} and indicates that the SM-like Higgs mass relies largely on loop corrections and not the tree-level enhancement due to a large $\lambda$.

\subsection{Prospects in different decay channels}

We look at some of the benchmarks in the three most sensitive decay channels $bb\gamma\gamma$, $bb\tau\tau$ and $bbbb$. In \Cref{tab:cuts} we show the preselection criteria applied before plotting each of the final states. However, note that the preselection only includes kinematical and geometrical criteria and does not include the expected reconstruction efficiency. Hence, the numbers shown in the plots should be multiplied by the reconstruction efficiency.
We think it can be useful to remind here that, like in the plots of the previous section, the only physical curves are the solid black and blue ones, representing the irreducible background and the sum of signal, background and their interference, respectively. All the other curves, and noticeably the solid red one, which represents the sum of all signal and interference contributions, do not represent physical observables, and can therefore contain bins with a negative number of events: these bins represent a dominant contribution by negative interference. It is also worth noticing that in the tails of some distribution, a negative number of events can be seen in the overall sums of the blue curves, but this clearly is an effect of statistical fluctuations in regions with very few MC events and has no physical meaning.

\begin{table}[h]
{\setlength\tabcolsep{10pt}
\begin{tabular}{ccc}
\hline\hline
\noalign{\vskip 3pt}
$bb\gamma\gamma$ & $bb\tau\tau$ & $bbbb$ \\
\hline
\noalign{\vskip 3pt}
$N(b)>1$ & $N(b)>1$ & $N(b)>3$ \\ 
$N(\gamma)>1$ & $N(\tau)>1$ & -- \\ 
$p_T(b)>45~(20)\; \mathrm{GeV}$ & $p_T(b)>45~(20) \;\mathrm{GeV}$ & $p_T(b)>40 \;\mathrm{GeV}$ \\
$|\eta(b)|<2.5$ & $|\eta(b)|<2.5$ & $|\eta(b)|<2.5$ \\
$|\eta(\gamma)|<2.5$ & $|\eta(\tau)|<2.5$ & -- \\
$120\; \mathrm{GeV} < M(\gamma\gamma) < 130\; \mathrm{GeV}$ & -- & -- \\[2pt]
\hline\hline
\end{tabular}
}
\caption{\label{tab:cuts} Selection and kinematic cuts for the three final states considered in the analysis.}
\end{table}

\subsubsection{Two b-jets and two photons}

The channel with two b-jets and two photons offers a low background and also a good mass resolution. The downside of this channel is the low event rate due to a BR$(hh\rightarrow bb\gamma\gamma)=0.26\%$ and a reconstruction efficiency of around 10\% after preselection. Therefore, there is limited sensitivity to most of our benchmarks in this channel at nominal HL-LHC integrated luminosity (3000~fb$^{-1}$), which is used to scale our results. The only exception is BP1 which has a large rate due to the low mass of the heavy Higgs, and likely is at the edge of being significant.

An interesting feature for many of the benchmarks presented is that they show destructive interference at low mass, where the $bb\gamma\gamma$ channel is a significant contributor to a potential detection of a LHC SM signal. This means that a first di-Higgs discovery might be delayed if alternative scenarios of this type are realized. This feature arises from an enhanced Higgs self-coupling, which would be needed for a first order electroweak phase transition \cite{Reichert:2017puo,Basler:2019iuu,Biekotter:2022kgf}.

\begin{figure}
    \centering
    \includegraphics[width=0.47\linewidth]{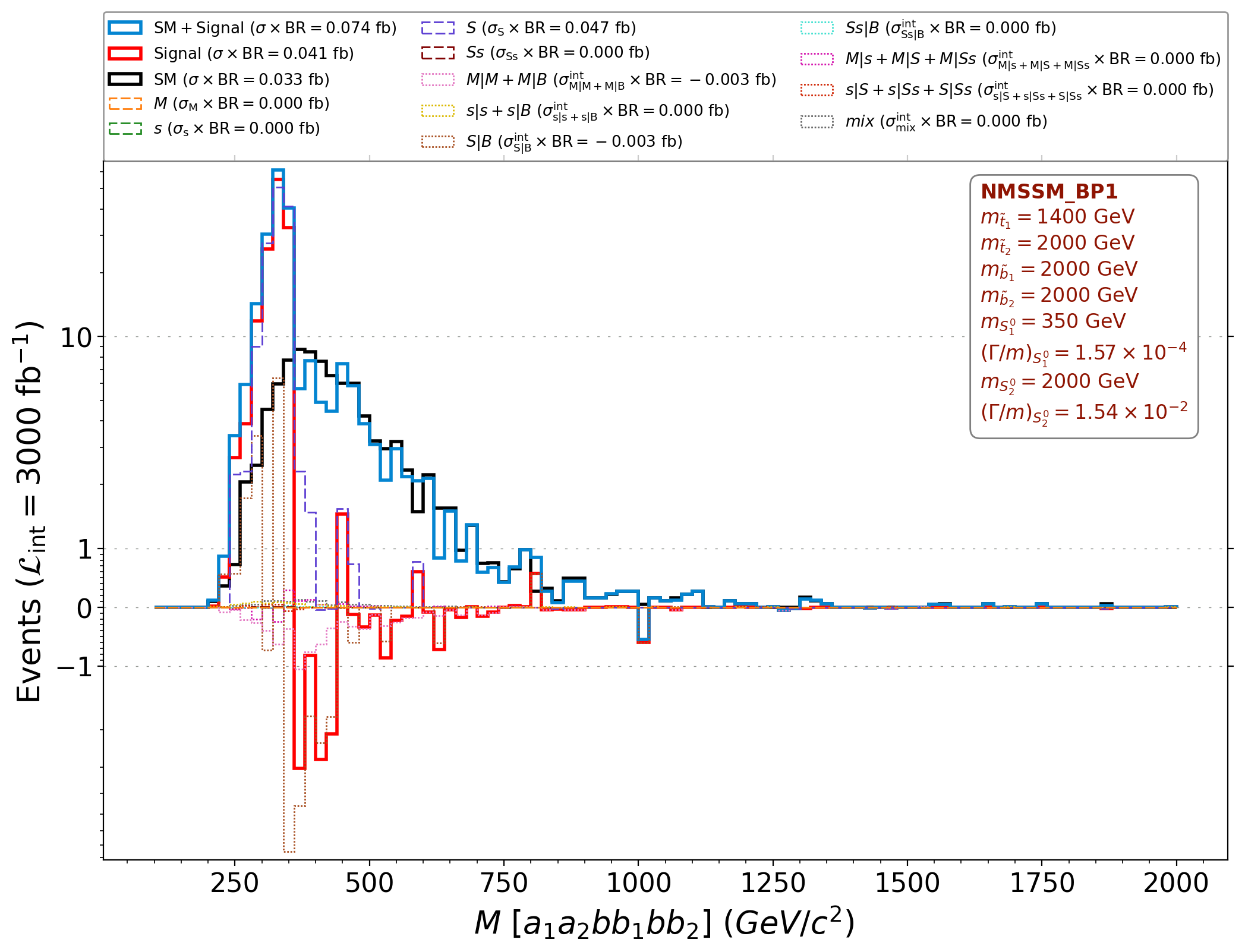}
    \includegraphics[width=0.47\linewidth]{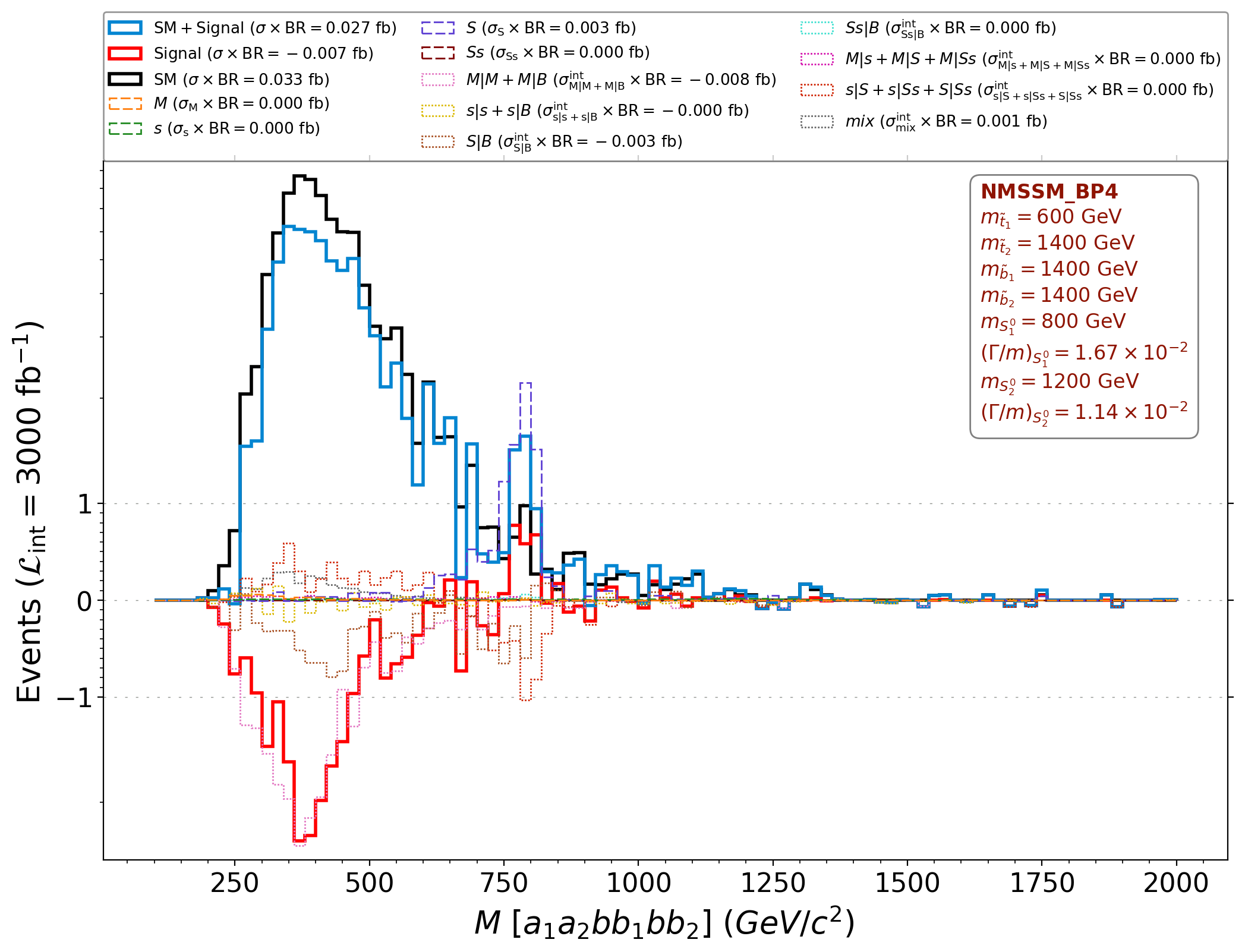}
    \caption{Differential distribution of the di-Higgs invariant mass in the $bb\gamma\gamma$ channel for BP1 and BP4 at 3000~fb$^{-1}$.}
    \label{fig:MH350_bbaa}
\end{figure}

The distribution of the $bb\gamma\gamma$ events after the preselection described in \Cref{tab:cuts} for benchmarks BP1 and BP4 are shown in \Cref{fig:MH350_bbaa}. Around a hundred of excess events will survive the preselection for BP1, but destructive interference at $m_{hh}>m_{S}$ is not as prominent.  As an example of the other benchmarks we show the distribution for BP4. For BP4 the resonant signals or their interference patterns are likely not resolvable even at HL-LHC.

\subsubsection{Two b-jets and two taus}

This final state has a larger rate than $bb\gamma\gamma$, but the mass resolution is not as good due to the neutrinos in the final state. However, the presence of the taus helps in reducing the QCD background. Usually this final state has the best sensitivity to intermediate invariant masses.

\begin{figure}
    \centering
    \includegraphics[width=0.47\linewidth]{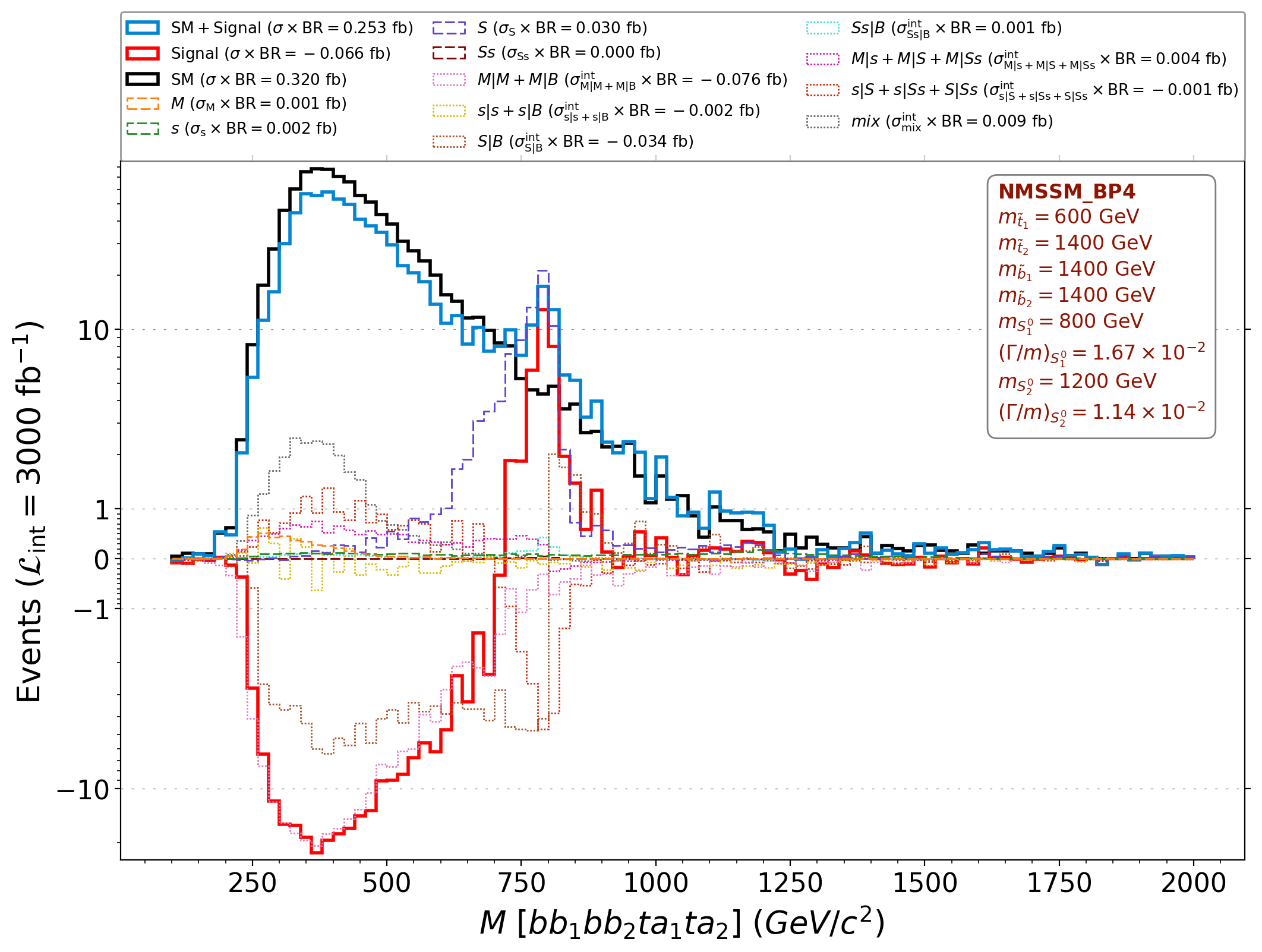}
    \includegraphics[width=0.47\linewidth]{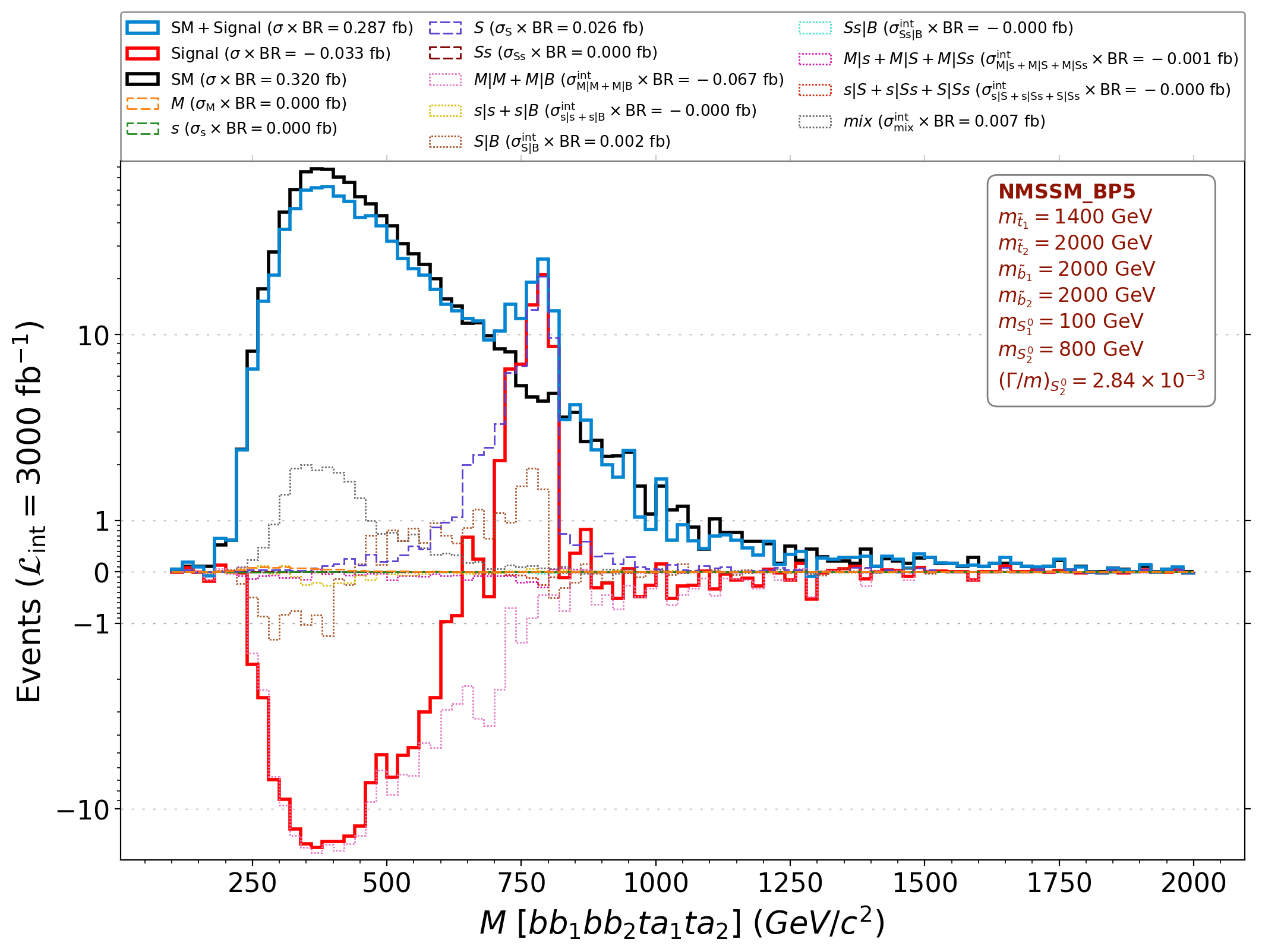}
    \caption{Differential distribution of the di-Higgs invariant mass in the $bb\tau\tau$ channel for BP4 and BP5 at 3000~fb$^{-1}$.}
    \label{fig:MH800_bbtata}
\end{figure}

In \Cref{fig:MH800_bbtata} we show the invariant mass distributions after the preselection of \Cref{tab:cuts} in the $bb\tau\tau$ channel for BP4 and BP5. Both of these have a heavy Higgs at $800$~GeV, but their interference patterns are opposite. There is an overall migration towards lower invariant masses in this channel due to the neutrinos in the final state. Therefore the constructive interference at $m_{hh}<m_{H}$ for BP5 is more visible than the destructive interference of BP4. Nevertheless, in comparison there is a difference between the interference patterns, but the event rate at $m_{H}=800$~GeV is so low that only HL-LHC can possibly distinguish the two cases.

\subsubsection{Four b-jets}

This final state has the largest event rate and the best sensitivity for high-mass resonances. However, due to low event yields in the high-mass tail, it will be more difficult to significantly disentangle any positive or negative 
interference compared to similar effects at lower mass.

\begin{figure}
    \centering
    \includegraphics[width=0.47\linewidth]{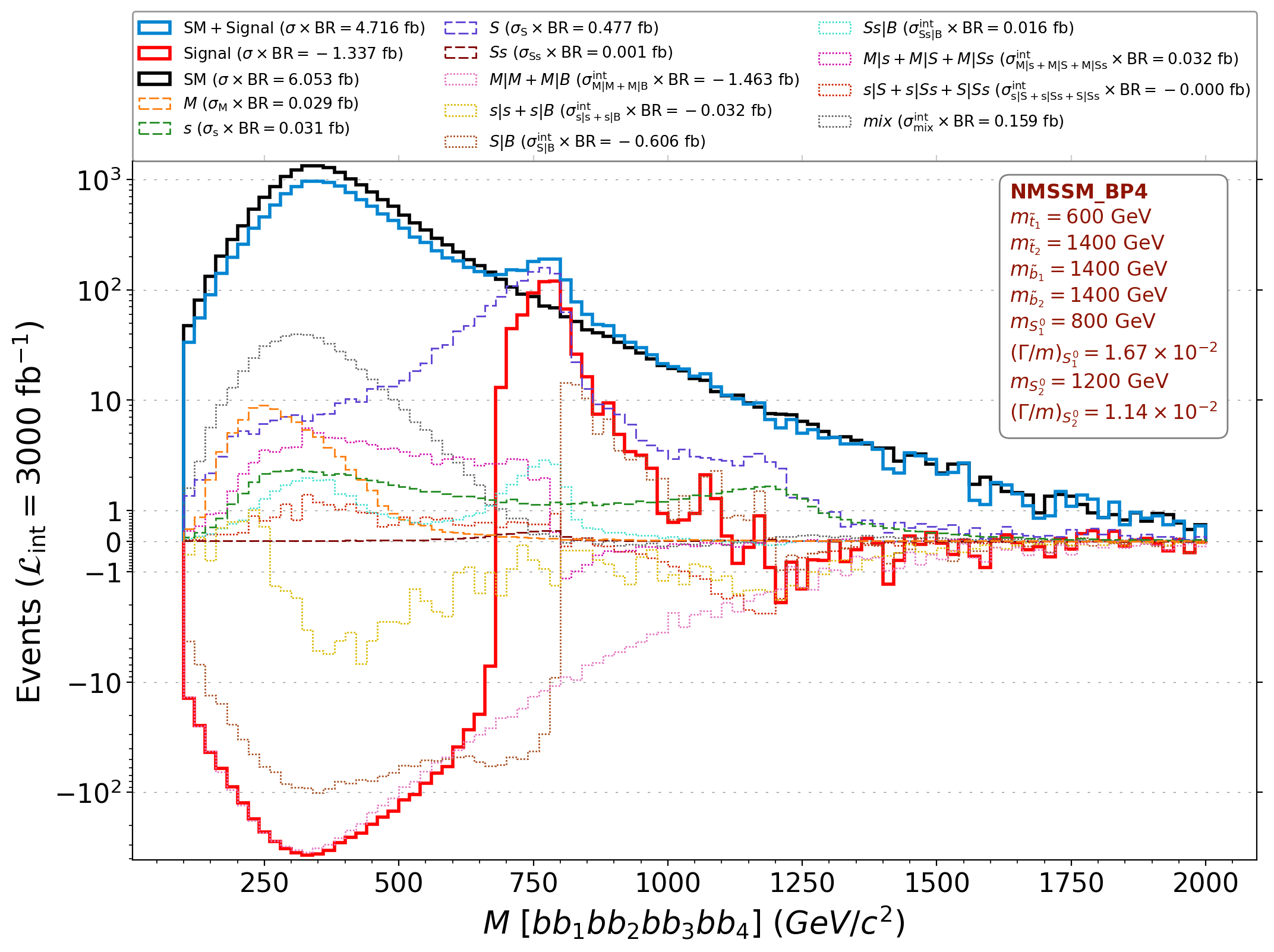}
    \includegraphics[width=0.47\linewidth]{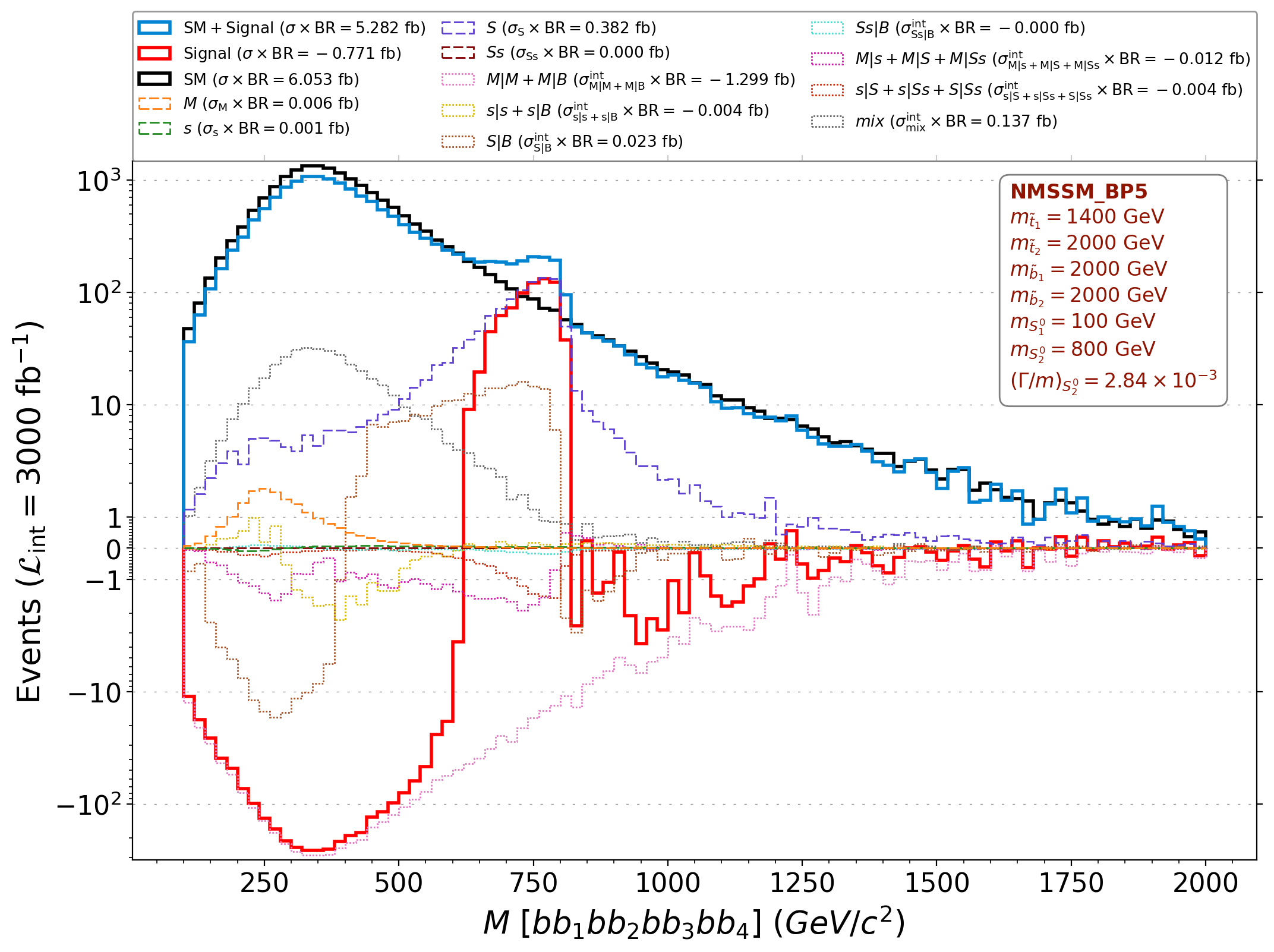}
    \caption{Differential distribution of the di-Higgs invariant mass in the $4b$ channel for BP4 and BP5 at 3000~fb$^{-1}$.}
    \label{fig:MH800_bbbb}
\end{figure}

In \Cref{fig:MH800_bbbb} we show the invariant mass distribution in the $4b$ final state for BP4 and BP5. The reconstruction efficiency for $b\bar bb\bar b$ is slightly below 10\% in the relevant mass range. The interference patterns are even less clear than in the $bb\tau\tau$ final state, although there is some difference in shapes. For constructive interference at $m_{hh}<m_{H}$ and destructive interference at $m_{hh}>m_{H}$ the distribution is nearly flat and drops sharply at $m_{H}$, while the distribution for the opposite case is smoother.

We may also notice that even though BP4 has a heavy Higgs at $1200$~GeV, which is visible in the parton level plot, the resolution of the $4b$ channel in the resolved jet analysis mode without any kinematical fit is not sufficient to produce any visible signal at that mass.

\subsection{Discussion}

The previous examples in NMSSM have shown the complex interplay between resonant and non-resonant contributions in Higgs pair production at the LHC, leading to different and potentially observable effects in the total cross-sections and differential distributions. 

The resonant signatures arising from a singlet-like or a doublet-like Higgs have peculiar features. The doublet state has a stronger coupling to quarks and hence a larger overall cross section. It also has a larger width, which makes the interference effects stronger. The singlet has an interference pattern with an excess at $m_{hh}<m_{S}$ and a deficit at $m_{hh}>m_{S}$ whenever the singlet is visible (\textit{i.e.}, when $A_{\lambda}$ is positive in the conventions of \Cref{eq:scalarpotential}). The doublet interference pattern tells us qualitatively about the contributions to the $125$~GeV mass. There is destructive interference at $m_{hh}<m_{H}$ if the mass is due to the tree-level superpotential contributions (the $\lambda$-term) and constructive interference at $m_{hh}<m_{H}$ if the mass is mainly due to loop-level top-stop contributions.

The presence of light squarks can modify these patterns, especially in the presence of neutral scalars with mass close to twice the mass of the squarks, where threshold effects play a significant role. 

If a qualitative understanding of the interference patterns can be achieved, one can gain understanding of the nature of the Higgs sector. The deconstruction of the signal provides precious insights in this respect, allowing to investigate the role of individual terms, cancellations and enhancements, and relate these to the relevant Lagrangian parameters, making it easy to connect the underlying physics to the corresponding features in the $m_{hh}$ distribution. Such an analysis can obviously be extended beyond our NMSSM example, though the restrictions that supersymmetry imposes on the quartic scalar couplings make the interpretation somewhat easier. 

At a practical level the mass resolution of the most sensitive final states is however limited in current experimental setups, so interference effects get partially smeared out due to migration of events between nearby bins. Constructive interference survives better to detector level than destructive interference, so much that we verified that even at detector level the distributions exhibit an asymmetric shape around the peak. Any definite conclusions of interference patterns will require the full integrated luminosity of the HL-LHC. Correlating more observables in presence of a larger statistics associated to higher luminosities will also clearly help.

At the moment our approach is limited to LO (one-loop) only. It is well known that di-Higgs production in the SM has large higher order corrections \cite{Dawson:1998py,deFlorian:2013uza,Frederix:2014hta,Borowka:2016ehy,Baglio:2018lrj}. In the heavy top limit this leads to a relatively constant K-factor close to $1.8$, but finite mass effects lead to a substantial dependence on the top mass scheme \cite{Baglio:2020wgt} and in different schemes the variation of the $K$-factor as a function of $m_{hh}$ varies. This can lead to shape errors of $20$--$30\%$, though local variation is milder. For the BSM part the NLO corrections are not always known. For squarks the NLO-QCD corrections have been computed in the limit of vanishing external momenta \cite{Agostini:2016vze} and the K-factor corrections are close to the SM ones. For the resonant part K-factors of heavy Higgs production in the MSSM are in the same ballpark as the ones for di-Higgs production in the SM \cite{Dawson:1996xz,Harlander:2003kf}, so our qualitative conclusions should hold.

\section{Conclusions}

We have shown that customary approaches to resonant SM-like di-Higgs production at the LHC that treat $s$-channel heavy resonances in 
Narrow Width Approximation (NWA), or even as Breit-Wigner (BW) distributions, neglecting the contributions to the signal due to interference effects can miss relevant effects. Crucially, for our purposes, we have shown that not only interferences between the heavy Higgs state graph and SM diagrams can be relevant but also those between such a graph and BSM diagrams carrying in the loop companion states to the top quark. We have proven this to be the case for a prevalent theory of the EW scale, chiefly, SUSY, in its NMSSM incarnation (wherein such top quark companion states are top squarks). In fact, in the latter SUSY construct, one could even have two heavy Higgs states as resonant objects in $s$-channel, which even more complicated pattern of interferences will be blurring further the validity of the aforementioned naive approaches.

This has been illustrated in the case of a variety of NMSSM BPs, wherein heavy Higgs states have been spaced in mass in such a way to enter the entire di-Higgs invariant mass spectrum from 100 GeV, to $2m_t$, to 500 GeV, to 800 GeV and to $2 m_{\tilde t}$, thereby capturing all relevant possible phenomenological features of the differential di-Higgs cross section in this BSM framework. We have also shown that the qualitative features of interferences and relative strengths of squark and heavy Higgs modifications have a connection to the SM-like Higgs mass generation.

Furthermore, we have shown that these interferences effects, originally studied at the parton level for on-shell SM-like Higgs states, survive to a tangible extent upon allowing for the decays of the latter into $b\bar b\gamma\gamma$, $b\bar b\tau^+\tau^-$ and $b\bar bb\bar b$ final states in the presence of parton shower, fragmentation/hadronisation and realistic detector effects. However, this largely depends upon the achievable di-Higgs mass resolution in the aforementioned decay channels: in the sense that, while not all individual  interference effects may  be visible in actual data, the  new signal definition  (including a globally interfering contribution to it) induces the generic feature of a visible distortion of the BW shape which needs to be accounted for in experimental analysis.

This work is a continuation of a previous one of ours, which tackled the non-resonant di-Higgs process, wherein we had introduced a dedicated framework to deconstruct the signal in a finite set of independent terms to be recombined using  specific coupling combinations as weights. This deconstruction technique is able to capture the individual contributions of interferences, finite widths or threshold effects, and has allowed us to map the full results of the di-Higgs process in terms of simplified model structures, providing a prompt interpretation of the underlying BSM dynamics. The modularity of this tecnique allows straightforward generalisations, such as the present case where resonant di-Higgs production has been included.  Indeed, such a framework can readily be extended to also accommodate other types of other coloured particles, such as fermionic top quark partners like, {\it e.g.}, VLQs in Compositeness, so as to generalise our approach to other theories of the EW scale with top companion states.

\section*{Acknowledgments}
SM is supported in part through the NExT Institute and the STFC Consolidated Grant No. ST/X000583/1. LP's work is supported by ICSC – Centro Nazionale di Ricerca in High Performance Computing, Big Data and Quantum Computing, funded by the European Union – NextGenerationEU. HW is supported by the Ruth and Nils-Erik Stenb\"ack’s Foundation. JS is in part supported by the Swedish Research Council grant no. 2023-04654. LP and HW acknowledge the use of the Fysikum HPC Cluster at Stockholm University.

\appendix

\section{Deconstructed cross sections}
\label{app:xsdec}

In this appendix we provide the complete set of contributions to the di-Higgs cross-section, \cref{eq:sigmahats}. The conventions adopted throughout the equations are the following:
\begin{itemize}
\item the reduced cross sections $\hat\sigma$ have been labelled according to the numbering scheme of \Cref{tab:deconstructedtopologies}, and their arguments are the masses (and widths for neutral scalars) of the particles propagating in the loops; factors of 2 from interference terms have been included into the $\hat\sigma$'s;
\item the SM irreducible background is labelled as "$\rm B$", the contribution of the topologies containing only modified SM couplings (\textbf{1} to \textbf{4} in \Cref{tab:deconstructedtopologies}) as "$\rm M$", the contribution of topologies associated with the propagation of coloured scalars (\textbf{5} to \textbf{8}) as "$\tilde s$", those associated with the neutral scalars (\textbf{9}) as $S$, and the contributions with both neutral and coloured scalar (\textbf{10}) as $S\tilde s$;
\item summations over SM quarks have index $f,f^\prime$, over squarks have index $i,j$, and over neutral scalars have index $I,J$; when summing over multiple indices, if the cross-section does not change by exchange of the indices, the summation is only for larger values of the second index ($\sum_i\sum_{j>i}$), otherwise the summation is just for different values  ($\sum_i\sum_{j \neq i}$)
\item $\hat\sigma_{x|y}^{\rm int}$ and $\hat\sigma_{x|B}^{\rm int}$ represent, respectively, the interference terms between the $x$ and $y$ classes of topologies and between $x$ and the SM background, while $\hat\sigma_{x|y-j|k}^{\rm int}$ or $\hat\sigma_{x|y-j|B}^{\rm int}$ signal combinations proportional to the same function of couplings;
\item for reduced cross-sections involving squarks or scalars, the summation index associated with the new particles is reported next to the topology class in the subscript of the $\hat\sigma$, {\it e.g.} $\sigma_{7i|7jk}^{\rm int}$ represent interference between two topologies of type \textbf{7} in which only the squark $i$ circulates in one and the squarks $j$ and $k$ circulate in the other;
\end{itemize}

The complete list of contributions is:
\begin{subequations}
\begin{align}
\sigma_{\rm M} &= 
\kappa_{hhh}^2 \hat\sigma_1 + 
\sum_{f=t,b}\Bigg[(\kappa_{hhh}\kappa_{hff})^2 \hat\sigma_{3f} + 
\kappa_{hff}^4 \hat\sigma_{4f} 
\Bigg] \;, \displaybreak[1]\\
\sigma_{\tilde s} &= \sum_{i} \Bigg[
\kappa_{h\tilde s \tilde s}^{ii} \sum_{j>i}\kappa_{h\tilde s \tilde s}^{jj}\hat\sigma_{5i|5j}^{\rm int}(m_{\tilde s_{i,j}}) + 
\kappa_{hhh}^2(\kappa_{h\tilde s \tilde s}^{ii})^2\hat\sigma_{6i}(m_{\tilde s_{i}}) + 
\kappa_{hhh}^2\kappa_{h\tilde s \tilde s}^{ii}\sum_{j>i}\kappa_{h\tilde s \tilde s}^{jj}\hat\sigma_{6i|6j}^{\rm int}(m_{\tilde s_{i,j}}) \nonumber\displaybreak[1]\\
&+
(\kappa_{h\tilde s \tilde s}^{ii})^4\hat\sigma_{7i}(m_{\tilde s_{i}}) + 
\sum_{j>i}(\kappa_{h\tilde s \tilde s}^{ij})^4\hat\sigma_{7ij}(m_{\tilde s_{i,j}}) + 
(\kappa_{h\tilde s \tilde s}^{ii})^2\sum_{j>i}(\kappa_{h\tilde s \tilde s}^{jj})^2\hat\sigma_{7i|7j}^{\rm int}(m_{\tilde s_{i,j}}) + 
(\kappa_{h\tilde s \tilde s}^{ii})^2\sum_{j\neq i}(\kappa_{h\tilde s \tilde s}^{ij})^2\hat\sigma_{7i|7ij}^{\rm int}(m_{\tilde s_{i,j}}) \nonumber\displaybreak[1]\\
&+ 
(\kappa_{h\tilde s \tilde s}^{ii})^2\sum_{j\neq i}\sum_{\scriptsize\setlength\arraycolsep{0pt}\begin{array}{c}k > j \\[-2pt] k \neq i\end{array}}(\kappa_{h\tilde s \tilde s}^{jk})^2\hat\sigma_{7i|7jk}^{\rm int}(m_{\tilde s_{i,j,k}}) +
\sum_{j>i} \sum_{\scriptsize\setlength\arraycolsep{0pt}\begin{array}{c}k>i \\[-2pt] k > j\end{array}} \kappa_{h\tilde s \tilde s}^{ij} \kappa_{h\tilde s \tilde s}^{ik} \hat\sigma_{7ij|7ik}^{\rm int}(m_{\tilde s_{i,j,k}}) + \sum_{j>i} \sum_{\scriptsize\setlength\arraycolsep{0pt}\begin{array}{c}k \neq i \\[-2pt] k \neq j\end{array}} \sum_{\scriptsize\setlength\arraycolsep{0pt}\begin{array}{c} l > k\\[-2pt] l \neq i \\[-2pt] l \neq j\end{array}} \kappa_{h\tilde s \tilde s}^{ij} \kappa_{h\tilde s \tilde s}^{kl} \hat\sigma_{7ij|7kl}^{\rm int}(m_{\tilde s_{i,j,k,l}})) \nonumber\displaybreak[1]\\
&+ 
(\kappa_{hh\tilde s \tilde s}^{ii})^2\hat\sigma_{8i}(m_{\tilde s_{i}}) +
\kappa_{hh\tilde s \tilde s}^{ii}\sum_{j>i}\kappa_{hh\tilde s \tilde s}^{jj}\hat\sigma_{8i|8j}^{\rm int}(m_{\tilde s_{i,j}}) \Bigg] \;, \label{eq:sigmaS}\displaybreak[1]\\
\sigma_{\rm S} &= 
\sum_{I} \Bigg[ \sum_{f=t,b} 
(\kappa_{Shh}^I \kappa_{Sff}^I)^2 \hat\sigma_{9If}(m_{S^0_I},\Gamma_{S^0_I}) + 
(\kappa_{Shh}^I)^2 \kappa_{Stt}^I \kappa_{Sbb}^I \hat\sigma_{9It|9Ib}^{\rm int}(m_{S^0_I},\Gamma_{S^0_I}) 
\nonumber\displaybreak[1]\\
&+ 
\sum_{J>I} \sum_{f=t,b} \kappa_{Shh}^I \kappa_{Shh}^J \kappa_{Sff}^I \kappa_{Sff}^J \hat\sigma_{9If|9Jf}^{\rm int}(m_{S^0_{I,J}},\Gamma_{S^0_{I,J}}) + 
\sum_{J\neq I} \kappa_{Shh}^I \kappa_{Shh}^J \kappa_{Stt}^I \kappa_{Sbb}^J \hat\sigma_{9It|9Jb}^{\rm int}(m_{S^0_{I,J}},\Gamma_{S^0_{I,J}})
\Bigg]\;, \displaybreak[1]\\
\sigma_{\rm S\tilde s} &= \sum_{i,I} \Bigg[ (\kappa_{Shh}^I \kappa_{S\tilde s \tilde s}^{Ii})^2 \hat\sigma_{10Ii}(m_{\tilde s_{i}},m_{S^0_I},\Gamma_{S^0_I}) + 
\sum_{J>I} \kappa_{Shh}^I \kappa_{Shh}^J \kappa_{S\tilde s \tilde s}^{Ii} \kappa_{S\tilde s \tilde s}^{Ji} \hat\sigma_{10Ii|10Ji}^{\rm int}(m_{\tilde s_{i}};m_{S^0_{I,J}},\Gamma_{S^0_{I,J}}) \nonumber\displaybreak[1]\\
&+ 
\sum_{j>i} (\kappa_{Shh}^I)^2 \kappa_{S\tilde s \tilde s}^{Ii} \kappa_{S\tilde s \tilde s}^{Ij} \hat\sigma_{10Ii|10Ij}^{\rm int}(m_{\tilde s_{i,j}};m_{S^0_I},\Gamma_{S^0_{I}}) + 
\sum_{J\neq I} \sum_{j \neq i} \kappa_{Shh}^I \kappa_{Shh}^J \kappa_{S\tilde s \tilde s}^{Ii} \kappa_{S\tilde s \tilde s}^{Jj} \hat\sigma_{10Ii|10Jj}^{\rm int}(m_{\tilde s_{i,j}};m_{S^0_{I,J}},\Gamma_{S^0_{I,J}})
\Bigg]\;, \displaybreak[1]\\
\sigma_{\rm M|B}^{\rm int} &=
\kappa_{hhh} \hat\sigma_{1|B} + 
\sum_{f=t,b} \kappa_{hff}\hat\sigma_{2f|B} ;, \displaybreak[1]\\
%
%
%
%
\sigma_{\tilde s|B}^{\rm int} &= \sum_{i} \Bigg[
\kappa_{h\tilde s \tilde s}^{ii}\hat\sigma_{5i|B}^{\rm int}(m_{\tilde s_i}) + 
\sum_{j>i}(\kappa_{h\tilde s \tilde s}^{ij})^2\hat\sigma_{7ij|B}^{\rm int}(m_{\tilde s_{i,j}}) + 
\kappa_{hh\tilde s \tilde s}^{ii}\hat\sigma_{8i|B}^{\rm int}(m_{\tilde s_i}) \Bigg]\;, \displaybreak[1]\\
\sigma_{\rm S|B}^{\rm int} &= 
\sum_{I} \sum_{f=t,b} 
\kappa_{Shh}^I \kappa_{Sff}^I \hat\sigma_{9If|B}^{\rm int}(m_{S^0_I},\Gamma_{S^0_I})
\;, \displaybreak[1]\\
\sigma_{\rm S\tilde s|B}^{\rm int} &= 
\sum_{i,I}
\kappa_{Shh}^I \kappa_{S\tilde s\tilde s}^i \hat\sigma_{10Ii|B}^{\rm int}(m_{\tilde s_i};m_{S^0_I},\Gamma_{S^0_I})
\;, \displaybreak[1]\\
\sigma_{\rm M|M}^{\rm int} &= 
\sum_{f=t,b} \Bigg[
\kappa_{hhh}^2\kappa_{hff}\hat\sigma_{1|3f}^{\rm int} + 
\kappa_{hhh}\kappa_{hff}^2\hat\sigma_{1|4f-2f|3f}^{\rm int} +
\kappa_{hff}^3\hat\sigma_{2f|4f}^{\rm int} + 
\kappa_{hhh}\kappa_{hff}^3\hat\sigma_{3f|4f}^{\rm int}
\nonumber\displaybreak[1]\\
&+
\sum_{f^\prime\neq f} \bigg(
\kappa_{hhh}\kappa_{hff}\kappa_{hf^\prime f^\prime}\hat\sigma_{2f|3f^\prime}^{\rm int} + 
\kappa_{hff}\kappa_{hf^\prime f^\prime}^2\hat\sigma_{2f|4f^\prime}^{\rm int} + 
\kappa_{hhh}\kappa_{hff}\kappa_{f^\prime f^\prime}^2\hat\sigma_{3f|4f^\prime}^{\rm int}
\bigg)\Bigg] \;, \displaybreak[1]\\
\sigma_{\tilde s|\tilde s}^{\rm int} &= \sum_{i} \Bigg[
\kappa_{hhh}\kappa_{h\tilde s \tilde s}^{ii}\sum_{j> i}\kappa_{h\tilde s \tilde s}^{jj}\hat\sigma_{5i|6j}^{\rm int}(m_{\tilde s_{i,j}}) +
(\kappa_{h\tilde s \tilde s}^{ii})^3\hat\sigma_{5i|7i}^{\rm int}(m_{\tilde s_{i}}) + 
\kappa_{h\tilde s \tilde s}^{ii}\sum_{j\neq i}(\kappa_{h\tilde s \tilde s}^{jj})^2\hat\sigma_{5i|7j}^{\rm int}(m_{\tilde s_{i,j}}) \nonumber\displaybreak[1]\\
&+
\kappa_{h\tilde s \tilde s}^{ii}\sum_{j\neq i}(\kappa_{h\tilde s \tilde s}^{ij})^2\hat\sigma_{5i|7ij}^{\rm int}(m_{\tilde s_{i,j}}) +
\kappa_{h\tilde s \tilde s}^{ii}\sum_{j\neq i}\sum_{\scriptsize\setlength\arraycolsep{0pt}\begin{array}{c}k >j \\[-2pt] k \neq i\end{array}}(\kappa_{h\tilde s \tilde s}^{jk})^2\hat\sigma_{5i|7jk}^{\rm int}(m_{\tilde s_{i,j,k}}) +
\kappa_{h\tilde s \tilde s}^{ii}\kappa_{hh\tilde s \tilde s}^{ii}\hat\sigma_{5i|8i}^{\rm int}(m_{\tilde s_{i}})\nonumber\displaybreak[1]\\
&+
\kappa_{h\tilde s \tilde s}^{ii}\sum_{j\neq i}\kappa_{hh\tilde s \tilde s}^{jj}\hat\sigma_{5i|8j}^{\rm int}(m_{\tilde s_{i,j}}) +
\kappa_{hhh}(\kappa_{h\tilde s \tilde s}^{ii})^3\hat\sigma_{6i|7i}^{\rm int}(m_{\tilde s_{i}}) +
\kappa_{hhh}\kappa_{h\tilde s \tilde s}^{ii}\sum_{j\neq i}(\kappa_{h\tilde s \tilde s}^{jj})^2\hat\sigma_{6i|7j}^{\rm int}(m_{\tilde s_{i,j}}) \nonumber\displaybreak[1]\\
&+ 
\kappa_{hhh}\kappa_{h\tilde s \tilde s}^{ii}\sum_{j\neq i}(\kappa_{h\tilde s \tilde s}^{ij})^2\hat\sigma_{6i|7ij}^{\rm int}(m_{\tilde s_{i,j}}) +
\kappa_{hhh}\kappa_{h\tilde s \tilde s}^{ii}\sum_{j\neq i}\sum_{\scriptsize\setlength\arraycolsep{0pt}\begin{array}{c}k >j \\[-2pt] k \neq i\end{array}}(\kappa_{h\tilde s \tilde s}^{jk})^2\hat\sigma_{6i|7jk}^{\rm int}(m_{\tilde s_{i,j,k}}) +
\kappa_{hhh}\kappa_{h\tilde s \tilde s}^{ii}\kappa_{hh\tilde s \tilde s}^{ii}\hat\sigma_{6i|8i}^{\rm int}(m_{\tilde s_{i}}) \nonumber\displaybreak[1]\\
&+ 
\kappa_{hhh}\kappa_{h\tilde s \tilde s}^{ii}\sum_{j\neq i}\kappa_{hh\tilde s \tilde s}^{jj}\hat\sigma_{6i|8j}^{\rm int}(m_{\tilde s_{i,j}}) + 
(\kappa_{h\tilde s \tilde s}^{ii})^2 \kappa_{hh\tilde s \tilde s}^{ii}\hat\sigma_{7i|8i}^{\rm int}(m_{\tilde s_{i}}) + 
(\kappa_{h\tilde s \tilde s}^{ii})^2 \sum_{j\neq i}\kappa_{hh\tilde s \tilde s}^{jj}\hat\sigma_{7i|8j}^{\rm int}(m_{\tilde s_{i,j}}) \nonumber\displaybreak[1]\\
&+ 
\sum_{j\neq i}(\kappa_{h\tilde s \tilde s}^{ij})^2 \kappa_{hh\tilde s \tilde s}^{ii}\hat\sigma_{7ij|8i}^{\rm int}(m_{\tilde s_{i,j}}) + 
\sum_{j>i}\sum_{\scriptsize\setlength\arraycolsep{0pt}\begin{array}{c}k\neq i \\[-2pt] k \neq j\end{array}}(\kappa_{h\tilde s \tilde s}^{ij})^2 \kappa_{hh\tilde s \tilde s}^{kk}\hat\sigma_{7ij|8k}^{\rm int}(m_{\tilde s_{i,j,k}})\Bigg] \label{eq:sigmaSS}\;,\displaybreak[1]\\
\sigma_{\rm M|\tilde s}^{\rm int} &= 
\sum_{i} \Bigg[
\kappa_{hhh}^2\kappa_{h\tilde s \tilde s}^{ii} \hat\sigma_{1|6i}^{\rm int}(m_{\tilde s_i}) + 
\kappa_{hhh}\sum_{j>i}(\kappa_{h\tilde s \tilde s}^{ij})^2 \hat\sigma_{1|7ij}^{\rm int}(m_{\tilde s_{i,j}}) + 
\kappa_{hhh}\kappa_{hh\tilde s \tilde s}^{ii}\hat\sigma_{1|8i}^{\rm int}(m_{\tilde s_i}) \nonumber\displaybreak[1]\\
&+ 
\sum_{f=t,b} \bigg(\kappa_{hff}\kappa_{h\tilde s \tilde s}^{ii} \hat\sigma_{2f|5i}^{\rm int}(m_{\tilde s_i}) +
\kappa_{hhh}\kappa_{hff}\kappa_{h\tilde s \tilde s}^{ii} \hat\sigma_{2f|6i-3f|5i}^{\rm int}(m_{\tilde s_i}) +
\kappa_{hff}(\kappa_{h\tilde s \tilde s}^{ii})^2 \hat\sigma_{2f|7o}^{\rm int}(m_{\tilde s_{i}}) \nonumber\displaybreak[1]\\
&+ 
\kappa_{hff}\sum_{j>i}(\kappa_{h\tilde s \tilde s}^{ij})^2 \hat\sigma_{2f|7ij}^{\rm int}(m_{\tilde s_{i,j}}) +
\kappa_{hff}\kappa_{hh\tilde s \tilde s}^{ii}\hat\sigma_{2f|8i}^{\rm int}(m_{\tilde s_i}) +
\kappa_{hhh}^2\kappa_{hff}\kappa_{h\tilde s \tilde s}^{ii} \hat\sigma_{3f|6i}^{\rm int}(m_{\tilde s_i}) +
\kappa_{hhh}\kappa_{hff}(\kappa_{h\tilde s \tilde s}^{ii})^2 \hat\sigma_{3f|7i}^{\rm int}(m_{\tilde s_i}) \nonumber\displaybreak[1]\\
&+
\kappa_{hhh}\kappa_{hff}\sum_{j>i}(\kappa_{h\tilde s \tilde s}^{ij})^2 \hat\sigma_{3f|7ij}^{\rm int}(m_{\tilde s_{i,j}}) +
\kappa_{hhh}\kappa_{hff}\kappa_{hh\tilde s \tilde s}^{ii} \hat\sigma_{3f|8i}^{\rm int}(m_{\tilde s_i}) +
\kappa_{hff}^2\kappa_{h\tilde s \tilde s}^{ii} \hat\sigma_{4f|5i}^{\rm int}(m_{\tilde s_i}) + 
\kappa_{hhh}\kappa_{hff}^2\kappa_{h\tilde s \tilde s}^{ii} \hat\sigma_{4f|6i}^{\rm int}(m_{\tilde s_i}) \nonumber\displaybreak[1]\\
&+ 
\kappa_{hff}^2(\kappa_{h\tilde s \tilde s}^{ii})^2 \hat\sigma_{4f|7i}^{\rm int}(m_{\tilde s_i}) +
\kappa_{hff}^2\sum_{j>i}(\kappa_{h\tilde s \tilde s}^{ij})^2 \hat\sigma_{4f|7ij}^{\rm int}(m_{\tilde s_{i,j}}) + 
\kappa_{hff}^2\kappa_{hh\tilde s \tilde s}^{ii} \hat\sigma_{4f|8i}^{\rm int}(m_{\tilde s_i})
\bigg)\Bigg] \;, \displaybreak[1]\\
\sigma_{\rm M|S}^{\rm int} &= 
\sum_{I} \sum_{f=t,b} \Bigg[
\kappa_{hhh}\kappa_{Shh}^I \kappa_{Sff}^I \hat\sigma_{1|9If}^{\rm int}(m_{S^0_I},\Gamma_{S^0_I}) + 
\kappa_{hff}\kappa_{Shh}^I \kappa_{Sff}^I \hat\sigma_{2f|9If}^{\rm int}(m_{S^0_I},\Gamma_{S^0_I}) \nonumber\displaybreak[1]\\
&+ 
\kappa_{hhh}\kappa_{hff}\kappa_{Shh}^I \kappa_{Sff}^I \hat\sigma_{3f|9If}^{\rm int}(m_{S^0_I},\Gamma_{S^0_I}) + 
\kappa_{hff}^2\kappa_{Shh}^I \kappa_{Sff}^I \hat\sigma_{4f|9If}^{\rm int}(m_{S^0_I},\Gamma_{S^0_I}) \nonumber\displaybreak[1]\\
&+  
\sum_{f^\prime\neq f} \bigg(
\kappa_{hff}\kappa_{Shh}^I \kappa_{Sf^\prime f^\prime}^I \hat\sigma_{2f|9If^\prime}^{\rm int}(m_{S^0_I},\Gamma_{S^0_I}) + 
\kappa_{hhh}\kappa_{hff}\kappa_{Shh}^I \kappa_{Sf^\prime f^\prime}^I \hat\sigma_{3f|9If^\prime}^{\rm int}(m_{S^0_I},\Gamma_{S^0_I})\nonumber\displaybreak[1]\\
&+ 
\kappa_{hff}^2\kappa_{Shh}^I \kappa_{Sf^\prime f^\prime}^I \hat\sigma_{4f|9If^\prime}^{\rm int}(m_{S^0_I},\Gamma_{S^0_I})
\bigg)\Bigg]
\displaybreak[1]\\
\sigma_{\rm M|S\tilde s}^{\rm int} &= 
\sum_{I,i} \Bigg[
\kappa_{hhh}\kappa_{Shh}^I \kappa_{S\tilde s\tilde s}^{Ii} \hat\sigma_{1|10Ii}^{\rm int}(m_{\tilde s_i};m_{S^0_I},\Gamma_{S^0_I}) + 
\sum_{f=t,b} \bigg(
\kappa_{hff}\kappa_{Shh}^I \kappa_{S\tilde s\tilde s}^{Ii} \hat\sigma_{2f|10Ii}^{\rm int}(m_{\tilde s_i};m_{S^0_I},\Gamma_{S^0_I}) \nonumber\displaybreak[1]\\
&+
\kappa_{hhh}\kappa_{hff}\kappa_{Shh}^I \kappa_{S\tilde s\tilde s}^{Ii} \hat\sigma_{3f|10Ii}^{\rm int}(m_{\tilde s_i};m_{S^0_I},\Gamma_{S^0_I}) + 
\kappa_{hff}^2 \kappa_{Shh}^I \kappa_{S\tilde s\tilde s}^{Ii} \hat\sigma_{4f|10Ii}^{\rm int}(m_{\tilde s_i};m_{S^0_I},\Gamma_{S^0_I}) 
\bigg)\Bigg]
\displaybreak[1]\\
\sigma_{\rm \tilde s|S}^{\rm int} &= 
\sum_{I,i} \sum_{f=t,b} \Bigg[
\kappa_{h\tilde s\tilde s}^{i}\kappa_{Shh}^I\kappa_{Sff}^I \hat\sigma_{5i|9If}^{\rm int}(m_{\tilde s_i};m_{S^0_I},\Gamma_{S^0_I}) +
\kappa_{hhh} \kappa_{h\tilde s\tilde s}^{i}\kappa_{Shh}^I\kappa_{Sff}^I \hat\sigma_{6i|9If}^{\rm int}(m_{\tilde s_i};m_{S^0_I},\Gamma_{S^0_I}) \nonumber\displaybreak[1]\\
&+
(\kappa_{h\tilde s\tilde s}^{i})^2\kappa_{Shh}^I\kappa_{Sff}^I \hat\sigma_{7i|9If}^{\rm int}(m_{\tilde s_i};m_{S^0_I},\Gamma_{S^0_I}) +
\sum_{j>i} \kappa_{h\tilde s\tilde s}^{ij}\kappa_{Shh}^I\kappa_{Sff}^I \hat\sigma_{7ij|9If}^{\rm int}(m_{\tilde s_{i,j}};m_{S^0_I},\Gamma_{S^0_I}) \nonumber\displaybreak[1]\\
&+
\kappa_{hh\tilde s\tilde s}^i\kappa_{Shh}^I\kappa_{Sff}^I \hat\sigma_{8i|9If}^{\rm int}(m_{\tilde s_i};m_{S^0_I},\Gamma_{S^0_I})
\Bigg]
\displaybreak[1]\\
\sigma_{\rm \tilde s|S\tilde s}^{\rm int} &= 
\sum_{I,i,j} \Bigg[
\kappa_{h\tilde s\tilde s}^{i}\kappa_{Shh}^I\kappa_{S\tilde s\tilde s}^{Ij} \hat\sigma_{5i|10Ij}^{\rm int}(m_{\tilde s_{i,j}};m_{S^0_I},\Gamma_{S^0_I}) +
\kappa_{hhh} \kappa_{h\tilde s\tilde s}^{i}\kappa_{Shh}^I\kappa_{S\tilde s\tilde s}^{Ij} \hat\sigma_{6i|10Ij}^{\rm int}(m_{\tilde s_{i,j}};m_{S^0_I},\Gamma_{S^0_I}) \nonumber\displaybreak[1]\\
&+
\sum_{k\geq i} (\kappa_{h\tilde s\tilde s}^{ik})^2\kappa_{Shh}^I\kappa_{S\tilde s\tilde s}^{Ij} \hat\sigma_{7ik|10Ij}^{\rm int}(m_{\tilde s_{i,j,k}};m_{S^0_I},\Gamma_{S^0_I}) +
\kappa_{hh\tilde s\tilde s}^{i}\kappa_{Shh}^I\kappa_{S\tilde s\tilde s}^{Ij} \hat\sigma_{8i|10Ij}^{\rm int}(m_{\tilde s_{i,j}};m_{S^0_I},\Gamma_{S^0_I})\Bigg]
\displaybreak[1]\\
\sigma_{\rm S|S\tilde s}^{\rm int} &= 
\sum_{I,J,i} \sum_{f=t,b} 
\kappa_{Shh}^I\kappa_{Sff}^I\kappa_{Shh}^J\kappa_{S\tilde s\tilde s}^{Ji} \hat\sigma_{9If|10Ji}^{\rm int}(m_{\tilde s_i};m_{S^0_{I,J}},\Gamma_{S^0_{I,J}})
\displaybreak[1]\\
\sigma_{\rm mix}^{\rm int} &=
\sum_{f=t,b}\Bigg[ \kappa_{hhh}\kappa_{hff} \hat\sigma_{1|2f-3f|B}^{\rm int} + 
\kappa_{hff}^2 \hat\sigma_{2f-4f|B}^{\rm int} \Bigg] \nonumber\displaybreak[1]\\
&+
\sum_{i}\Bigg[\kappa_{hhh} \kappa_{h\tilde s \tilde s}^{ii}\hat\sigma_{1|5i-6i|B}^{\rm int}(m_{\tilde s_i}) + 
\kappa_{hhh}(\kappa_{h\tilde s \tilde s}^{ii})^2 \hat\sigma_{1|7i-5i|6i}^{\rm int}(m_{\tilde s_{i}}) +
(\kappa_{h\tilde s \tilde s}^{ii})^2\hat\sigma_{5i-7i|B}^{\rm int}(m_{\tilde s_i})\Bigg]
\;. 
\label{eq:sigmaMSB}
\end{align}
\label{eq:sigmahats}
\end{subequations}

\section{UFO model content}
\label{app:UFOmodel}
The UFO model provided with this analysis can be downloaded at \url{https://github.com/FeynRules/Models/tree/main/DiHiggs-BSM-Simplified} contains the minimal particle content to generate any signal and interference term for di-Higgs studies involving the propagation of new coloured or neutral particles, at LO, using {\sc MG5\_aMC}.

As mentioned in \Cref{sec:deconstruction}, this amount in 4 coloured scalars of unspecified charge, labelled as \cbc{\verb~sq1, sq2, sq3~} and \cbc{\verb~sq4~} (already present in the UFO model used for our previous analysis~\cite{Moretti:2023dlx}), 2 neutral scalars labelled as \cbc{\verb~s01~} and \cbc{\verb~s02~}, 4 vector-like quarks with charge 2/3 (\cbc{\verb~tp1, tp2, tp3~} and \cbc{\verb~tp4~}) and 3 vector-like quarks with charge -1/3 ((\cbc{\verb~bp1, bp2~} and \cbc{\verb~bp3~}). The new fermions have not been used in this analysis but will be included in a future complete treatment of the process. For fermions the charge is important, as they can mix with the SM third generation fermions, and the different number of top and bottom partners is again related to the minimal number of particles necessary to describe all combinations appearing when squaring amplitudes in the deconstruction. The couplings, coupling orders associated to the new particles, and simulation syntax for deconstructed elements follow the same rationale described in Appendix A of~\cite{Moretti:2023dlx}, to which we refer for further details.

The model contains three restriction cards. They are for the two Higgs doublet model (2HDM), the MSSM and NMSSM particle contents. Naturally these can be used for any other model with the same relevant particle content, \textit{e.g.} the 2HDM restriction card would be appropriate also for the SM extended with a singlet Higgs. In order to use the model, one will need to calculate the couplings using the conventions of \Cref{eq:LagNP}.

\bibliography{resonantdihiggs}

\begin{thebibliography}{68}%
\makeatletter
\providecommand \@ifxundefined [1]{%
 \@ifx{#1\undefined}
}%
\providecommand \@ifnum [1]{%
 \ifnum #1\expandafter \@firstoftwo
 \else \expandafter \@secondoftwo
 \fi
}%
\providecommand \@ifx [1]{%
 \ifx #1\expandafter \@firstoftwo
 \else \expandafter \@secondoftwo
 \fi
}%
\providecommand \natexlab [1]{#1}%
\providecommand \enquote  [1]{``#1''}%
\providecommand \bibnamefont  [1]{#1}%
\providecommand \bibfnamefont [1]{#1}%
\providecommand \citenamefont [1]{#1}%
\providecommand \href@noop [0]{\@secondoftwo}%
\providecommand \href [0]{\begingroup \@sanitize@url \@href}%
\providecommand \@href[1]{\@@startlink{#1}\@@href}%
\providecommand \@@href[1]{\endgroup#1\@@endlink}%
\providecommand \@sanitize@url [0]{\catcode `\\12\catcode `\$12\catcode
  `\&12\catcode `\#12\catcode `\^12\catcode `\_12\catcode `\%12\relax}%
\providecommand \@@startlink[1]{}%
\providecommand \@@endlink[0]{}%
\providecommand \url  [0]{\begingroup\@sanitize@url \@url }%
\providecommand \@url [1]{\endgroup\@href {#1}{\urlprefix }}%
\providecommand \urlprefix  [0]{URL }%
\providecommand \Eprint [0]{\href }%
\providecommand \doibase [0]{https://doi.org/}%
\providecommand \selectlanguage [0]{\@gobble}%
\providecommand \bibinfo  [0]{\@secondoftwo}%
\providecommand \bibfield  [0]{\@secondoftwo}%
\providecommand \translation [1]{[#1]}%
\providecommand \BibitemOpen [0]{}%
\providecommand \bibitemStop [0]{}%
\providecommand \bibitemNoStop [0]{.\EOS\space}%
\providecommand \EOS [0]{\spacefactor3000\relax}%
\providecommand \BibitemShut  [1]{\csname bibitem#1\endcsname}%
\let\auto@bib@innerbib\@empty
\bibitem [{\citenamefont {Glover}\ and\ \citenamefont {van~der
  Bij}(1988)}]{Glover:1987nx}%
  \BibitemOpen
  \bibfield  {author} {\bibinfo {author} {\bibfnamefont {E.~W.~N.}\
  \bibnamefont {Glover}}\ and\ \bibinfo {author} {\bibfnamefont {J.~J.}\
  \bibnamefont {van~der Bij}},\ }\bibfield  {title} {\bibinfo {title} {{HIGGS
  BOSON PAIR PRODUCTION VIA GLUON FUSION}},\ }\href
  {https://doi.org/10.1016/0550-3213(88)90083-1} {\bibfield  {journal}
  {\bibinfo  {journal} {Nucl. Phys. B}\ }\textbf {\bibinfo {volume} {309}},\
  \bibinfo {pages} {282} (\bibinfo {year} {1988})}\BibitemShut {NoStop}%
\bibitem [{\citenamefont {Dicus}\ \emph {et~al.}(1988)\citenamefont {Dicus},
  \citenamefont {Kao},\ and\ \citenamefont {Willenbrock}}]{Dicus:1987ic}%
  \BibitemOpen
  \bibfield  {author} {\bibinfo {author} {\bibfnamefont {D.~A.}\ \bibnamefont
  {Dicus}}, \bibinfo {author} {\bibfnamefont {C.}~\bibnamefont {Kao}},\ and\
  \bibinfo {author} {\bibfnamefont {S.~S.~D.}\ \bibnamefont {Willenbrock}},\
  }\bibfield  {title} {\bibinfo {title} {{Higgs Boson Pair Production From
  Gluon Fusion}},\ }\href {https://doi.org/10.1016/0370-2693(88)90202-X}
  {\bibfield  {journal} {\bibinfo  {journal} {Phys. Lett. B}\ }\textbf
  {\bibinfo {volume} {203}},\ \bibinfo {pages} {457} (\bibinfo {year}
  {1988})}\BibitemShut {NoStop}%
\bibitem [{\citenamefont {Kanemura}\ \emph {et~al.}(2003)\citenamefont
  {Kanemura}, \citenamefont {Kiyoura}, \citenamefont {Okada}, \citenamefont
  {Senaha},\ and\ \citenamefont {Yuan}}]{Kanemura:2002vm}%
  \BibitemOpen
  \bibfield  {author} {\bibinfo {author} {\bibfnamefont {S.}~\bibnamefont
  {Kanemura}}, \bibinfo {author} {\bibfnamefont {S.}~\bibnamefont {Kiyoura}},
  \bibinfo {author} {\bibfnamefont {Y.}~\bibnamefont {Okada}}, \bibinfo
  {author} {\bibfnamefont {E.}~\bibnamefont {Senaha}},\ and\ \bibinfo {author}
  {\bibfnamefont {C.~P.}\ \bibnamefont {Yuan}},\ }\bibfield  {title} {\bibinfo
  {title} {{New physics effect on the Higgs selfcoupling}},\ }\href
  {https://doi.org/10.1016/S0370-2693(03)00268-5} {\bibfield  {journal}
  {\bibinfo  {journal} {Phys. Lett. B}\ }\textbf {\bibinfo {volume} {558}},\
  \bibinfo {pages} {157} (\bibinfo {year} {2003})},\ \Eprint
  {https://arxiv.org/abs/hep-ph/0211308} {arXiv:hep-ph/0211308} \BibitemShut
  {NoStop}%
\bibitem [{\citenamefont {Noble}\ and\ \citenamefont
  {Perelstein}(2008)}]{Noble:2007kk}%
  \BibitemOpen
  \bibfield  {author} {\bibinfo {author} {\bibfnamefont {A.}~\bibnamefont
  {Noble}}\ and\ \bibinfo {author} {\bibfnamefont {M.}~\bibnamefont
  {Perelstein}},\ }\bibfield  {title} {\bibinfo {title} {{Higgs self-coupling
  as a probe of electroweak phase transition}},\ }\href
  {https://doi.org/10.1103/PhysRevD.78.063518} {\bibfield  {journal} {\bibinfo
  {journal} {Phys. Rev. D}\ }\textbf {\bibinfo {volume} {78}},\ \bibinfo
  {pages} {063518} (\bibinfo {year} {2008})},\ \Eprint
  {https://arxiv.org/abs/0711.3018} {arXiv:0711.3018 [hep-ph]} \BibitemShut
  {NoStop}%
\bibitem [{\citenamefont {Rodejohann}\ and\ \citenamefont
  {Zhang}(2012)}]{Rodejohann:2012px}%
  \BibitemOpen
  \bibfield  {author} {\bibinfo {author} {\bibfnamefont {W.}~\bibnamefont
  {Rodejohann}}\ and\ \bibinfo {author} {\bibfnamefont {H.}~\bibnamefont
  {Zhang}},\ }\bibfield  {title} {\bibinfo {title} {{Impact of massive
  neutrinos on the Higgs self-coupling and electroweak vacuum stability}},\
  }\href {https://doi.org/10.1007/JHEP06(2012)022} {\bibfield  {journal}
  {\bibinfo  {journal} {JHEP}\ }\textbf {\bibinfo {volume} {06}},\ \bibinfo
  {pages} {022}},\ \Eprint {https://arxiv.org/abs/1203.3825} {arXiv:1203.3825
  [hep-ph]} \BibitemShut {NoStop}%
\bibitem [{\citenamefont {Wu}\ \emph {et~al.}(2015)\citenamefont {Wu},
  \citenamefont {Yang}, \citenamefont {Yuan},\ and\ \citenamefont
  {Zhang}}]{Wu:2015nba}%
  \BibitemOpen
  \bibfield  {author} {\bibinfo {author} {\bibfnamefont {L.}~\bibnamefont
  {Wu}}, \bibinfo {author} {\bibfnamefont {J.~M.}\ \bibnamefont {Yang}},
  \bibinfo {author} {\bibfnamefont {C.-P.}\ \bibnamefont {Yuan}},\ and\
  \bibinfo {author} {\bibfnamefont {M.}~\bibnamefont {Zhang}},\ }\bibfield
  {title} {\bibinfo {title} {{Higgs self-coupling in the MSSM and NMSSM after
  the LHC Run 1}},\ }\href {https://doi.org/10.1016/j.physletb.2015.06.020}
  {\bibfield  {journal} {\bibinfo  {journal} {Phys. Lett. B}\ }\textbf
  {\bibinfo {volume} {747}},\ \bibinfo {pages} {378} (\bibinfo {year}
  {2015})},\ \Eprint {https://arxiv.org/abs/1504.06932} {arXiv:1504.06932
  [hep-ph]} \BibitemShut {NoStop}%
\bibitem [{\citenamefont {Di~Luzio}\ \emph {et~al.}(2017)\citenamefont
  {Di~Luzio}, \citenamefont {Gr\"ober},\ and\ \citenamefont
  {Spannowsky}}]{DiLuzio:2017tfn}%
  \BibitemOpen
  \bibfield  {author} {\bibinfo {author} {\bibfnamefont {L.}~\bibnamefont
  {Di~Luzio}}, \bibinfo {author} {\bibfnamefont {R.}~\bibnamefont {Gr\"ober}},\
  and\ \bibinfo {author} {\bibfnamefont {M.}~\bibnamefont {Spannowsky}},\
  }\bibfield  {title} {\bibinfo {title} {{Maxi-sizing the trilinear Higgs
  self-coupling: how large could it be?}},\ }\href
  {https://doi.org/10.1140/epjc/s10052-017-5361-0} {\bibfield  {journal}
  {\bibinfo  {journal} {Eur. Phys. J. C}\ }\textbf {\bibinfo {volume} {77}},\
  \bibinfo {pages} {788} (\bibinfo {year} {2017})},\ \Eprint
  {https://arxiv.org/abs/1704.02311} {arXiv:1704.02311 [hep-ph]} \BibitemShut
  {NoStop}%
\bibitem [{\citenamefont {Bahl}\ \emph {et~al.}(2022)\citenamefont {Bahl},
  \citenamefont {Braathen},\ and\ \citenamefont {Weiglein}}]{Bahl:2022jnx}%
  \BibitemOpen
  \bibfield  {author} {\bibinfo {author} {\bibfnamefont {H.}~\bibnamefont
  {Bahl}}, \bibinfo {author} {\bibfnamefont {J.}~\bibnamefont {Braathen}},\
  and\ \bibinfo {author} {\bibfnamefont {G.}~\bibnamefont {Weiglein}},\
  }\bibfield  {title} {\bibinfo {title} {{New Constraints on Extended Higgs
  Sectors from the Trilinear Higgs Coupling}},\ }\href
  {https://doi.org/10.1103/PhysRevLett.129.231802} {\bibfield  {journal}
  {\bibinfo  {journal} {Phys. Rev. Lett.}\ }\textbf {\bibinfo {volume} {129}},\
  \bibinfo {pages} {231802} (\bibinfo {year} {2022})},\ \Eprint
  {https://arxiv.org/abs/2202.03453} {arXiv:2202.03453 [hep-ph]} \BibitemShut
  {NoStop}%
\bibitem [{\citenamefont {Batell}\ \emph {et~al.}(2015)\citenamefont {Batell},
  \citenamefont {McCullough}, \citenamefont {Stolarski},\ and\ \citenamefont
  {Verhaaren}}]{Batell:2015koa}%
  \BibitemOpen
  \bibfield  {author} {\bibinfo {author} {\bibfnamefont {B.}~\bibnamefont
  {Batell}}, \bibinfo {author} {\bibfnamefont {M.}~\bibnamefont {McCullough}},
  \bibinfo {author} {\bibfnamefont {D.}~\bibnamefont {Stolarski}},\ and\
  \bibinfo {author} {\bibfnamefont {C.~B.}\ \bibnamefont {Verhaaren}},\
  }\bibfield  {title} {\bibinfo {title} {{Putting a Stop to di-Higgs
  Modifications}},\ }\href {https://doi.org/10.1007/JHEP09(2015)216} {\bibfield
   {journal} {\bibinfo  {journal} {JHEP}\ }\textbf {\bibinfo {volume} {09}},\
  \bibinfo {pages} {216}},\ \Eprint {https://arxiv.org/abs/1508.01208}
  {arXiv:1508.01208 [hep-ph]} \BibitemShut {NoStop}%
\bibitem [{\citenamefont {Huang}\ \emph {et~al.}(2018)\citenamefont {Huang},
  \citenamefont {Joglekar}, \citenamefont {Li},\ and\ \citenamefont
  {Wagner}}]{Huang:2017nnw}%
  \BibitemOpen
  \bibfield  {author} {\bibinfo {author} {\bibfnamefont {P.}~\bibnamefont
  {Huang}}, \bibinfo {author} {\bibfnamefont {A.}~\bibnamefont {Joglekar}},
  \bibinfo {author} {\bibfnamefont {M.}~\bibnamefont {Li}},\ and\ \bibinfo
  {author} {\bibfnamefont {C.~E.~M.}\ \bibnamefont {Wagner}},\ }\bibfield
  {title} {\bibinfo {title} {{Corrections to di-Higgs boson production with
  light stops and modified Higgs couplings}},\ }\href
  {https://doi.org/10.1103/PhysRevD.97.075001} {\bibfield  {journal} {\bibinfo
  {journal} {Phys. Rev. D}\ }\textbf {\bibinfo {volume} {97}},\ \bibinfo
  {pages} {075001} (\bibinfo {year} {2018})},\ \Eprint
  {https://arxiv.org/abs/1711.05743} {arXiv:1711.05743 [hep-ph]} \BibitemShut
  {NoStop}%
\bibitem [{\citenamefont {Moretti}\ \emph {et~al.}(2023)\citenamefont
  {Moretti}, \citenamefont {Panizzi}, \citenamefont {Sj\"olin},\ and\
  \citenamefont {Waltari}}]{Moretti:2023dlx}%
  \BibitemOpen
  \bibfield  {author} {\bibinfo {author} {\bibfnamefont {S.}~\bibnamefont
  {Moretti}}, \bibinfo {author} {\bibfnamefont {L.}~\bibnamefont {Panizzi}},
  \bibinfo {author} {\bibfnamefont {J.}~\bibnamefont {Sj\"olin}},\ and\
  \bibinfo {author} {\bibfnamefont {H.}~\bibnamefont {Waltari}},\ }\bibfield
  {title} {\bibinfo {title} {{Deconstructing squark contributions to di-Higgs
  production at the LHC}},\ }\href
  {https://doi.org/10.1103/PhysRevD.107.115010} {\bibfield  {journal} {\bibinfo
   {journal} {Phys. Rev. D}\ }\textbf {\bibinfo {volume} {107}},\ \bibinfo
  {pages} {115010} (\bibinfo {year} {2023})},\ \Eprint
  {https://arxiv.org/abs/2302.03401} {arXiv:2302.03401 [hep-ph]} \BibitemShut
  {NoStop}%
\bibitem [{\citenamefont {De~Curtis}\ \emph {et~al.}(2024)\citenamefont
  {De~Curtis}, \citenamefont {Delle~Rose}, \citenamefont {Egle}, \citenamefont
  {Moretti}, \citenamefont {M\"uhlleitner},\ and\ \citenamefont
  {Sakurai}}]{DeCurtis:2023pus}%
  \BibitemOpen
  \bibfield  {author} {\bibinfo {author} {\bibfnamefont {S.}~\bibnamefont
  {De~Curtis}}, \bibinfo {author} {\bibfnamefont {L.}~\bibnamefont
  {Delle~Rose}}, \bibinfo {author} {\bibfnamefont {F.}~\bibnamefont {Egle}},
  \bibinfo {author} {\bibfnamefont {S.}~\bibnamefont {Moretti}}, \bibinfo
  {author} {\bibfnamefont {M.}~\bibnamefont {M\"uhlleitner}},\ and\ \bibinfo
  {author} {\bibfnamefont {K.}~\bibnamefont {Sakurai}},\ }\bibfield  {title}
  {\bibinfo {title} {{Composite 2-Higgs doublet model: strong effects on Higgs
  pair production}},\ }\href {https://doi.org/10.1007/JHEP06(2024)063}
  {\bibfield  {journal} {\bibinfo  {journal} {JHEP}\ }\textbf {\bibinfo
  {volume} {06}},\ \bibinfo {pages} {063}},\ \Eprint
  {https://arxiv.org/abs/2310.10471} {arXiv:2310.10471 [hep-ph]} \BibitemShut
  {NoStop}%
\bibitem [{\citenamefont {Plehn}\ \emph {et~al.}(1996)\citenamefont {Plehn},
  \citenamefont {Spira},\ and\ \citenamefont {Zerwas}}]{Plehn:1996wb}%
  \BibitemOpen
  \bibfield  {author} {\bibinfo {author} {\bibfnamefont {T.}~\bibnamefont
  {Plehn}}, \bibinfo {author} {\bibfnamefont {M.}~\bibnamefont {Spira}},\ and\
  \bibinfo {author} {\bibfnamefont {P.~M.}\ \bibnamefont {Zerwas}},\ }\bibfield
   {title} {\bibinfo {title} {{Pair production of neutral Higgs particles in
  gluon-gluon collisions}},\ }\href
  {https://doi.org/10.1016/0550-3213(96)00418-X} {\bibfield  {journal}
  {\bibinfo  {journal} {Nucl. Phys. B}\ }\textbf {\bibinfo {volume} {479}},\
  \bibinfo {pages} {46} (\bibinfo {year} {1996})},\ \bibinfo {note} {[Erratum:
  Nucl.Phys.B 531, 655--655 (1998)]},\ \Eprint
  {https://arxiv.org/abs/hep-ph/9603205} {arXiv:hep-ph/9603205} \BibitemShut
  {NoStop}%
\bibitem [{\citenamefont {Dolan}\ \emph {et~al.}(2013)\citenamefont {Dolan},
  \citenamefont {Englert},\ and\ \citenamefont {Spannowsky}}]{Dolan:2012ac}%
  \BibitemOpen
  \bibfield  {author} {\bibinfo {author} {\bibfnamefont {M.~J.}\ \bibnamefont
  {Dolan}}, \bibinfo {author} {\bibfnamefont {C.}~\bibnamefont {Englert}},\
  and\ \bibinfo {author} {\bibfnamefont {M.}~\bibnamefont {Spannowsky}},\
  }\bibfield  {title} {\bibinfo {title} {{New Physics in LHC Higgs boson pair
  production}},\ }\href {https://doi.org/10.1103/PhysRevD.87.055002} {\bibfield
   {journal} {\bibinfo  {journal} {Phys. Rev. D}\ }\textbf {\bibinfo {volume}
  {87}},\ \bibinfo {pages} {055002} (\bibinfo {year} {2013})},\ \Eprint
  {https://arxiv.org/abs/1210.8166} {arXiv:1210.8166 [hep-ph]} \BibitemShut
  {NoStop}%
\bibitem [{\citenamefont {No}\ and\ \citenamefont
  {Ramsey-Musolf}(2014)}]{No:2013wsa}%
  \BibitemOpen
  \bibfield  {author} {\bibinfo {author} {\bibfnamefont {J.~M.}\ \bibnamefont
  {No}}\ and\ \bibinfo {author} {\bibfnamefont {M.}~\bibnamefont
  {Ramsey-Musolf}},\ }\bibfield  {title} {\bibinfo {title} {{Probing the Higgs
  Portal at the LHC Through Resonant di-Higgs Production}},\ }\href
  {https://doi.org/10.1103/PhysRevD.89.095031} {\bibfield  {journal} {\bibinfo
  {journal} {Phys. Rev. D}\ }\textbf {\bibinfo {volume} {89}},\ \bibinfo
  {pages} {095031} (\bibinfo {year} {2014})},\ \Eprint
  {https://arxiv.org/abs/1310.6035} {arXiv:1310.6035 [hep-ph]} \BibitemShut
  {NoStop}%
\bibitem [{\citenamefont {Dawson}\ and\ \citenamefont
  {Lewis}(2015)}]{Dawson:2015haa}%
  \BibitemOpen
  \bibfield  {author} {\bibinfo {author} {\bibfnamefont {S.}~\bibnamefont
  {Dawson}}\ and\ \bibinfo {author} {\bibfnamefont {I.~M.}\ \bibnamefont
  {Lewis}},\ }\bibfield  {title} {\bibinfo {title} {{NLO corrections to double
  Higgs boson production in the Higgs singlet model}},\ }\href
  {https://doi.org/10.1103/PhysRevD.92.094023} {\bibfield  {journal} {\bibinfo
  {journal} {Phys. Rev. D}\ }\textbf {\bibinfo {volume} {92}},\ \bibinfo
  {pages} {094023} (\bibinfo {year} {2015})},\ \Eprint
  {https://arxiv.org/abs/1508.05397} {arXiv:1508.05397 [hep-ph]} \BibitemShut
  {NoStop}%
\bibitem [{\citenamefont {Sirunyan}\ \emph {et~al.}(2020)\citenamefont
  {Sirunyan} \emph {et~al.}}]{CMS:2020cga}%
  \BibitemOpen
  \bibfield  {author} {\bibinfo {author} {\bibfnamefont {A.~M.}\ \bibnamefont
  {Sirunyan}} \emph {et~al.} (\bibinfo {collaboration} {CMS}),\ }\bibfield
  {title} {\bibinfo {title} {{Measurements of $\mathrm{t\bar{t}}H$ Production
  and the CP Structure of the Yukawa Interaction between the Higgs Boson and
  Top Quark in the Diphoton Decay Channel}},\ }\href
  {https://doi.org/10.1103/PhysRevLett.125.061801} {\bibfield  {journal}
  {\bibinfo  {journal} {Phys. Rev. Lett.}\ }\textbf {\bibinfo {volume} {125}},\
  \bibinfo {pages} {061801} (\bibinfo {year} {2020})},\ \Eprint
  {https://arxiv.org/abs/2003.10866} {arXiv:2003.10866 [hep-ex]} \BibitemShut
  {NoStop}%
\bibitem [{\citenamefont {Aad}\ \emph {et~al.}(2020{\natexlab{a}})\citenamefont
  {Aad} \emph {et~al.}}]{ATLAS:2020ior}%
  \BibitemOpen
  \bibfield  {author} {\bibinfo {author} {\bibfnamefont {G.}~\bibnamefont
  {Aad}} \emph {et~al.} (\bibinfo {collaboration} {ATLAS}),\ }\bibfield
  {title} {\bibinfo {title} {{$CP$ Properties of Higgs Boson Interactions with
  Top Quarks in the $t\bar{t}H$ and $tH$ Processes Using $H \rightarrow
  \gamma\gamma$ with the ATLAS Detector}},\ }\href
  {https://doi.org/10.1103/PhysRevLett.125.061802} {\bibfield  {journal}
  {\bibinfo  {journal} {Phys. Rev. Lett.}\ }\textbf {\bibinfo {volume} {125}},\
  \bibinfo {pages} {061802} (\bibinfo {year} {2020}{\natexlab{a}})},\ \Eprint
  {https://arxiv.org/abs/2004.04545} {arXiv:2004.04545 [hep-ex]} \BibitemShut
  {NoStop}%
\bibitem [{\citenamefont {Aad}\ \emph {et~al.}(2025{\natexlab{a}})\citenamefont
  {Aad} \emph {et~al.}}]{ATLAS:2024gth}%
  \BibitemOpen
  \bibfield  {author} {\bibinfo {author} {\bibfnamefont {G.}~\bibnamefont
  {Aad}} \emph {et~al.} (\bibinfo {collaboration} {ATLAS}),\ }\bibfield
  {title} {\bibinfo {title} {{Measurement of the associated production of a
  top-antitop-quark pair and a Higgs boson decaying into a $b\bar{b}$ pair in
  $pp$ collisions at $\sqrt{s}=13$ TeV using the ATLAS detector at the LHC}},\
  }\href {https://doi.org/10.1140/epjc/s10052-025-13740-x} {\bibfield
  {journal} {\bibinfo  {journal} {Eur. Phys. J. C}\ }\textbf {\bibinfo {volume}
  {85}},\ \bibinfo {pages} {210} (\bibinfo {year} {2025}{\natexlab{a}})},\
  \Eprint {https://arxiv.org/abs/2407.10904} {arXiv:2407.10904 [hep-ex]}
  \BibitemShut {NoStop}%
\bibitem [{\citenamefont {Dawson}\ and\ \citenamefont
  {Lewis}(2017)}]{Dawson:2016ugw}%
  \BibitemOpen
  \bibfield  {author} {\bibinfo {author} {\bibfnamefont {S.}~\bibnamefont
  {Dawson}}\ and\ \bibinfo {author} {\bibfnamefont {I.~M.}\ \bibnamefont
  {Lewis}},\ }\bibfield  {title} {\bibinfo {title} {{Singlet Model Interference
  Effects with High Scale UV Physics}},\ }\href
  {https://doi.org/10.1103/PhysRevD.95.015004} {\bibfield  {journal} {\bibinfo
  {journal} {Phys. Rev. D}\ }\textbf {\bibinfo {volume} {95}},\ \bibinfo
  {pages} {015004} (\bibinfo {year} {2017})},\ \Eprint
  {https://arxiv.org/abs/1605.04944} {arXiv:1605.04944 [hep-ph]} \BibitemShut
  {NoStop}%
\bibitem [{\citenamefont {Carena}\ \emph {et~al.}(2018)\citenamefont {Carena},
  \citenamefont {Liu},\ and\ \citenamefont {Riembau}}]{Carena:2018vpt}%
  \BibitemOpen
  \bibfield  {author} {\bibinfo {author} {\bibfnamefont {M.}~\bibnamefont
  {Carena}}, \bibinfo {author} {\bibfnamefont {Z.}~\bibnamefont {Liu}},\ and\
  \bibinfo {author} {\bibfnamefont {M.}~\bibnamefont {Riembau}},\ }\bibfield
  {title} {\bibinfo {title} {{Probing the electroweak phase transition via
  enhanced di-Higgs boson production}},\ }\href
  {https://doi.org/10.1103/PhysRevD.97.095032} {\bibfield  {journal} {\bibinfo
  {journal} {Phys. Rev. D}\ }\textbf {\bibinfo {volume} {97}},\ \bibinfo
  {pages} {095032} (\bibinfo {year} {2018})},\ \Eprint
  {https://arxiv.org/abs/1801.00794} {arXiv:1801.00794 [hep-ph]} \BibitemShut
  {NoStop}%
\bibitem [{\citenamefont {Feuerstake}\ \emph {et~al.}(2025)\citenamefont
  {Feuerstake}, \citenamefont {Fuchs}, \citenamefont {Robens},\ and\
  \citenamefont {Winterbottom}}]{Feuerstake:2024uxs}%
  \BibitemOpen
  \bibfield  {author} {\bibinfo {author} {\bibfnamefont {F.}~\bibnamefont
  {Feuerstake}}, \bibinfo {author} {\bibfnamefont {E.}~\bibnamefont {Fuchs}},
  \bibinfo {author} {\bibfnamefont {T.}~\bibnamefont {Robens}},\ and\ \bibinfo
  {author} {\bibfnamefont {D.}~\bibnamefont {Winterbottom}},\ }\bibfield
  {title} {\bibinfo {title} {{Interference effects in resonant di-Higgs
  production at the LHC in the Higgs singlet extension}},\ }\href
  {https://doi.org/10.1007/JHEP04(2025)094} {\bibfield  {journal} {\bibinfo
  {journal} {JHEP}\ }\textbf {\bibinfo {volume} {04}},\ \bibinfo {pages}
  {094}},\ \Eprint {https://arxiv.org/abs/2409.06651} {arXiv:2409.06651
  [hep-ph]} \BibitemShut {NoStop}%
\bibitem [{\citenamefont {Ellwanger}\ \emph {et~al.}(2010)\citenamefont
  {Ellwanger}, \citenamefont {Hugonie},\ and\ \citenamefont
  {Teixeira}}]{Ellwanger:2009dp}%
  \BibitemOpen
  \bibfield  {author} {\bibinfo {author} {\bibfnamefont {U.}~\bibnamefont
  {Ellwanger}}, \bibinfo {author} {\bibfnamefont {C.}~\bibnamefont {Hugonie}},\
  and\ \bibinfo {author} {\bibfnamefont {A.~M.}\ \bibnamefont {Teixeira}},\
  }\bibfield  {title} {\bibinfo {title} {{The Next-to-Minimal Supersymmetric
  Standard Model}},\ }\href {https://doi.org/10.1016/j.physrep.2010.07.001}
  {\bibfield  {journal} {\bibinfo  {journal} {Phys. Rept.}\ }\textbf {\bibinfo
  {volume} {496}},\ \bibinfo {pages} {1} (\bibinfo {year} {2010})},\ \Eprint
  {https://arxiv.org/abs/0910.1785} {arXiv:0910.1785 [hep-ph]} \BibitemShut
  {NoStop}%
\bibitem [{\citenamefont {Moretti}\ and\ \citenamefont
  {Khalil}(2019)}]{Moretti:2019ulc}%
  \BibitemOpen
  \bibfield  {author} {\bibinfo {author} {\bibfnamefont {S.}~\bibnamefont
  {Moretti}}\ and\ \bibinfo {author} {\bibfnamefont {S.}~\bibnamefont
  {Khalil}},\ }\href@noop {} {\emph {\bibinfo {title} {{Supersymmetry Beyond
  Minimality: From Theory to Experiment}}}}\ (\bibinfo  {publisher} {CRC
  Press},\ \bibinfo {year} {2019})\BibitemShut {NoStop}%
\bibitem [{\citenamefont {Aad}\ \emph {et~al.}(2020{\natexlab{b}})\citenamefont
  {Aad} \emph {et~al.}}]{ATLAS:2020zms}%
  \BibitemOpen
  \bibfield  {author} {\bibinfo {author} {\bibfnamefont {G.}~\bibnamefont
  {Aad}} \emph {et~al.} (\bibinfo {collaboration} {ATLAS}),\ }\bibfield
  {title} {\bibinfo {title} {{Search for heavy Higgs bosons decaying into two
  tau leptons with the ATLAS detector using $pp$ collisions at $\sqrt{s}=13$
  TeV}},\ }\href {https://doi.org/10.1103/PhysRevLett.125.051801} {\bibfield
  {journal} {\bibinfo  {journal} {Phys. Rev. Lett.}\ }\textbf {\bibinfo
  {volume} {125}},\ \bibinfo {pages} {051801} (\bibinfo {year}
  {2020}{\natexlab{b}})},\ \Eprint {https://arxiv.org/abs/2002.12223}
  {arXiv:2002.12223 [hep-ex]} \BibitemShut {NoStop}%
\bibitem [{\citenamefont {Tumasyan}\ \emph {et~al.}(2023)\citenamefont
  {Tumasyan} \emph {et~al.}}]{CMS:2022goy}%
  \BibitemOpen
  \bibfield  {author} {\bibinfo {author} {\bibfnamefont {A.}~\bibnamefont
  {Tumasyan}} \emph {et~al.} (\bibinfo {collaboration} {CMS}),\ }\bibfield
  {title} {\bibinfo {title} {{Searches for additional Higgs bosons and for
  vector leptoquarks in $\tau\tau$ final states in proton-proton collisions at
  $\sqrt{s}$ = 13 TeV}},\ }\href {https://doi.org/10.1007/JHEP07(2023)073}
  {\bibfield  {journal} {\bibinfo  {journal} {JHEP}\ }\textbf {\bibinfo
  {volume} {07}},\ \bibinfo {pages} {073}},\ \Eprint
  {https://arxiv.org/abs/2208.02717} {arXiv:2208.02717 [hep-ex]} \BibitemShut
  {NoStop}%
\bibitem [{\citenamefont {Drees}(1989)}]{Drees:1988fc}%
  \BibitemOpen
  \bibfield  {author} {\bibinfo {author} {\bibfnamefont {M.}~\bibnamefont
  {Drees}},\ }\bibfield  {title} {\bibinfo {title} {{Supersymmetric Models with
  Extended Higgs Sector}},\ }\href {https://doi.org/10.1142/S0217751X89001448}
  {\bibfield  {journal} {\bibinfo  {journal} {Int. J. Mod. Phys. A}\ }\textbf
  {\bibinfo {volume} {4}},\ \bibinfo {pages} {3635} (\bibinfo {year}
  {1989})}\BibitemShut {NoStop}%
\bibitem [{\citenamefont {Ellwanger}(2013)}]{Ellwanger:2013ova}%
  \BibitemOpen
  \bibfield  {author} {\bibinfo {author} {\bibfnamefont {U.}~\bibnamefont
  {Ellwanger}},\ }\bibfield  {title} {\bibinfo {title} {{Higgs pair production
  in the NMSSM at the LHC}},\ }\href {https://doi.org/10.1007/JHEP08(2013)077}
  {\bibfield  {journal} {\bibinfo  {journal} {JHEP}\ }\textbf {\bibinfo
  {volume} {08}},\ \bibinfo {pages} {077}},\ \Eprint
  {https://arxiv.org/abs/1306.5541} {arXiv:1306.5541 [hep-ph]} \BibitemShut
  {NoStop}%
\bibitem [{\citenamefont {Heng}\ \emph {et~al.}(2018)\citenamefont {Heng},
  \citenamefont {Gong},\ and\ \citenamefont {Zhou}}]{Heng:2018kyd}%
  \BibitemOpen
  \bibfield  {author} {\bibinfo {author} {\bibfnamefont {Z.}~\bibnamefont
  {Heng}}, \bibinfo {author} {\bibfnamefont {X.}~\bibnamefont {Gong}},\ and\
  \bibinfo {author} {\bibfnamefont {H.}~\bibnamefont {Zhou}},\ }\bibfield
  {title} {\bibinfo {title} {{Pair production of Higgs boson in NMSSM at the
  LHC with the next-to-lightest CP-even Higgs boson being SM-like}},\ }\href
  {https://doi.org/10.1088/1674-1137/42/7/073103} {\bibfield  {journal}
  {\bibinfo  {journal} {Chin. Phys. C}\ }\textbf {\bibinfo {volume} {42}},\
  \bibinfo {pages} {073103} (\bibinfo {year} {2018})},\ \Eprint
  {https://arxiv.org/abs/1805.01598} {arXiv:1805.01598 [hep-ph]} \BibitemShut
  {NoStop}%
\bibitem [{\citenamefont {Cao}\ \emph {et~al.}(2013)\citenamefont {Cao},
  \citenamefont {Heng}, \citenamefont {Shang}, \citenamefont {Wan},\ and\
  \citenamefont {Yang}}]{Cao:2013si}%
  \BibitemOpen
  \bibfield  {author} {\bibinfo {author} {\bibfnamefont {J.}~\bibnamefont
  {Cao}}, \bibinfo {author} {\bibfnamefont {Z.}~\bibnamefont {Heng}}, \bibinfo
  {author} {\bibfnamefont {L.}~\bibnamefont {Shang}}, \bibinfo {author}
  {\bibfnamefont {P.}~\bibnamefont {Wan}},\ and\ \bibinfo {author}
  {\bibfnamefont {J.~M.}\ \bibnamefont {Yang}},\ }\bibfield  {title} {\bibinfo
  {title} {{Pair Production of a 125 GeV Higgs Boson in MSSM and NMSSM at the
  LHC}},\ }\href {https://doi.org/10.1007/JHEP04(2013)134} {\bibfield
  {journal} {\bibinfo  {journal} {JHEP}\ }\textbf {\bibinfo {volume} {04}},\
  \bibinfo {pages} {134}},\ \Eprint {https://arxiv.org/abs/1301.6437}
  {arXiv:1301.6437 [hep-ph]} \BibitemShut {NoStop}%
\bibitem [{\citenamefont {Heng}\ \emph {et~al.}(2013)\citenamefont {Heng},
  \citenamefont {Shang},\ and\ \citenamefont {Wan}}]{Heng:2013wia}%
  \BibitemOpen
  \bibfield  {author} {\bibinfo {author} {\bibfnamefont {Z.}~\bibnamefont
  {Heng}}, \bibinfo {author} {\bibfnamefont {L.}~\bibnamefont {Shang}},\ and\
  \bibinfo {author} {\bibfnamefont {P.}~\bibnamefont {Wan}},\ }\bibfield
  {title} {\bibinfo {title} {{Pair production of a 125 GeV Higgs boson in MSSM
  and NMSSM at the ILC}},\ }\href {https://doi.org/10.1007/JHEP10(2013)047}
  {\bibfield  {journal} {\bibinfo  {journal} {JHEP}\ }\textbf {\bibinfo
  {volume} {10}},\ \bibinfo {pages} {047}},\ \Eprint
  {https://arxiv.org/abs/1306.0279} {arXiv:1306.0279 [hep-ph]} \BibitemShut
  {NoStop}%
\bibitem [{\citenamefont {Cao}\ \emph {et~al.}(2014)\citenamefont {Cao},
  \citenamefont {Li}, \citenamefont {Shang}, \citenamefont {Wu},\ and\
  \citenamefont {Zhang}}]{Cao:2014kya}%
  \BibitemOpen
  \bibfield  {author} {\bibinfo {author} {\bibfnamefont {J.}~\bibnamefont
  {Cao}}, \bibinfo {author} {\bibfnamefont {D.}~\bibnamefont {Li}}, \bibinfo
  {author} {\bibfnamefont {L.}~\bibnamefont {Shang}}, \bibinfo {author}
  {\bibfnamefont {P.}~\bibnamefont {Wu}},\ and\ \bibinfo {author}
  {\bibfnamefont {Y.}~\bibnamefont {Zhang}},\ }\bibfield  {title} {\bibinfo
  {title} {{Exploring the Higgs Sector of a Most Natural NMSSM and its
  Prediction on Higgs Pair Production at the LHC}},\ }\href
  {https://doi.org/10.1007/JHEP12(2014)026} {\bibfield  {journal} {\bibinfo
  {journal} {JHEP}\ }\textbf {\bibinfo {volume} {12}},\ \bibinfo {pages}
  {026}},\ \Eprint {https://arxiv.org/abs/1409.8431} {arXiv:1409.8431 [hep-ph]}
  \BibitemShut {NoStop}%
\bibitem [{\citenamefont {Huang}\ and\ \citenamefont
  {Ng}(2020)}]{Huang:2019bcs}%
  \BibitemOpen
  \bibfield  {author} {\bibinfo {author} {\bibfnamefont {P.}~\bibnamefont
  {Huang}}\ and\ \bibinfo {author} {\bibfnamefont {Y.~H.}\ \bibnamefont {Ng}},\
  }\bibfield  {title} {\bibinfo {title} {{Di-Higgs Production in SUSY models at
  the LHC}},\ }\href {https://doi.org/10.1140/epjp/s13360-020-00677-1}
  {\bibfield  {journal} {\bibinfo  {journal} {Eur. Phys. J. Plus}\ }\textbf
  {\bibinfo {volume} {135}},\ \bibinfo {pages} {660} (\bibinfo {year}
  {2020})},\ \Eprint {https://arxiv.org/abs/1910.13968} {arXiv:1910.13968
  [hep-ph]} \BibitemShut {NoStop}%
\bibitem [{\citenamefont {Degrande}\ \emph {et~al.}(2012)\citenamefont
  {Degrande}, \citenamefont {Duhr}, \citenamefont {Fuks}, \citenamefont
  {Grellscheid}, \citenamefont {Mattelaer},\ and\ \citenamefont
  {Reiter}}]{Degrande:2011ua}%
  \BibitemOpen
  \bibfield  {author} {\bibinfo {author} {\bibfnamefont {C.}~\bibnamefont
  {Degrande}}, \bibinfo {author} {\bibfnamefont {C.}~\bibnamefont {Duhr}},
  \bibinfo {author} {\bibfnamefont {B.}~\bibnamefont {Fuks}}, \bibinfo {author}
  {\bibfnamefont {D.}~\bibnamefont {Grellscheid}}, \bibinfo {author}
  {\bibfnamefont {O.}~\bibnamefont {Mattelaer}},\ and\ \bibinfo {author}
  {\bibfnamefont {T.}~\bibnamefont {Reiter}},\ }\bibfield  {title} {\bibinfo
  {title} {{UFO - The Universal FeynRules Output}},\ }\href
  {https://doi.org/10.1016/j.cpc.2012.01.022} {\bibfield  {journal} {\bibinfo
  {journal} {Comput. Phys. Commun.}\ }\textbf {\bibinfo {volume} {183}},\
  \bibinfo {pages} {1201} (\bibinfo {year} {2012})},\ \Eprint
  {https://arxiv.org/abs/1108.2040} {arXiv:1108.2040 [hep-ph]} \BibitemShut
  {NoStop}%
\bibitem [{\citenamefont {Alwall}\ \emph {et~al.}(2014)\citenamefont {Alwall},
  \citenamefont {Frederix}, \citenamefont {Frixione}, \citenamefont {Hirschi},
  \citenamefont {Maltoni}, \citenamefont {Mattelaer}, \citenamefont {Shao},
  \citenamefont {Stelzer}, \citenamefont {Torrielli},\ and\ \citenamefont
  {Zaro}}]{Alwall:2014hca}%
  \BibitemOpen
  \bibfield  {author} {\bibinfo {author} {\bibfnamefont {J.}~\bibnamefont
  {Alwall}}, \bibinfo {author} {\bibfnamefont {R.}~\bibnamefont {Frederix}},
  \bibinfo {author} {\bibfnamefont {S.}~\bibnamefont {Frixione}}, \bibinfo
  {author} {\bibfnamefont {V.}~\bibnamefont {Hirschi}}, \bibinfo {author}
  {\bibfnamefont {F.}~\bibnamefont {Maltoni}}, \bibinfo {author} {\bibfnamefont
  {O.}~\bibnamefont {Mattelaer}}, \bibinfo {author} {\bibfnamefont {H.~S.}\
  \bibnamefont {Shao}}, \bibinfo {author} {\bibfnamefont {T.}~\bibnamefont
  {Stelzer}}, \bibinfo {author} {\bibfnamefont {P.}~\bibnamefont {Torrielli}},\
  and\ \bibinfo {author} {\bibfnamefont {M.}~\bibnamefont {Zaro}},\ }\bibfield
  {title} {\bibinfo {title} {{The automated computation of tree-level and
  next-to-leading order differential cross sections, and their matching to
  parton shower simulations}},\ }\href
  {https://doi.org/10.1007/JHEP07(2014)079} {\bibfield  {journal} {\bibinfo
  {journal} {JHEP}\ }\textbf {\bibinfo {volume} {07}},\ \bibinfo {pages}
  {079}},\ \Eprint {https://arxiv.org/abs/1405.0301} {arXiv:1405.0301 [hep-ph]}
  \BibitemShut {NoStop}%
\bibitem [{\citenamefont {Frederix}\ \emph {et~al.}(2018)\citenamefont
  {Frederix}, \citenamefont {Frixione}, \citenamefont {Hirschi}, \citenamefont
  {Pagani}, \citenamefont {Shao},\ and\ \citenamefont
  {Zaro}}]{Frederix:2018nkq}%
  \BibitemOpen
  \bibfield  {author} {\bibinfo {author} {\bibfnamefont {R.}~\bibnamefont
  {Frederix}}, \bibinfo {author} {\bibfnamefont {S.}~\bibnamefont {Frixione}},
  \bibinfo {author} {\bibfnamefont {V.}~\bibnamefont {Hirschi}}, \bibinfo
  {author} {\bibfnamefont {D.}~\bibnamefont {Pagani}}, \bibinfo {author}
  {\bibfnamefont {H.~S.}\ \bibnamefont {Shao}},\ and\ \bibinfo {author}
  {\bibfnamefont {M.}~\bibnamefont {Zaro}},\ }\bibfield  {title} {\bibinfo
  {title} {{The automation of next-to-leading order electroweak
  calculations}},\ }\href {https://doi.org/10.1007/JHEP11(2021)085} {\bibfield
  {journal} {\bibinfo  {journal} {JHEP}\ }\textbf {\bibinfo {volume} {07}},\
  \bibinfo {pages} {185}},\ \bibinfo {note} {[Erratum: JHEP 11, 085 (2021)]},\
  \Eprint {https://arxiv.org/abs/1804.10017} {arXiv:1804.10017 [hep-ph]}
  \BibitemShut {NoStop}%
\bibitem [{\citenamefont {Ball}\ \emph {et~al.}(2015)\citenamefont {Ball} \emph
  {et~al.}}]{NNPDF:2014otw}%
  \BibitemOpen
  \bibfield  {author} {\bibinfo {author} {\bibfnamefont {R.~D.}\ \bibnamefont
  {Ball}} \emph {et~al.} (\bibinfo {collaboration} {NNPDF}),\ }\bibfield
  {title} {\bibinfo {title} {{Parton distributions for the LHC Run II}},\
  }\href {https://doi.org/10.1007/JHEP04(2015)040} {\bibfield  {journal}
  {\bibinfo  {journal} {JHEP}\ }\textbf {\bibinfo {volume} {04}},\ \bibinfo
  {pages} {040}},\ \Eprint {https://arxiv.org/abs/1410.8849} {arXiv:1410.8849
  [hep-ph]} \BibitemShut {NoStop}%
\bibitem [{\citenamefont {Sj\"ostrand}\ \emph {et~al.}(2015)\citenamefont
  {Sj\"ostrand}, \citenamefont {Ask}, \citenamefont {Christiansen},
  \citenamefont {Corke}, \citenamefont {Desai}, \citenamefont {Ilten},
  \citenamefont {Mrenna}, \citenamefont {Prestel}, \citenamefont {Rasmussen},\
  and\ \citenamefont {Skands}}]{Sjostrand:2014zea}%
  \BibitemOpen
  \bibfield  {author} {\bibinfo {author} {\bibfnamefont {T.}~\bibnamefont
  {Sj\"ostrand}}, \bibinfo {author} {\bibfnamefont {S.}~\bibnamefont {Ask}},
  \bibinfo {author} {\bibfnamefont {J.~R.}\ \bibnamefont {Christiansen}},
  \bibinfo {author} {\bibfnamefont {R.}~\bibnamefont {Corke}}, \bibinfo
  {author} {\bibfnamefont {N.}~\bibnamefont {Desai}}, \bibinfo {author}
  {\bibfnamefont {P.}~\bibnamefont {Ilten}}, \bibinfo {author} {\bibfnamefont
  {S.}~\bibnamefont {Mrenna}}, \bibinfo {author} {\bibfnamefont
  {S.}~\bibnamefont {Prestel}}, \bibinfo {author} {\bibfnamefont {C.~O.}\
  \bibnamefont {Rasmussen}},\ and\ \bibinfo {author} {\bibfnamefont {P.~Z.}\
  \bibnamefont {Skands}},\ }\bibfield  {title} {\bibinfo {title} {{An
  introduction to PYTHIA 8.2}},\ }\href
  {https://doi.org/10.1016/j.cpc.2015.01.024} {\bibfield  {journal} {\bibinfo
  {journal} {Comput. Phys. Commun.}\ }\textbf {\bibinfo {volume} {191}},\
  \bibinfo {pages} {159} (\bibinfo {year} {2015})},\ \Eprint
  {https://arxiv.org/abs/1410.3012} {arXiv:1410.3012 [hep-ph]} \BibitemShut
  {NoStop}%
\bibitem [{\citenamefont {Cacciari}\ \emph {et~al.}(2008)\citenamefont
  {Cacciari}, \citenamefont {Salam},\ and\ \citenamefont
  {Soyez}}]{Cacciari:2008gp}%
  \BibitemOpen
  \bibfield  {author} {\bibinfo {author} {\bibfnamefont {M.}~\bibnamefont
  {Cacciari}}, \bibinfo {author} {\bibfnamefont {G.~P.}\ \bibnamefont
  {Salam}},\ and\ \bibinfo {author} {\bibfnamefont {G.}~\bibnamefont {Soyez}},\
  }\bibfield  {title} {\bibinfo {title} {{The anti-$k_t$ jet clustering
  algorithm}},\ }\href {https://doi.org/10.1088/1126-6708/2008/04/063}
  {\bibfield  {journal} {\bibinfo  {journal} {JHEP}\ }\textbf {\bibinfo
  {volume} {04}},\ \bibinfo {pages} {063}},\ \Eprint
  {https://arxiv.org/abs/0802.1189} {arXiv:0802.1189 [hep-ph]} \BibitemShut
  {NoStop}%
\bibitem [{\citenamefont {Cacciari}\ \emph {et~al.}(2012)\citenamefont
  {Cacciari}, \citenamefont {Salam},\ and\ \citenamefont
  {Soyez}}]{Cacciari:2011ma}%
  \BibitemOpen
  \bibfield  {author} {\bibinfo {author} {\bibfnamefont {M.}~\bibnamefont
  {Cacciari}}, \bibinfo {author} {\bibfnamefont {G.~P.}\ \bibnamefont
  {Salam}},\ and\ \bibinfo {author} {\bibfnamefont {G.}~\bibnamefont {Soyez}},\
  }\bibfield  {title} {\bibinfo {title} {{FastJet User Manual}},\ }\href
  {https://doi.org/10.1140/epjc/s10052-012-1896-2} {\bibfield  {journal}
  {\bibinfo  {journal} {Eur. Phys. J. C}\ }\textbf {\bibinfo {volume} {72}},\
  \bibinfo {pages} {1896} (\bibinfo {year} {2012})},\ \Eprint
  {https://arxiv.org/abs/1111.6097} {arXiv:1111.6097 [hep-ph]} \BibitemShut
  {NoStop}%
\bibitem [{\citenamefont {Conte}\ \emph {et~al.}(2013)\citenamefont {Conte},
  \citenamefont {Fuks},\ and\ \citenamefont {Serret}}]{Conte:2012fm}%
  \BibitemOpen
  \bibfield  {author} {\bibinfo {author} {\bibfnamefont {E.}~\bibnamefont
  {Conte}}, \bibinfo {author} {\bibfnamefont {B.}~\bibnamefont {Fuks}},\ and\
  \bibinfo {author} {\bibfnamefont {G.}~\bibnamefont {Serret}},\ }\bibfield
  {title} {\bibinfo {title} {{MadAnalysis 5, A User-Friendly Framework for
  Collider Phenomenology}},\ }\href {https://doi.org/10.1016/j.cpc.2012.09.009}
  {\bibfield  {journal} {\bibinfo  {journal} {Comput. Phys. Commun.}\ }\textbf
  {\bibinfo {volume} {184}},\ \bibinfo {pages} {222} (\bibinfo {year}
  {2013})},\ \Eprint {https://arxiv.org/abs/1206.1599} {arXiv:1206.1599
  [hep-ph]} \BibitemShut {NoStop}%
\bibitem [{\citenamefont {Tumasyan}\ \emph {et~al.}(2022)\citenamefont
  {Tumasyan} \emph {et~al.}}]{CMS:2022dwd}%
  \BibitemOpen
  \bibfield  {author} {\bibinfo {author} {\bibfnamefont {A.}~\bibnamefont
  {Tumasyan}} \emph {et~al.} (\bibinfo {collaboration} {CMS}),\ }\bibfield
  {title} {\bibinfo {title} {{A portrait of the Higgs boson by the CMS
  experiment ten years after the discovery.}},\ }\href
  {https://doi.org/10.1038/s41586-022-04892-x} {\bibfield  {journal} {\bibinfo
  {journal} {Nature}\ }\textbf {\bibinfo {volume} {607}},\ \bibinfo {pages}
  {60} (\bibinfo {year} {2022})},\ \bibinfo {note} {[Erratum: Nature 623,
  (2023)]},\ \Eprint {https://arxiv.org/abs/2207.00043} {arXiv:2207.00043
  [hep-ex]} \BibitemShut {NoStop}%
\bibitem [{\citenamefont {Aad}\ \emph {et~al.}(2024)\citenamefont {Aad} \emph
  {et~al.}}]{ATLAS:2024lyh}%
  \BibitemOpen
  \bibfield  {author} {\bibinfo {author} {\bibfnamefont {G.}~\bibnamefont
  {Aad}} \emph {et~al.} (\bibinfo {collaboration} {ATLAS}),\ }\bibfield
  {title} {\bibinfo {title} {{Interpretations of the ATLAS measurements of
  Higgs boson production and decay rates and differential cross-sections in pp
  collisions at $ \sqrt{s} $ = 13 TeV}},\ }\href
  {https://doi.org/10.1007/JHEP11(2024)097} {\bibfield  {journal} {\bibinfo
  {journal} {JHEP}\ }\textbf {\bibinfo {volume} {11}},\ \bibinfo {pages}
  {097}},\ \Eprint {https://arxiv.org/abs/2402.05742} {arXiv:2402.05742
  [hep-ex]} \BibitemShut {NoStop}%
\bibitem [{\citenamefont {Carena}\ \emph {et~al.}(2016)\citenamefont {Carena},
  \citenamefont {Haber}, \citenamefont {Low}, \citenamefont {Shah},\ and\
  \citenamefont {Wagner}}]{Carena:2015moc}%
  \BibitemOpen
  \bibfield  {author} {\bibinfo {author} {\bibfnamefont {M.}~\bibnamefont
  {Carena}}, \bibinfo {author} {\bibfnamefont {H.~E.}\ \bibnamefont {Haber}},
  \bibinfo {author} {\bibfnamefont {I.}~\bibnamefont {Low}}, \bibinfo {author}
  {\bibfnamefont {N.~R.}\ \bibnamefont {Shah}},\ and\ \bibinfo {author}
  {\bibfnamefont {C.~E.~M.}\ \bibnamefont {Wagner}},\ }\bibfield  {title}
  {\bibinfo {title} {{Alignment limit of the NMSSM Higgs sector}},\ }\href
  {https://doi.org/10.1103/PhysRevD.93.035013} {\bibfield  {journal} {\bibinfo
  {journal} {Phys. Rev. D}\ }\textbf {\bibinfo {volume} {93}},\ \bibinfo
  {pages} {035013} (\bibinfo {year} {2016})},\ \Eprint
  {https://arxiv.org/abs/1510.09137} {arXiv:1510.09137 [hep-ph]} \BibitemShut
  {NoStop}%
\bibitem [{\citenamefont {Porod}(2003)}]{Porod:2003um}%
  \BibitemOpen
  \bibfield  {author} {\bibinfo {author} {\bibfnamefont {W.}~\bibnamefont
  {Porod}},\ }\bibfield  {title} {\bibinfo {title} {{SPheno, a program for
  calculating supersymmetric spectra, SUSY particle decays and SUSY particle
  production at e+ e- colliders}},\ }\href
  {https://doi.org/10.1016/S0010-4655(03)00222-4} {\bibfield  {journal}
  {\bibinfo  {journal} {Comput. Phys. Commun.}\ }\textbf {\bibinfo {volume}
  {153}},\ \bibinfo {pages} {275} (\bibinfo {year} {2003})},\ \Eprint
  {https://arxiv.org/abs/hep-ph/0301101} {arXiv:hep-ph/0301101} \BibitemShut
  {NoStop}%
\bibitem [{\citenamefont {Porod}\ and\ \citenamefont
  {Staub}(2012)}]{Porod:2011nf}%
  \BibitemOpen
  \bibfield  {author} {\bibinfo {author} {\bibfnamefont {W.}~\bibnamefont
  {Porod}}\ and\ \bibinfo {author} {\bibfnamefont {F.}~\bibnamefont {Staub}},\
  }\bibfield  {title} {\bibinfo {title} {{SPheno 3.1: Extensions including
  flavour, CP-phases and models beyond the MSSM}},\ }\href
  {https://doi.org/10.1016/j.cpc.2012.05.021} {\bibfield  {journal} {\bibinfo
  {journal} {Comput. Phys. Commun.}\ }\textbf {\bibinfo {volume} {183}},\
  \bibinfo {pages} {2458} (\bibinfo {year} {2012})},\ \Eprint
  {https://arxiv.org/abs/1104.1573} {arXiv:1104.1573 [hep-ph]} \BibitemShut
  {NoStop}%
\bibitem [{\citenamefont {Staub}(2014)}]{Staub:2013tta}%
  \BibitemOpen
  \bibfield  {author} {\bibinfo {author} {\bibfnamefont {F.}~\bibnamefont
  {Staub}},\ }\bibfield  {title} {\bibinfo {title} {{SARAH 4 : A tool for (not
  only SUSY) model builders}},\ }\href
  {https://doi.org/10.1016/j.cpc.2014.02.018} {\bibfield  {journal} {\bibinfo
  {journal} {Comput. Phys. Commun.}\ }\textbf {\bibinfo {volume} {185}},\
  \bibinfo {pages} {1773} (\bibinfo {year} {2014})},\ \Eprint
  {https://arxiv.org/abs/1309.7223} {arXiv:1309.7223 [hep-ph]} \BibitemShut
  {NoStop}%
\bibitem [{\citenamefont {Barate}\ \emph {et~al.}(2003)\citenamefont {Barate}
  \emph {et~al.}}]{LEPWorkingGroupforHiggsbosonsearches:2003ing}%
  \BibitemOpen
  \bibfield  {author} {\bibinfo {author} {\bibfnamefont {R.}~\bibnamefont
  {Barate}} \emph {et~al.} (\bibinfo {collaboration} {LEP Working Group for
  Higgs boson searches, ALEPH, DELPHI, L3, OPAL}),\ }\bibfield  {title}
  {\bibinfo {title} {{Search for the standard model Higgs boson at LEP}},\
  }\href {https://doi.org/10.1016/S0370-2693(03)00614-2} {\bibfield  {journal}
  {\bibinfo  {journal} {Phys. Lett. B}\ }\textbf {\bibinfo {volume} {565}},\
  \bibinfo {pages} {61} (\bibinfo {year} {2003})},\ \Eprint
  {https://arxiv.org/abs/hep-ex/0306033} {arXiv:hep-ex/0306033} \BibitemShut
  {NoStop}%
\bibitem [{\citenamefont {Aad}\ \emph {et~al.}(2025{\natexlab{b}})\citenamefont
  {Aad} \emph {et~al.}}]{ATLAS:2024bjr}%
  \BibitemOpen
  \bibfield  {author} {\bibinfo {author} {\bibfnamefont {G.}~\bibnamefont
  {Aad}} \emph {et~al.} (\bibinfo {collaboration} {ATLAS}),\ }\bibfield
  {title} {\bibinfo {title} {{Search for diphoton resonances in the 66 to 110
  GeV mass range using pp collisions at $ \sqrt{s} $ = 13 TeV with the ATLAS
  detector}},\ }\href {https://doi.org/10.1007/JHEP01(2025)053} {\bibfield
  {journal} {\bibinfo  {journal} {JHEP}\ }\textbf {\bibinfo {volume} {01}},\
  \bibinfo {pages} {053}},\ \Eprint {https://arxiv.org/abs/2407.07546}
  {arXiv:2407.07546 [hep-ex]} \BibitemShut {NoStop}%
\bibitem [{\citenamefont {Sirunyan}\ \emph {et~al.}(2019)\citenamefont
  {Sirunyan} \emph {et~al.}}]{CMS:2018cyk}%
  \BibitemOpen
  \bibfield  {author} {\bibinfo {author} {\bibfnamefont {A.~M.}\ \bibnamefont
  {Sirunyan}} \emph {et~al.} (\bibinfo {collaboration} {CMS}),\ }\bibfield
  {title} {\bibinfo {title} {{Search for a standard model-like Higgs boson in
  the mass range between 70 and 110 GeV in the diphoton final state in
  proton-proton collisions at $\sqrt{s}=$ 8 and 13 TeV}},\ }\href
  {https://doi.org/10.1016/j.physletb.2019.03.064} {\bibfield  {journal}
  {\bibinfo  {journal} {Phys. Lett. B}\ }\textbf {\bibinfo {volume} {793}},\
  \bibinfo {pages} {320} (\bibinfo {year} {2019})},\ \Eprint
  {https://arxiv.org/abs/1811.08459} {arXiv:1811.08459 [hep-ex]} \BibitemShut
  {NoStop}%
\bibitem [{\citenamefont {Biek\"otter}\ \emph {et~al.}(2022)\citenamefont
  {Biek\"otter}, \citenamefont {Heinemeyer},\ and\ \citenamefont
  {Weiglein}}]{Biekotter:2022jyr}%
  \BibitemOpen
  \bibfield  {author} {\bibinfo {author} {\bibfnamefont {T.}~\bibnamefont
  {Biek\"otter}}, \bibinfo {author} {\bibfnamefont {S.}~\bibnamefont
  {Heinemeyer}},\ and\ \bibinfo {author} {\bibfnamefont {G.}~\bibnamefont
  {Weiglein}},\ }\bibfield  {title} {\bibinfo {title} {{Mounting evidence for a
  95 GeV Higgs boson}},\ }\href {https://doi.org/10.1007/JHEP08(2022)201}
  {\bibfield  {journal} {\bibinfo  {journal} {JHEP}\ }\textbf {\bibinfo
  {volume} {08}},\ \bibinfo {pages} {201}},\ \Eprint
  {https://arxiv.org/abs/2203.13180} {arXiv:2203.13180 [hep-ph]} \BibitemShut
  {NoStop}%
\bibitem [{\citenamefont {Azevedo}\ \emph {et~al.}(2023)\citenamefont
  {Azevedo}, \citenamefont {Biek\"otter},\ and\ \citenamefont
  {Ferreira}}]{Azevedo:2023zkg}%
  \BibitemOpen
  \bibfield  {author} {\bibinfo {author} {\bibfnamefont {D.}~\bibnamefont
  {Azevedo}}, \bibinfo {author} {\bibfnamefont {T.}~\bibnamefont
  {Biek\"otter}},\ and\ \bibinfo {author} {\bibfnamefont {P.~M.}\ \bibnamefont
  {Ferreira}},\ }\bibfield  {title} {\bibinfo {title} {{2HDM interpretations of
  the CMS diphoton excess at 95 GeV}},\ }\href
  {https://doi.org/10.1007/JHEP11(2023)017} {\bibfield  {journal} {\bibinfo
  {journal} {JHEP}\ }\textbf {\bibinfo {volume} {11}},\ \bibinfo {pages}
  {017}},\ \Eprint {https://arxiv.org/abs/2305.19716} {arXiv:2305.19716
  [hep-ph]} \BibitemShut {NoStop}%
\bibitem [{\citenamefont {Belyaev}\ \emph {et~al.}(2024)\citenamefont
  {Belyaev}, \citenamefont {Benbrik}, \citenamefont {Boukidi}, \citenamefont
  {Chakraborti}, \citenamefont {Moretti},\ and\ \citenamefont
  {Semlali}}]{Belyaev:2023xnv}%
  \BibitemOpen
  \bibfield  {author} {\bibinfo {author} {\bibfnamefont {A.}~\bibnamefont
  {Belyaev}}, \bibinfo {author} {\bibfnamefont {R.}~\bibnamefont {Benbrik}},
  \bibinfo {author} {\bibfnamefont {M.}~\bibnamefont {Boukidi}}, \bibinfo
  {author} {\bibfnamefont {M.}~\bibnamefont {Chakraborti}}, \bibinfo {author}
  {\bibfnamefont {S.}~\bibnamefont {Moretti}},\ and\ \bibinfo {author}
  {\bibfnamefont {S.}~\bibnamefont {Semlali}},\ }\bibfield  {title} {\bibinfo
  {title} {{Explanation of the hints for a 95 GeV Higgs boson within a 2-Higgs
  Doublet Model}},\ }\href {https://doi.org/10.1007/JHEP05(2024)209} {\bibfield
   {journal} {\bibinfo  {journal} {JHEP}\ }\textbf {\bibinfo {volume} {05}},\
  \bibinfo {pages} {209}},\ \Eprint {https://arxiv.org/abs/2306.09029}
  {arXiv:2306.09029 [hep-ph]} \BibitemShut {NoStop}%
\bibitem [{\citenamefont {Ellwanger}\ and\ \citenamefont
  {Hugonie}(2023)}]{Ellwanger:2023zjc}%
  \BibitemOpen
  \bibfield  {author} {\bibinfo {author} {\bibfnamefont {U.}~\bibnamefont
  {Ellwanger}}\ and\ \bibinfo {author} {\bibfnamefont {C.}~\bibnamefont
  {Hugonie}},\ }\bibfield  {title} {\bibinfo {title} {{Additional Higgs Bosons
  near 95 and 650 GeV in the NMSSM}},\ }\href
  {https://doi.org/10.1140/epjc/s10052-023-12315-y} {\bibfield  {journal}
  {\bibinfo  {journal} {Eur. Phys. J. C}\ }\textbf {\bibinfo {volume} {83}},\
  \bibinfo {pages} {1138} (\bibinfo {year} {2023})},\ \Eprint
  {https://arxiv.org/abs/2309.07838} {arXiv:2309.07838 [hep-ph]} \BibitemShut
  {NoStop}%
\bibitem [{\citenamefont {Cao}\ \emph {et~al.}(2024)\citenamefont {Cao},
  \citenamefont {Jia}, \citenamefont {Lian},\ and\ \citenamefont
  {Meng}}]{Cao:2023gkc}%
  \BibitemOpen
  \bibfield  {author} {\bibinfo {author} {\bibfnamefont {J.}~\bibnamefont
  {Cao}}, \bibinfo {author} {\bibfnamefont {X.}~\bibnamefont {Jia}}, \bibinfo
  {author} {\bibfnamefont {J.}~\bibnamefont {Lian}},\ and\ \bibinfo {author}
  {\bibfnamefont {L.}~\bibnamefont {Meng}},\ }\bibfield  {title} {\bibinfo
  {title} {{95~GeV diphoton and bb\textasciimacron{} excesses in the general
  next-to-minimal supersymmetric standard model}},\ }\href
  {https://doi.org/10.1103/PhysRevD.109.075001} {\bibfield  {journal} {\bibinfo
   {journal} {Phys. Rev. D}\ }\textbf {\bibinfo {volume} {109}},\ \bibinfo
  {pages} {075001} (\bibinfo {year} {2024})},\ \Eprint
  {https://arxiv.org/abs/2310.08436} {arXiv:2310.08436 [hep-ph]} \BibitemShut
  {NoStop}%
\bibitem [{\citenamefont {Ellwanger}\ \emph {et~al.}(2024)\citenamefont
  {Ellwanger}, \citenamefont {Hugonie}, \citenamefont {King},\ and\
  \citenamefont {Moretti}}]{Ellwanger:2024vvs}%
  \BibitemOpen
  \bibfield  {author} {\bibinfo {author} {\bibfnamefont {U.}~\bibnamefont
  {Ellwanger}}, \bibinfo {author} {\bibfnamefont {C.}~\bibnamefont {Hugonie}},
  \bibinfo {author} {\bibfnamefont {S.~F.}\ \bibnamefont {King}},\ and\
  \bibinfo {author} {\bibfnamefont {S.}~\bibnamefont {Moretti}},\ }\bibfield
  {title} {\bibinfo {title} {{NMSSM explanation for excesses in the search for
  neutralinos and charginos and a 95 GeV Higgs boson}},\ }\href
  {https://doi.org/10.1140/epjc/s10052-024-13129-2} {\bibfield  {journal}
  {\bibinfo  {journal} {Eur. Phys. J. C}\ }\textbf {\bibinfo {volume} {84}},\
  \bibinfo {pages} {788} (\bibinfo {year} {2024})},\ \Eprint
  {https://arxiv.org/abs/2404.19338} {arXiv:2404.19338 [hep-ph]} \BibitemShut
  {NoStop}%
\bibitem [{\citenamefont {Reichert}\ \emph {et~al.}(2018)\citenamefont
  {Reichert}, \citenamefont {Eichhorn}, \citenamefont {Gies}, \citenamefont
  {Pawlowski}, \citenamefont {Plehn},\ and\ \citenamefont
  {Scherer}}]{Reichert:2017puo}%
  \BibitemOpen
  \bibfield  {author} {\bibinfo {author} {\bibfnamefont {M.}~\bibnamefont
  {Reichert}}, \bibinfo {author} {\bibfnamefont {A.}~\bibnamefont {Eichhorn}},
  \bibinfo {author} {\bibfnamefont {H.}~\bibnamefont {Gies}}, \bibinfo {author}
  {\bibfnamefont {J.~M.}\ \bibnamefont {Pawlowski}}, \bibinfo {author}
  {\bibfnamefont {T.}~\bibnamefont {Plehn}},\ and\ \bibinfo {author}
  {\bibfnamefont {M.~M.}\ \bibnamefont {Scherer}},\ }\bibfield  {title}
  {\bibinfo {title} {{Probing baryogenesis through the Higgs boson
  self-coupling}},\ }\href {https://doi.org/10.1103/PhysRevD.97.075008}
  {\bibfield  {journal} {\bibinfo  {journal} {Phys. Rev. D}\ }\textbf {\bibinfo
  {volume} {97}},\ \bibinfo {pages} {075008} (\bibinfo {year} {2018})},\
  \Eprint {https://arxiv.org/abs/1711.00019} {arXiv:1711.00019 [hep-ph]}
  \BibitemShut {NoStop}%
\bibitem [{\citenamefont {Basler}\ \emph {et~al.}(2020)\citenamefont {Basler},
  \citenamefont {M\"uhlleitner},\ and\ \citenamefont
  {M\"uller}}]{Basler:2019iuu}%
  \BibitemOpen
  \bibfield  {author} {\bibinfo {author} {\bibfnamefont {P.}~\bibnamefont
  {Basler}}, \bibinfo {author} {\bibfnamefont {M.}~\bibnamefont
  {M\"uhlleitner}},\ and\ \bibinfo {author} {\bibfnamefont {J.}~\bibnamefont
  {M\"uller}},\ }\bibfield  {title} {\bibinfo {title} {{Electroweak Phase
  Transition in Non-Minimal Higgs Sectors}},\ }\href
  {https://doi.org/10.1007/JHEP05(2020)016} {\bibfield  {journal} {\bibinfo
  {journal} {JHEP}\ }\textbf {\bibinfo {volume} {05}},\ \bibinfo {pages}
  {016}},\ \Eprint {https://arxiv.org/abs/1912.10477} {arXiv:1912.10477
  [hep-ph]} \BibitemShut {NoStop}%
\bibitem [{\citenamefont {Biek\"otter}\ \emph {et~al.}(2023)\citenamefont
  {Biek\"otter}, \citenamefont {Heinemeyer}, \citenamefont {No}, \citenamefont
  {Olea-Romacho},\ and\ \citenamefont {Weiglein}}]{Biekotter:2022kgf}%
  \BibitemOpen
  \bibfield  {author} {\bibinfo {author} {\bibfnamefont {T.}~\bibnamefont
  {Biek\"otter}}, \bibinfo {author} {\bibfnamefont {S.}~\bibnamefont
  {Heinemeyer}}, \bibinfo {author} {\bibfnamefont {J.~M.}\ \bibnamefont {No}},
  \bibinfo {author} {\bibfnamefont {M.~O.}\ \bibnamefont {Olea-Romacho}},\ and\
  \bibinfo {author} {\bibfnamefont {G.}~\bibnamefont {Weiglein}},\ }\bibfield
  {title} {\bibinfo {title} {{The trap in the early Universe: impact on the
  interplay between gravitational waves and LHC physics in the 2HDM}},\ }\href
  {https://doi.org/10.1088/1475-7516/2023/03/031} {\bibfield  {journal}
  {\bibinfo  {journal} {JCAP}\ }\textbf {\bibinfo {volume} {03}},\ \bibinfo
  {pages} {031}},\ \Eprint {https://arxiv.org/abs/2208.14466} {arXiv:2208.14466
  [hep-ph]} \BibitemShut {NoStop}%
\bibitem [{\citenamefont {Dawson}\ \emph {et~al.}(1998)\citenamefont {Dawson},
  \citenamefont {Dittmaier},\ and\ \citenamefont {Spira}}]{Dawson:1998py}%
  \BibitemOpen
  \bibfield  {author} {\bibinfo {author} {\bibfnamefont {S.}~\bibnamefont
  {Dawson}}, \bibinfo {author} {\bibfnamefont {S.}~\bibnamefont {Dittmaier}},\
  and\ \bibinfo {author} {\bibfnamefont {M.}~\bibnamefont {Spira}},\ }\bibfield
   {title} {\bibinfo {title} {{Neutral Higgs boson pair production at hadron
  colliders: QCD corrections}},\ }\href
  {https://doi.org/10.1103/PhysRevD.58.115012} {\bibfield  {journal} {\bibinfo
  {journal} {Phys. Rev. D}\ }\textbf {\bibinfo {volume} {58}},\ \bibinfo
  {pages} {115012} (\bibinfo {year} {1998})},\ \Eprint
  {https://arxiv.org/abs/hep-ph/9805244} {arXiv:hep-ph/9805244} \BibitemShut
  {NoStop}%
\bibitem [{\citenamefont {de~Florian}\ and\ \citenamefont
  {Mazzitelli}(2013)}]{deFlorian:2013uza}%
  \BibitemOpen
  \bibfield  {author} {\bibinfo {author} {\bibfnamefont {D.}~\bibnamefont
  {de~Florian}}\ and\ \bibinfo {author} {\bibfnamefont {J.}~\bibnamefont
  {Mazzitelli}},\ }\bibfield  {title} {\bibinfo {title} {{Two-loop virtual
  corrections to Higgs pair production}},\ }\href
  {https://doi.org/10.1016/j.physletb.2013.06.046} {\bibfield  {journal}
  {\bibinfo  {journal} {Phys. Lett. B}\ }\textbf {\bibinfo {volume} {724}},\
  \bibinfo {pages} {306} (\bibinfo {year} {2013})},\ \Eprint
  {https://arxiv.org/abs/1305.5206} {arXiv:1305.5206 [hep-ph]} \BibitemShut
  {NoStop}%
\bibitem [{\citenamefont {Frederix}\ \emph {et~al.}(2014)\citenamefont
  {Frederix}, \citenamefont {Frixione}, \citenamefont {Hirschi}, \citenamefont
  {Maltoni}, \citenamefont {Mattelaer}, \citenamefont {Torrielli},
  \citenamefont {Vryonidou},\ and\ \citenamefont {Zaro}}]{Frederix:2014hta}%
  \BibitemOpen
  \bibfield  {author} {\bibinfo {author} {\bibfnamefont {R.}~\bibnamefont
  {Frederix}}, \bibinfo {author} {\bibfnamefont {S.}~\bibnamefont {Frixione}},
  \bibinfo {author} {\bibfnamefont {V.}~\bibnamefont {Hirschi}}, \bibinfo
  {author} {\bibfnamefont {F.}~\bibnamefont {Maltoni}}, \bibinfo {author}
  {\bibfnamefont {O.}~\bibnamefont {Mattelaer}}, \bibinfo {author}
  {\bibfnamefont {P.}~\bibnamefont {Torrielli}}, \bibinfo {author}
  {\bibfnamefont {E.}~\bibnamefont {Vryonidou}},\ and\ \bibinfo {author}
  {\bibfnamefont {M.}~\bibnamefont {Zaro}},\ }\bibfield  {title} {\bibinfo
  {title} {{Higgs pair production at the LHC with NLO and parton-shower
  effects}},\ }\href {https://doi.org/10.1016/j.physletb.2014.03.026}
  {\bibfield  {journal} {\bibinfo  {journal} {Phys. Lett. B}\ }\textbf
  {\bibinfo {volume} {732}},\ \bibinfo {pages} {142} (\bibinfo {year}
  {2014})},\ \Eprint {https://arxiv.org/abs/1401.7340} {arXiv:1401.7340
  [hep-ph]} \BibitemShut {NoStop}%
\bibitem [{\citenamefont {Borowka}\ \emph {et~al.}(2016)\citenamefont
  {Borowka}, \citenamefont {Greiner}, \citenamefont {Heinrich}, \citenamefont
  {Jones}, \citenamefont {Kerner}, \citenamefont {Schlenk}, \citenamefont
  {Schubert},\ and\ \citenamefont {Zirke}}]{Borowka:2016ehy}%
  \BibitemOpen
  \bibfield  {author} {\bibinfo {author} {\bibfnamefont {S.}~\bibnamefont
  {Borowka}}, \bibinfo {author} {\bibfnamefont {N.}~\bibnamefont {Greiner}},
  \bibinfo {author} {\bibfnamefont {G.}~\bibnamefont {Heinrich}}, \bibinfo
  {author} {\bibfnamefont {S.~P.}\ \bibnamefont {Jones}}, \bibinfo {author}
  {\bibfnamefont {M.}~\bibnamefont {Kerner}}, \bibinfo {author} {\bibfnamefont
  {J.}~\bibnamefont {Schlenk}}, \bibinfo {author} {\bibfnamefont
  {U.}~\bibnamefont {Schubert}},\ and\ \bibinfo {author} {\bibfnamefont
  {T.}~\bibnamefont {Zirke}},\ }\bibfield  {title} {\bibinfo {title} {{Higgs
  Boson Pair Production in Gluon Fusion at Next-to-Leading Order with Full
  Top-Quark Mass Dependence}},\ }\href
  {https://doi.org/10.1103/PhysRevLett.117.079901} {\bibfield  {journal}
  {\bibinfo  {journal} {Phys. Rev. Lett.}\ }\textbf {\bibinfo {volume} {117}},\
  \bibinfo {pages} {012001} (\bibinfo {year} {2016})},\ \bibinfo {note}
  {[Erratum: Phys.Rev.Lett. 117, 079901 (2016)]},\ \Eprint
  {https://arxiv.org/abs/1604.06447} {arXiv:1604.06447 [hep-ph]} \BibitemShut
  {NoStop}%
\bibitem [{\citenamefont {Baglio}\ \emph {et~al.}(2019)\citenamefont {Baglio},
  \citenamefont {Campanario}, \citenamefont {Glaus}, \citenamefont
  {M\"uhlleitner}, \citenamefont {Spira},\ and\ \citenamefont
  {Streicher}}]{Baglio:2018lrj}%
  \BibitemOpen
  \bibfield  {author} {\bibinfo {author} {\bibfnamefont {J.}~\bibnamefont
  {Baglio}}, \bibinfo {author} {\bibfnamefont {F.}~\bibnamefont {Campanario}},
  \bibinfo {author} {\bibfnamefont {S.}~\bibnamefont {Glaus}}, \bibinfo
  {author} {\bibfnamefont {M.}~\bibnamefont {M\"uhlleitner}}, \bibinfo {author}
  {\bibfnamefont {M.}~\bibnamefont {Spira}},\ and\ \bibinfo {author}
  {\bibfnamefont {J.}~\bibnamefont {Streicher}},\ }\bibfield  {title} {\bibinfo
  {title} {{Gluon fusion into Higgs pairs at NLO QCD and the top mass
  scheme}},\ }\href {https://doi.org/10.1140/epjc/s10052-019-6973-3} {\bibfield
   {journal} {\bibinfo  {journal} {Eur. Phys. J. C}\ }\textbf {\bibinfo
  {volume} {79}},\ \bibinfo {pages} {459} (\bibinfo {year} {2019})},\ \Eprint
  {https://arxiv.org/abs/1811.05692} {arXiv:1811.05692 [hep-ph]} \BibitemShut
  {NoStop}%
\bibitem [{\citenamefont {Baglio}\ \emph {et~al.}(2021)\citenamefont {Baglio},
  \citenamefont {Campanario}, \citenamefont {Glaus}, \citenamefont
  {M\"uhlleitner}, \citenamefont {Ronca},\ and\ \citenamefont
  {Spira}}]{Baglio:2020wgt}%
  \BibitemOpen
  \bibfield  {author} {\bibinfo {author} {\bibfnamefont {J.}~\bibnamefont
  {Baglio}}, \bibinfo {author} {\bibfnamefont {F.}~\bibnamefont {Campanario}},
  \bibinfo {author} {\bibfnamefont {S.}~\bibnamefont {Glaus}}, \bibinfo
  {author} {\bibfnamefont {M.}~\bibnamefont {M\"uhlleitner}}, \bibinfo {author}
  {\bibfnamefont {J.}~\bibnamefont {Ronca}},\ and\ \bibinfo {author}
  {\bibfnamefont {M.}~\bibnamefont {Spira}},\ }\bibfield  {title} {\bibinfo
  {title} {{$gg\to HH$ : Combined uncertainties}},\ }\href
  {https://doi.org/10.1103/PhysRevD.103.056002} {\bibfield  {journal} {\bibinfo
   {journal} {Phys. Rev. D}\ }\textbf {\bibinfo {volume} {103}},\ \bibinfo
  {pages} {056002} (\bibinfo {year} {2021})},\ \Eprint
  {https://arxiv.org/abs/2008.11626} {arXiv:2008.11626 [hep-ph]} \BibitemShut
  {NoStop}%
\bibitem [{\citenamefont {Agostini}\ \emph {et~al.}(2016)\citenamefont
  {Agostini}, \citenamefont {Degrassi}, \citenamefont {Gr\"ober},\ and\
  \citenamefont {Slavich}}]{Agostini:2016vze}%
  \BibitemOpen
  \bibfield  {author} {\bibinfo {author} {\bibfnamefont {A.}~\bibnamefont
  {Agostini}}, \bibinfo {author} {\bibfnamefont {G.}~\bibnamefont {Degrassi}},
  \bibinfo {author} {\bibfnamefont {R.}~\bibnamefont {Gr\"ober}},\ and\
  \bibinfo {author} {\bibfnamefont {P.}~\bibnamefont {Slavich}},\ }\bibfield
  {title} {\bibinfo {title} {{NLO-QCD corrections to Higgs pair production in
  the MSSM}},\ }\href {https://doi.org/10.1007/JHEP04(2016)106} {\bibfield
  {journal} {\bibinfo  {journal} {JHEP}\ }\textbf {\bibinfo {volume} {04}},\
  \bibinfo {pages} {106}},\ \Eprint {https://arxiv.org/abs/1601.03671}
  {arXiv:1601.03671 [hep-ph]} \BibitemShut {NoStop}%
\bibitem [{\citenamefont {Dawson}\ \emph {et~al.}(1996)\citenamefont {Dawson},
  \citenamefont {Djouadi},\ and\ \citenamefont {Spira}}]{Dawson:1996xz}%
  \BibitemOpen
  \bibfield  {author} {\bibinfo {author} {\bibfnamefont {S.}~\bibnamefont
  {Dawson}}, \bibinfo {author} {\bibfnamefont {A.}~\bibnamefont {Djouadi}},\
  and\ \bibinfo {author} {\bibfnamefont {M.}~\bibnamefont {Spira}},\ }\bibfield
   {title} {\bibinfo {title} {{QCD corrections to SUSY Higgs production: The
  Role of squark loops}},\ }\href {https://doi.org/10.1103/PhysRevLett.77.16}
  {\bibfield  {journal} {\bibinfo  {journal} {Phys. Rev. Lett.}\ }\textbf
  {\bibinfo {volume} {77}},\ \bibinfo {pages} {16} (\bibinfo {year} {1996})},\
  \Eprint {https://arxiv.org/abs/hep-ph/9603423} {arXiv:hep-ph/9603423}
  \BibitemShut {NoStop}%
\bibitem [{\citenamefont {Harlander}\ and\ \citenamefont
  {Steinhauser}(2003)}]{Harlander:2003kf}%
  \BibitemOpen
  \bibfield  {author} {\bibinfo {author} {\bibfnamefont {R.}~\bibnamefont
  {Harlander}}\ and\ \bibinfo {author} {\bibfnamefont {M.}~\bibnamefont
  {Steinhauser}},\ }\bibfield  {title} {\bibinfo {title} {{Effects of SUSY QCD
  in hadronic Higgs production at next-to-next-to-leading order}},\ }\href
  {https://doi.org/10.1103/PhysRevD.68.111701} {\bibfield  {journal} {\bibinfo
  {journal} {Phys. Rev. D}\ }\textbf {\bibinfo {volume} {68}},\ \bibinfo
  {pages} {111701} (\bibinfo {year} {2003})},\ \Eprint
  {https://arxiv.org/abs/hep-ph/0308210} {arXiv:hep-ph/0308210} \BibitemShut
  {NoStop}%
\end{thebibliography}%

\end{document}